\documentclass[english]{article}
\usepackage[T1]{fontenc}
\usepackage[latin1]{inputenc}
\usepackage{geometry}
\usepackage{amsmath}
\usepackage{amssymb}

\makeatletter

\providecommand{\LyX}{L\kern-.1667em\lower.25em\hbox{Y}\kern-.125emX\@}

\newcounter{draft}
\usepackage{amssymb} 
\usepackage{ifthen}
\ifthenelse{\thedraft=1}{\usepackage[notref,notcite]{showkeys}}{}

\usepackage{hyperref}
\usepackage{cite} 

\newcommand{\Ram}[2]{\centerline{\framebox[#1\textwidth]{\parbox{\textwidth}{#2}}}}
\newcommand{\Ramm}[2]{\framebox[#1\textwidth]{\parbox{\textwidth}{$#2$}}}

\numberwithin{equation}{section}

\newcommand{\bref}[1]{\def\theequation{\ref{#1}={\bf \thesection}.\arabic{equation}}}
\newcommand{\eref}{\def\theequation{{\bf \thesection}.\arabic{equation}}}
\newcommand{\varLabel}[1]{\newcounter{#1}\setcounter{#1}{\arabic{equation}}
       \newcounter{#1Sec}\setcounter{#1Sec}{\thesection}\addtocounter{#1}{1}
       \hypertarget{#1}{}}
\newcommand{\varRef}[1]{\hyperlink{#1}{{\bf\arabic{#1Sec}}.\arabic{#1}}} 

\newcommand{\rem}[1]{\ifthenelse{\thedraft=1}{\textsc{\texttt{<< #1 >>}}}{}}

\newcommand{\bmul}{\begin{multicols}{2}[\vspace{-1cm}]}
\newcommand{\emul}{\end{multicols}\vspace{-0.5cm}}

\date{\rem{\today}}
\newcommand{\Teil}[2]{#2} 

\usepackage{babel}
\makeatother
\begin{document}
\Teil{Macros}{
\newcommand{\bs}[1]{\boldsymbol{#1}}

\newcommand{\mf}[1]{\mathfrak{#1} }

\newcommand{\mc}[1]{\mathcal{#1}}

\newcommand{\norm}[1]{{\parallel#1 \parallel}}

\newcommand{\Norm}[1]{\left\Vert #1 \right\Vert }

\newcommand{\partiell}[2]{\frac{\partial#1 }{\partial#2 }}

\newcommand{\Partiell}[2]{\left( \frac{\partial#1 }{\partial#2 }\right) }

\newcommand{\ola}[1]{\overleftarrow{#1}}

\newcommand{\lpartial}{\overleftarrow{\partial}}

\newcommand{\partl}[1]{\frac{\partial}{\partial#1}}

\newcommand{\partr}[1]{\frac{\lpartial}{\partial#1}}

\newcommand{\funktional}[2]{\frac{\delta#1 }{\delta#2 }}

\newcommand{\funktl}[1]{\frac{\delta}{\delta#1}}

\newcommand{\funktr}[1]{\frac{\ola{\delta}}{\delta#1}}

\newcommand{\de}{{\bf d}\!}

\newcommand{\es}{{\bf s}\!}

\newcommand{\dew}{\bs{d}^{\textrm{w}}\!}

\newcommand{\Lie}{\bs{\mc{L}}}

\newcommand{\Dorf}{\bs{\mc{D}}}

\newcommand{\pe}{\bs{\partial}}

\newcommand{\De}{\textrm{D}\!}

\newcommand{\total}[2]{\frac{\de#1 }{\de#2 }}

\newcommand{\Frac}[2]{\left( \frac{#1 }{#2 }\right) }

\newcommand{\To}{\rightarrow}
 
\newcommand{\ket}[1]{|#1 >}

\newcommand{\bra}[1]{<#1 |}

\newcommand{\Ket}[1]{\left| #1 \right\rangle }

\newcommand{\Bra}[1]{\left\langle #1 \right| }
 
\newcommand{\braket}[2]{<#1 |#2 >}

\newcommand{\Braket}[2]{\Bra{#1 }\left. #2 \right\rangle }

\newcommand{\kom}[2]{[#1 ,#2 ]}

\newcommand{\Kom}[2]{\left[ #1 ,#2 \right] }

\newcommand{\abs}[1]{\mid#1 \mid}

\newcommand{\Abs}[1]{\left| #1 \right| }

\newcommand{\erw}[1]{\langle#1\rangle}

\newcommand{\Erw}[1]{\left\langle #1 \right\rangle }

\newcommand{\bei}[2]{\left. #1 \right| _{#2 }}

\newcommand{\dann}{\Rightarrow}

\newcommand{\q}[1]{\underline{#1 }}

\newcommand{\hoch}[1]{{}^{#1 }}

\newcommand{\tief}[1]{{}_{#1 }}

\newcommand{\lqn}[1]{\lefteqn{#1}}

\newcommand{\os}[2]{\overset{\lqn{#1}}{#2}}

\newcommand{\us}[2]{\underset{\lqn{{\scriptstyle #2}}}{#1}}

\newcommand{\ous}[3]{\underset{#3}{\os{#1}{#2}}}

\newcommand{\zwek}[2]{\begin{array}{c}
#1\\
#2\end{array}}

\newcommand{\drek}[3]{\begin{array}{c}
#1\\
#2\\
#3\end{array}}

\newcommand{\UB}[2]{\underbrace{#1 }_{\le{#2 }}}

\newcommand{\OB}[2]{\overbrace{#1 }^{\le{#2 }}}

\newcommand{\tr}{\textrm{tr}\,}

\newcommand{\Tr}{\textrm{Tr}\,}

\newcommand{\Det}{\textrm{Det}\,}

\newcommand{\diag}{\textrm{diag}\,}

\newcommand{\Diag}{\textrm{Diag}\,}

\newcommand{\one}{1\!\!1}

\newcommand{\fussend}{\diamond}

\newcommand{\eps}{\varepsilon}

\newcommand{\dali}{\Box}
\newcommand{\choice}[2]{\ifthenelse{\thechoice=1}{#1}{\ifthenelse{\thechoice=2}{#2}{\left\{  \begin{array}{c}

 #1\\
#2\end{array}\right\}  }}}
 
\newcommand{\lchoice}[2]{\ifthenelse{\thechoice=1}{#1}{\ifthenelse{\thechoice=2}{#2}{\left\{  \begin{array}{c}

 #1\\
#2\end{array}\right.}}}
 
\newcommand{\lcsign}{\ifthenelse{\thechoice=1}{+}{\ifthenelse{\thechoice=2}{-}{\pm}}}
 
\newcommand{\lcmsign}{\ifthenelse{\thechoice=1}{-}{\ifthenelse{\thechoice=2}{+}{\mp}}}

\newcommand{\lcconst}{c}
{}
 
\newcommand{\weyl}{\alpha}

\newcommand{\greq}{=_{g}}

\newcommand{\grequiv}{\equiv_{g}}

\newcommand{\grdef}{:=_{g}}

\newcommand{\Greq}{=_{G}}

\newcommand{\Grequiv}{\equiv_{G}}

\newcommand{\Grdef}{:=_{G}}

\newcommand{\Ggreq}{=_{Gg}}

\newcommand{\Greqornot}{=_{(G)}}

\newcommand{\Grequivornot}{\equiv_{(G)}}

\newcommand{\Grdefornot}{:=_{(G)}}

\newcommand{\greqornot}{=_{(g)}}

\newcommand{\grequivornot}{\equiv_{(g)}}

\newcommand{\grdefornot}{:=_{(g)}}

\newcommand{\fatkomma}{\textrm{{\bf ,}}}

\newcommand{\basis}{\boldsymbol{\mf{t}}}

\newcommand{\ip}{\imath}

\newcommand{\Beta}{\textrm{\Large$\beta$}}

\newcommand{\sBeta}{\textrm{\large$\beta$}}

\newcommand{\be}{\bs{b}}

\newcommand{\ce}{\bs{c}}

\newcommand{\Q}{\bs{Q}}

\newcommand{\mm}{\bs{m}\ldots\bs{m}}

\newcommand{\nn}{\bs{n}\ldots\bs{n}}

\newcommand{\kk}{k\ldots k}

\newcommand{\OO}{\bs{\Omega}}

\newcommand{\oo}{\bs{o}}

\newcommand{\tet}{\bs{\theta}}

\newcommand{\Es}{\bs{S}}

\newcommand{\Ce}{\bs{C}}

\newcommand{\lam}{\bs{\lambda}}

\newcommand{\ro}{\bs{\rho}}

\newcommand{\cov}{\textrm{D}_{\tet}}

\newcommand{\feps}{\bs{\eps}}

\newcommand{\qu}{\textrm{Q}_{\tet}}

\newcommand{\dimw}{d_{\textrm{w}}}

\newcommand{\mteta}{\mu(\tet)}

\newcommand{\msig}{d^{\lqn{\hoch{\dimw}}}\sigma}

\newcommand{\msigp}{d^{\hoch{{\scriptscriptstyle \dimw}\lqn{{\scriptscriptstyle -1}}}}\sigma}

\newcommand{\backtilde}{\!\!\tilde{}\,\,}

\title{Brackets, Sigma Models and Integrability of Generalized Complex
Structures}

\author{\begin{picture}(0,0)\unitlength=1mm\put(80,40){SPhT-T06/095}\put(80,35){TUW-06-06}\end{picture} \textbf{Sebastian
Guttenberg}%
\thanks{basti@hep.itp.tuwien.ac.at%
}\emph{\vspace{.5cm}} \emph{}\\
\emph{Service de Physique Th\'eorique, CEA/Saclay}\\
\emph{91191 Gif-sur-Yvette Cedex, France\vspace{.5cm}} \emph{}\\
\emph{Institut f\"ur Theoretische Physik, Technische Universit\"at
Wien,}\\
\emph{1040 Wien, Austria}}

\maketitle
\begin{abstract}
It is shown how derived brackets naturally arise in sigma-models via
Poisson- or antibracket\rem{or by the quantum-commutator}, generalizing
a recent observation by Alekseev and Strobl. On the way to a precise
formulation of this relation, an explicit coordinate expression for
the derived bracket is obtained. The generalized Nijenhuis tensor
of generalized complex geometry is shown to coincide up to a de-Rham
closed term with the derived bracket of the structure with itself
and a new coordinate expression for this tensor is presented. The
insight is applied to two known two-dimensional sigma models in a
background with generalized complex structure. Introductions to geometric
brackets on the one hand and to generalized complex geometry on the
other hand are given in the appendix. \noindent \vspace{2cm}
\end{abstract}
}

\Teil{A}{\rem{To do:\\
Zitate: Michael?\\
 H-twisting\\
Quantum-(J,J)-bracket\\
r-wedge <-> $\ip^{(p,q)}$\\
(pure) Spinors\\
Zucc: missing last condition\\
Hull: foot:dual-coord}\newpage 

\tableofcontents{}\newpage\eref

\section{Introduction}

There are quite a lot of different geometric brackets floating around
in the literature, like Schouten bracket, Nijenhuis bracket or in
generalized complex geometry the Dorfman bracket and Courant bracket,
to list just some of them. They are often related to integrability
conditions for some structures on manifolds. The vanishing of the
Nijenhuis bracket of a complex structure with itself, for example,
is equivalent to its integrability. The same is true for the Schouten
bracket and a Poisson structure. The above brackets can be unified
with the concept of derived brackets \cite{Kosmann-Schwarzbach:2003en}.
Within this concept, they are all just natural extensions of the Lie-bracket
of vector fields to higher rank tensor fields. 

It is well known that the antibracket appearing in the Lagrangian
formalism for sigma models is closely related to the Schouten-bracket
in target space. In addition it was recently observed by Alekseev
and Strobl that the Dorfman bracket for sums of vectors and one-forms
appears naturally in two dimensional sigma models%
\footnote{In \cite{Alekseev:2004np}, the non-symmetric bracket is called 'Courant
bracket'. Following e.g. Gualtieri \cite{Gualtieri:007} or \cite{Kosmann-Schwarzbach:2003en},
it will be called 'Dorfman bracket' in this paper, while 'Courant
bracket' is reserved for its antisymmetrization (see (\ref{eq:Dorfman-bracket})
and (\ref{eq:Courant-bracket})).$\quad\fussend$%
} \cite{Alekseev:2004np}. This was generalized by Bonelli and Zabzine
\cite{Bonelli:2005ti} to a derived bracket for sums of vectors and
$p$-forms on a $p$-brane%
\footnote{The \emph{Vinogradov bracket} appearing in \cite{Bonelli:2005ti}
is just the antisymmetrization of a derived bracket (see footnote\rem{spaeter vielleicht nimmer}
\ref{Vinogradov-bracket} on page \pageref{Vinogradov-bracket}).$\qquad\fussend$%
}. These observations lead to the natural question whether there is
a general relation between the sigma-model Poisson bracket or antibracket
and derived brackets in target space. Working out the precise relation
for sigma models with a special field content but undetermined dimension
and dynamics, is the major subject of the present paper.

One of the motivations for this article, was the application to generalized
complex geometry. The importance of the latter in string theory is
due to the observation that effective spacetime supersymmetry after
compactification requires the compactification manifold to be a generalized
Calabi-Yau manifold \cite{Hitchin:2004ut,Gualtieri:007,Grana:2004bg,Grana:2004??,Grana:2005ny,Grana:2005jc}.
Deviations from an ordinary Calabi Yau manifold are due to fluxes
and also the concept of mirror symmetry can be generalized in this
context. There are numerous other important articles on the subject,
like e.g. \cite{Kapustin:2004gv,Pestun:2005rp,Pestun:2006rj,Jeschek:2004je,Jeschek:2005ek}
and many more. A more complete list of references can be found in
\cite{Grana:2005jc}. A major part of the considerations so far was
done from the supergravity point of view. Target space supersymmetry
is, however, related to an $N=2$ supersymmetry on the worldsheet.
For this reason the relation between an extended worldsheet supersymmetry
and the presence of an integrable generalized complex structure (GCS)
was studied in \cite{Lindstrom:2004iw} (the reviews \cite{Zabzine:2006uz,Lindstrom:2006ee}
on generalized complex geometry have this relation in mind). Zabzine
clarified in \cite{Zabzine:2005qf} the relation in a model independent
way in a Hamiltonian description and showed that the existence of
a second non-manifest worldsheet supersymmetry $\Q_{2}$ in an $N=1$
sigma-model is equivalent to the existence of an integrable GCS $\mc{J}$.
It is the observation that the integrability of the GCS $\mc{J}$
can be written as the vanishing of a generalized bracket $\left[\mc{J}\bs{,}\mc{J}\right]_{B}=0$
which leads to the natural question, whether there is a direct mapping
between $\left[\mc{J},\mc{J}\right]_{B}=0\,\&\,\mc{J}^{2}=-1$ on
the one side and $\{\Q_{2},\Q_{2}\}=2P$ on the other side. This will
be a natural application in subsection \ref{sub:Zabzine} of the more
general preceding considerations about the relation between (super-)Poisson
brackets in sigma models with special field content and derived brackets
in the target space. 

A second interesting application is Zucchini's Hitchin-sigma-model
\cite{Zucchini:2004ta}. There are two more papers on that subject
\cite{Zucchini:2005rh,Zucchini:2005cq}\rem{noch ein ganz neues!},
but the present discussion refers only to the first one. Zucchini's
model is a two dimensional sigma-model in a target space with a generalized
complex structure (GCS). The sigma-model is topological when the GCS
is integrable, while the inverse does not hold. The condition for
the sigma model to be topological is the master equation $(S\bs{,}S)=0$.
Again we might wonder whether there is a direct mapping between the
antibracket and $S$ on the one hand and the geometric bracket and
$\mc{J}$ on the other hand and it will be shown in subsection \ref{sub:Zucchini}
how this mapping works as an application of the considerations in
subsection \ref{sub:antibracket}. In order to understand more about
geometric brackets in general, however, it was necessary to dive into
\rem{??} Kosmann-Schwarzbach's review on derived brackets \cite{Kosmann-Schwarzbach:2003en}
which led to observations that go beyond the application to the integrability
of a GCS . \rem{An introduction into some aspects of derived brackets, based on , is given in the appendix .}

The structure of the paper is as follows: The general relation between
sigma models and derived brackets in target space will be studied
in the next section. The necessary geometric setup will be established
in \ref{sub:bc-phase-space}. Although there are no new deep insights
in \ref{sub:bc-phase-space}, the unconventional idea to extend the
exterior derivative on forms to multivector valued forms (see (\ref{eq:dK-coord})
and (\ref{eq:d-auf-partial})) will provide a tool to write down a
coordinate expression for the general derived bracket between multivector
valued forms (\ref{eq:bc-derived-bracket-coord}) which to my knowledge
does not yet exist in literature. The main results in section \ref{sec:sigma-model-induced},
however, are the propositions 1 on page \pageref{eq:Proposition1}
and 1b on page \pageref{eq:Proposition1b} for the relation between
the Poisson-bracket in a sigma-model with special field content and
the derived bracket in the target space, and the proposition 3b on
page \pageref{eq:PropositionIIIb} for the relation between the antibracket
in a sigma-model and the derived bracket in target space. Proposition
2 on page \pageref{eq:Proposition2} is just a short quantum consideration
which only works for the particle case. In section \ref{sec:Applications-in-string}
the propositions 1b and 3b are finally applied to the two examples
which were mentioned above. 

Another result is the relation between the generalized Nijenhuis tensor
and the derived bracket of $\mc{J}$ with itself, given in (\ref{eq:relation-between-derived-and-Nij-Tens-in-main-part}).
The derivation of this can be found in the appendix on page \pageref{sub:Derivation-via-derived-bracket}.
In addition to this, there is a new coordinate form of the generalized
Nijenhuis tensor presented in (\ref{eq:generalized-integrabilityIII})
on page \pageref{eq:generalized-integrabilityIII}, which might be
easier to memorize than the known ones. There is also a short comment
in footnote \ref{foot:dual-coord} on page \pageref{foot:dual-coord}
on a possible relation to Hull's doubled geometry. 

Appendix \ref{sec:Conventions} summarizes the used conventions, while
appendix \ref{sec:bracket-review} is an introduction to geometric
brackets. Finally, appendix \ref{sec:Generalized-complex-geometry}
provides some aspects of generalized complex geometry which might
be necessary to understand the two applications of above. 

\rem{Hier waere noch eine section ueber normale komplexe Struktur als Motivation. Probleme mit der Leibnizregel...}

\section{Sigma-model-induced brackets}

\label{sec:sigma-model-induced}

\subsection{Geometric brackets in phase space formulation}

\label{sub:bc-phase-space} In the following some basic geometric
ingredients which are necessary to formulate derived brackets will
be given. Although there is no sigma model and no physics explicitly
involved in this first subsection, the presentation and the techniques
will be very suggestive, s.th. there is visually no big change when
we proceed after that with considerations on sigma-models.

\subsubsection{Algebraic brackets}

\label{sub:Algebraic-brackets} Consider a real differentiable manifold
$M$. The interior product with a vector field $v=v^{k}\pe_{k}$ (in
a local coordinate basis) acting on a differential form $\rho$ is
a differential operator in the sense that it differentiates with respect
to the basis elements of the cotangent space:%
\footnote{Note, that a convention is used, were the prefactor $\frac{1}{r!}$
which usually comes along with an $r$-form is absorbed into the definition
of the wedge-product. The common conventions can for all equations
easily be recovered by redefining all coefficients appropriately,
e.g. $\rho_{m_{1}\ldots m_{r}}\To\frac{1}{r!}\rho_{m_{1}\ldots m_{r}}.\qquad\fussend$%
}\begin{eqnarray}
\ip_{v}\rho^{(r)} & = & r\cdot v^{k}\rho_{km_{1}\ldots m_{r-1}}^{(r)}(x)\:\de x^{m_{1}}\cdots\de x^{m_{r-1}}=v^{k}\partl{(\de x^{k})}\left(\rho_{m_{1}\ldots m_{r}}\de x^{m_{1}}\cdots\de x^{m_{r}}\right)\end{eqnarray}
Let us rename%
\footnote{The similarity with ghosts is of course no accident. It is well known
(see e.g. \cite{Henneaux:1992ig}) that ghosts in a gauge theory can
be seen as 1-forms dual to the gauge-vector fields and the BRST differential
as the sum of the Koszul-Tate differential (whose homology implements
the restriction to the constraint surface) and the longitudinal exterior
derivative along the constraint surface. In that sense the present
description corresponds to a topological theory, where all degrees
of freedom are gauged away. But we will not necessarily always view
$\bs{c}^{m}$ as ghosts in the following. So let us in the beginning
see $\bs{c}^{m}$ just as another name for $\de x^{m}$. We do not
yet assume an underlying sigma-model, i.e. $\bs{b}_{m}$ and $\bs{c}^{m}$
do not necessarily depend on a worldsheet variable.$\quad\fussend$%
} \begin{eqnarray}
\ce^{m} & \equiv & \de x^{m}\\
\be_{m} & \equiv & \pe_{m}\end{eqnarray}
The vector $v$ takes locally the form $v=v^{m}\be_{m}$ and when
we introduce a canonical graded Poisson bracket between $\ce^{m}$
and $\be_{m}$ via $\left\{ \be_{m},\ce^{n}\right\} =\delta_{m}^{n}$
, we get \begin{eqnarray}
\ip_{v}\rho & = & \left\{ v,\rho\right\} \end{eqnarray}
\rem{Es gilt auch $\ip_{v}K=\left\{ v,K\right\} $, aber i.a. $\ip_{v}T\neq\left\{ v,T\right\} $,
wenn $T$ $p$ enthaelt. Ferner gilt fuer 1-Formen $\left\{ \omega,\rho\right\} =\ip_{\omega}^{(1)}\rho=0\neq\ip_{\omega}^{(0)}\rho=\ip_{\omega}\rho$.
Entsprechend gilt fuer generalized vectors $\left\{ \mf{a},\rho\right\} =\ip_{\mf{a}}^{(1)}\rho=\ip_{a}\rho\neq\ip_{\mf{a}}\rho$
und $\left\{ \mf{a},K\right\} =\ip_{\mf{a}}^{(1)}K-(-)^{k-k'}\ip_{K}^{(1)}\mf{a}=\ip_{a}^{(1)}K-(-)^{k-k'}\ip_{K}^{(1)}\alpha$} 
Extending also the local $x$-coordinate-space to a phase space by
introducing the conjugate momentum $p_{m}$ (whose geometric interpretation
we will discover soon), we have altogether the (graded) Poisson bracket
\begin{eqnarray}
\left\{ \be_{m},\ce^{n}\right\}  & = & \delta_{m}^{n}=\left\{ \ce^{n},\be_{m}\right\} \label{eq:Poisson-bracket-bc}\\
\left\{ p_{m},x^{n}\right\}  & = & \delta_{m}^{n}=-\left\{ x^{n},p_{m}\right\} \label{eq:Poisson-bracket-xp}\\
\left\{ A,B\right\}  & = & A\partr{\be_{k}}\partl{\ce^{k}}B+A\partr{p_{k}}\partl{x^{k}}B-(-)^{AB}\left(B\partr{\be_{k}}\partl{\ce^{k}}A+B\partr{p_{k}}\partl{x^{k}}A\right)\label{eq:Poisson-bracket}\end{eqnarray}
and can write the exterior derivative acting on forms as generated
via the Poisson-bracket by an odd phase-space function $\oo(\ce,p)$
\begin{eqnarray}
\oo & \equiv & \oo(\ce,p)\equiv\ce^{k}p_{k}\label{eq:BRST-op}\\
\left\{ \oo,\rho^{(r)}\right\}  & = & \ce^{k}\left\{ p_{k},\rho_{m_{1}\ldots m_{r}}(x)\right\} \ce^{m_{1}}\cdots\ce^{m_{r}}=\de\rho^{(r)}\label{eq:exterior-derivative-via-BRST}\end{eqnarray}
The variables $\ce^{m}$,$\be_{m}$,$x^{m}$ and $p_{m}$ can be seen
as coordinates of $T^{*}(\Pi TM)$, the cotangent bundle of the tangent
bundle with parity inversed fiber. \rem{$\textrm{Fun}\left(\Pi TM\right)=\Gamma\left(\bigwedge^{\bullet}T^{*}M\right)$,
$\textrm{Fun}\left(T^{*}(\Pi TM)\right)\stackrel{?}{=}\Gamma\left(\textrm{End}(\Omega^{\bullet}(M))\right)$.
In the present article, however, we will stick to the interpretation
of $\ce^{m}$ and $\be_{m}$ as coordinate basis elements of $T^{*}M$
and $TM$ and we will assign an interpretation of $p_{m}$ in \ref{sub:Extended-exterior-derivative}
only via an embedding into the space of differential operators acting
on forms.}

\subsubsection*{Interior product and {}``quantization''}

Given a multivector valued form $K^{(k,k')}$ of form degree $k$
and multivector degree $k'$, it reads in the local coordinate patch
with the new symbols \begin{eqnarray}
K^{(k,k')}\equiv K^{(k,k')}(x,\ce,\be) & \equiv & K_{m_{1}\ldots m_{k}}\hoch{n_{1}\ldots n_{k'}}(x)\,\bs{c}^{m_{1}}\cdots\ce^{m_{k}}\be_{n_{1}}\cdots\be_{n_{k'}}\equiv K_{\mm}\hoch{\nn}\label{eq:multivector-valued-form-K}\end{eqnarray}
The notation $K(x,\ce,\be)$ should stress, that $K$ is locally a
(smooth on a $C^{\infty}$ manifold) function of the phase space variables
which will later be used for analytic continuation ($x$ will be allowed
to take c-number values of a superfunction). The last expression in
the above equation introduces a \textbf{schematic index notation}
which is useful to write down the explicit coordinate form for lengthy
expressions. See in the appendix \ref{fat-index} at page \pageref{fat-index}
for a more detailed description of its definition. It should, however,
be self-explanatory enough for a first reading of the article

One can define a natural generalization of the interior product with
a vector $\ip_{v}$ to an \textbf{interior product} with a multivector
valued form $\ip_{K}$ acting on some $r$-form (in fact, it is more
like a combination of an interior and an exterior product -- see footnote
\ref{foot:interior-product} on page \pageref{foot:interior-product}\rem{Anhang}
--, but we will stick to this name)\begin{eqnarray}
\ip_{K^{(k,k')}}\rho^{(r)} & \equiv & (k')!\left(\zwek{r}{k'}\right)K_{\bs{m}\ldots\bs{m}}\hoch{l_{1}\ldots l_{k'}}\rho_{\underbrace{{\scriptstyle l_{k'}\ldots l_{1}\bs{m}\ldots\bs{m}}}_{r}}=\label{eq:bc-interior-product}\\
 & = & K_{m_{1}\ldots m_{k}}\hoch{n_{1}\ldots n_{k'}}\bs{c}^{m_{1}}\cdots\ce^{m_{k}}\left\{ \be_{n_{1}},\left\{ \cdots,\left\{ \be_{n_{k'}},\rho^{(r)}\right\} \right\} \right\} \label{eq:bc-interior-product-I}\\
 & = & K_{m_{1}\ldots m_{k}}\hoch{n_{1}\ldots n_{k'}}\bs{c}^{m_{1}}\cdots\ce^{m_{k}}\partl{\ce^{n_{1}}}\cdots\partl{\ce^{n_{k'}}}\rho^{(r)}\label{eq:bc-interior-product-II}\end{eqnarray}
It is a derivative of order $k'$ and thus not a derivative in the
usual sense like $\ip_{v}$. The third line shows the reason for the
normalization of the first line, while the second line is added for
later convenience. The interior product is commonly used as an \textbf{embedding}
of the multivector valued forms in the space of differential operators
acting on forms, i.e. $K\To\ip_{K}$, s.th. structures of the latter
can be induced on the space of multivector valued forms. In (\ref{eq:bc-interior-product-II})
the interior product $\ip_{K}$ can be seen, up to a factor of $\hbar/i$,
as the quantum operator corresponding to $K$, where the form $\rho$
plays the role of a wave function. The natural ordering is here to
put the conjugate momenta to the right. We can therefore fix the following
{}``\textbf{quantization}'' rule (corresponding to $\hat{\be}=\frac{\hbar}{i}\partl{\ce}$)\begin{eqnarray}
\hat{K}^{(k,k')} & \equiv & \left(\frac{\hbar}{i}\right)^{k'}\ip_{K^{(k,k')}}\label{eq:quantization}\\
\textrm{with }\ip_{K^{(k,k')}} & = & K_{\mm}\hoch{n_{1}\ldots n_{k'}}\frac{\partial^{k'}}{\partial\ce^{n_{1}}\cdots\partial\ce^{n_{k'}}}\label{eq:bc-interior-product-III}\end{eqnarray}
 The (graded) commutator of two interior products induces an algebraic
bracket due to Buttin \cite{Buttin:1974}\rem{(see also in the appendix,
in subsection \ref{sub:multivector-form-brackets})}, which is defined
via\varLabel{bcAlgebraicBracket}
\begin{eqnarray}
\left[\ip_{K^{(k,k')}},\ip_{L^{(l,,l')}}\right] & \equiv & \ip_{\left[K,L\right]^{\Delta}}\label{eq:bc-algebraic-bracket}\end{eqnarray}
 A short calculation, using the obvious generalization of $\partial_{x}^{n}(f(x)g(x))=\sum_{p=0}^{n}\left(\zwek{n}{p}\right)\partial_{x}^{p}f(x)\partial_{x}^{n-p}g(x)$
leads to \begin{eqnarray}
\ip_{K}\ip_{L} & = & \sum_{p\geq0}\ip_{\ip_{K}^{(p)}L}=\ip_{K\wedge L}+\sum_{p\geq1}\ip_{\ip_{K}^{(p)}L}\label{eq:bc-product-of-interior-products}\end{eqnarray}
where we introduced the \textbf{interior product of order $p$} \rem{(see
(\ref{eq:interior-productIII}))}\begin{eqnarray}
\ip_{K^{(k,k')}}^{(p)} & \equiv & \left(\zwek{k'}{p}\right)K_{\bs{m}\ldots\bs{m}}\hoch{\bs{n}\ldots\bs{n}l_{1}\ldots l_{p}}\frac{\partial^{p}}{\partial\ce^{n_{1}}\cdots\partial\ce^{n_{p}}}=\label{eq:bc-interior-pproduct-I}\\
 & = & \frac{1}{p!}K\frac{\lpartial^{p}}{\partial\be_{n_{p}}\cdots\partial\be_{n_{1}}}\frac{\partial^{p}}{\partial\ce^{n_{1}}\cdots\partial\ce^{n_{p}}}\label{eq:bc-interior-pproductII}\\
\dann\,\ip_{K^{(k,k')}}^{(p)}L^{(l,l')} & = & (-)^{(k'-p)(l-p)}p!\left(\zwek{k'}{p}\right)\left(\zwek{l}{p}\right)K_{\bs{m}\ldots\bs{m}}\hoch{\bs{n}\ldots\bs{n}l_{1}\ldots l_{p}}L_{l_{p}\ldots l_{1}\bs{m}\ldots\bs{m}}\hoch{\bs{n}\ldots\bs{n}}\label{eq:bc-interior-pproduct-on-L}\end{eqnarray}
which contracts only $p$ of all $k'$ upper indices and therefore
coincides with the interior product of above when acting on forms
for $p=k'$ and with the wedge product for $p=0$.\begin{equation}
\ip_{K^{(k,k')}}^{(k')}\rho=\ip_{K^{(k,k')}}\rho,\qquad\ip_{K}^{(0)}L=K\wedge L\end{equation}
\rem{\[
K\wedge L=(-)^{k'l}K_{\bs{m}\ldots\bs{m}}\hoch{\bs{n}\ldots\bs{n}}L_{\bs{m}\ldots\bs{m}}\hoch{\bs{n}\ldots\bs{n}}\]
}\rem{Kann man noch weiter verallgemeinern (haengt mit r-wedge zusammen) zu
(\ref{eq:interior-productIV})\begin{eqnarray*}
\ip_{K^{(k,k')}}^{(p,q)} & \equiv & \frac{1}{p!q!}K\frac{\lpartial^{p}}{\partial\be_{n_{p}}\cdots\partial\be_{n_{1}}}\frac{\lpartial^{q}}{\partial\ce^{k_{q}}\cdots\partial\ce^{k_{1}}}\frac{\partial^{q}}{\partial\be_{k_{1}}\cdots\partial\be_{k_{q}}}\frac{\partial^{p}}{\partial\ce^{n_{1}}\cdots\partial\ce^{n_{p}}}\end{eqnarray*}
} Using (\ref{eq:bc-product-of-interior-products}) the \textbf{algebraic
bracket} $[\:,\:]^{\Delta}$ defined in (\varRef{bcAlgebraicBracket})
can thus be written as \varLabel{bcAlgebraicBracketI}
\begin{eqnarray}
[K^{(k,k')},L^{(l,l')}]^{\Delta} & = & \sum_{p\geq1}\underbrace{\ip_{K}^{(p)}L-(-)^{(k-k')(l-l')}\ip_{L}^{(p)}K}_{\equiv[K,L]_{(p)}^{\Delta}}\label{eq:bc-algebraic-bracketI}\end{eqnarray}
 (\ref{eq:bc-interior-pproduct-on-L}) provides the explicit coordinate
form of this algebraic bracket. From (\ref{eq:bc-interior-pproductII})
we recover the known fact that the $p=1$ term of the algebraic bracket
is induced by the Poisson-bracket and therefore is itself an algebraic
bracket, called the \textbf{big bracket} \cite{Kosmann-Schwarzbach:2003en}
or \textbf{Buttin's algebraic bracket} \cite{Buttin:1974}\rem{ (\ref{eq:bigbracket})}\begin{eqnarray}
\hspace{-0.5cm}\lqn{\Ramm{.59}{\Big.}}\quad[K,L]_{(1)}^{\Delta} & = & \ip_{K}^{(1)}L-(-)^{(k-k')(l-l')}\ip_{L}^{(1)}K\quad\stackrel{(\ref{eq:bc-interior-pproductII})}{=}\left\{ K,L\right\} \quad=\label{eq:bc-big-bracket}\\
 & \stackrel{(\ref{eq:bc-interior-pproduct-on-L})}{=} & (-)^{(k'-1)(l-1)}k'l\, K_{\bs{m}\ldots\bs{m}}\hoch{\bs{n}\ldots\bs{n}l_{1}}L_{l_{1}\bs{m}\ldots\bs{m}}\hoch{\bs{n}\ldots\bs{n}}+\label{eq:bc-big-bracket-coord}\\
 &  & -(-)^{(k-k')(l-l')}(-)^{(l'-1)(k-1)}l'k\, L_{\bs{m}\ldots\bs{m}}\hoch{\bs{n}\ldots\bs{n}l_{1}}K_{l_{1}\bs{m}\ldots\bs{m}}\hoch{\bs{n}\ldots\bs{n}}\qquad\nonumber \end{eqnarray}
For $k'=l'=1$ it reduces to the Richardson-Nijenhuis bracket (\ref{eq:Richardson-Nijenhuis-bracket-coord})
for vector valued forms. In \cite{Kosmann-Schwarzbach:2003en} the
big bracket is described as the canonical Poisson structure on $\bigwedge^{\bullet}(T\oplus T^{*})$
which matches with the observation in (\ref{eq:bc-big-bracket}).
The bracket takes an especially pleasant coordinate form for generalized
multivectors as is presented in equation (\ref{eq:multvec-bigbrack})
on page \pageref{eq:multvec-bigbrack}.

The multivector-degree of the $p$-th term of the complete algebraic
bracket (\varRef{bcAlgebraicBracketI}) is $(k'+l'-p)$, so that we
can rewrite (\varRef{bcAlgebraicBracket}) in terms of {}``quantum''-operators
(\ref{eq:quantization}) in the following way: \begin{eqnarray}
\left[\hat{K}^{(k,k')},\hat{L}^{(l,l')}\right] & = & \sum_{p\geq1}\left(\frac{\hbar}{i}\right)^{p}\widehat{\left[K,L\right]_{(p)}^{\Delta}}=\label{eq:quantum-commutator1}\\
 & = & \left(\frac{\hbar}{i}\right)\widehat{\left\{ K,L\right\} }+\sum_{p\geq2}\left(\frac{\hbar}{i}\right)^{p}\widehat{\left[K,L\right]_{(p)}^{\Delta}}\label{eq:quantum-commutatorII}\end{eqnarray}
 The Poisson bracket is, as it should be, the leading order of the
quantum bracket.

\subsubsection{Extended exterior derivative and the derived bracket of the commutator}

\label{sub:Extended-exterior-derivative} In the previous subsection
the commutator of differential operators induced (via the interior
product as embedding) an algebraic bracket on the embedded tensors.
Also other structures from the operator space can be induced on the
tensors. Having the commutator at hand, one can build the \textbf{derived
bracket} (see footnote \ref{foot-derived-bracket} on page \pageref{foot-derived-bracket}\rem{(\ref{eq:derived-bracketI})})\rem{Verweis auf hoffentl eine subsection im Anhang}
of the commutator by additionally commuting the first argument with
the exterior derivative. Being interested in the induced structure
on multivector valued forms, we consider as arguments only interior
products with those multivector valued forms\begin{eqnarray}
\left[\ip_{K},_{\de}\ip_{L}\right] & \equiv & \left[\left[\ip_{K},\de\,\right],\ip_{L}\right]\label{eq:bc-derived-bracket}\end{eqnarray}
One can likewise use other differentials to build a derived bracket,
e.g. the twisted differential $\left[\de+H,\ldots\right]$ with an
odd closed form $H$, which leads to so called twisted brackets. Let
us restrict to $\de$ for the moment.\rem{Kommt da noch was? Querverweis?}
The derived bracket is in general not skew-symmetric but it obeys
a graded Jacobi-identity \rem{(Leibniz rule, when acting on itself from the left, due to )}
and is therefore what one calls a Loday bracket. When looking for
new brackets, the Jacobi identity is the property which is hardest
to check. A mechanism like above, which automatically provides it
is therefore very powerful. The above derived bracket will induce
brackets like the Schouten bracket or even the Dorfman bracket of
generalized complex geometry on the tensors. In general, however,
the interior products are not closed under its action, i.e. the result
of the bracket cannot necessarily be written as $\ip_{\tilde{K}}$
for some $\tilde{K}$. An expression for a general bracket on the
tensor level, which reduces in the corresponding cases to the well
known brackets therefore does not exist. Instead one normally has
to derive the brackets in the special cases separately. In the following,
however, a natural approach is discussed including the new variable
$p_{m}$, introduced in (\ref{eq:Poisson-bracket-xp}), which leads
to an explicit coordinate expression for the general bracket. This
expression is of course tensorial only in the mentioned special cases,
that is when terms with $p_{m}$ vanish. This is not an artificial
procedure, as the conjugate variable $p_{m}$ to $x^{m}$ is always
present in sigma-models, and it will in turn explain the geometric
meaning of $p_{m}$.

The exterior derivative $\de$ acting on forms is usually not defined
acting on multivector valued forms (otherwise we could build the derived
bracket of the algebraic bracket (\varRef{bcAlgebraicBracketI}) by
$\de\,$ without lifting everything to operators via the interior
product). But via $\{\oo,K^{(k,k')}\}$ we can, at least formally,
define a differential on multivector valued forms. The result, however,
contains the variable $p_{k}$ which we have not yet interpreted geometrically.
After extending the definition of the interior product to objects
containing $p_{m}$, we will get the relation $\left[\de\,,\ip_{K}\right]=\ip_{\left\{ \oo,K\right\} }$,
i.e. $\{\oo,\ldots\}$ can be seen as an induced differential from
the space of operators. For forms $\omega^{(q)}$, this simply reads
$\left[\de\,,\ip_{\omega}\right]=\ip_{\de\omega}$. The definition
$\de K\equiv\left\{ \oo,K\right\} $ thus seems to be a reasonable
extension of the exterior derivative to multivector valued forms.
Let us first provide the necessary definitions.

Consider a phase space function, which is of arbitrary order in the
variable $p_{k}$\begin{eqnarray}
T^{(t,t',t'')}(x,\ce,\be,p) & \equiv & T_{m_{1}\ldots m_{t}}\hoch{n_{1}\ldots n_{t'}k_{1}\ldots k_{t''}}(x)\,\ce^{m_{1}}\cdots\ce^{m_{t}}\be_{m_{1}}\cdots\be_{m_{t'}}p_{k_{1}}\cdots p_{k_{t''}}\label{eq:Tcbp}\end{eqnarray}
$T$ is symmetrized in $k_{1}\ldots k_{t''}\,$,while it is antisymmetrized
in the remaining indices. Using the usual quantization rules $\be\To\frac{\hbar}{i}\partl{\ce}$
and $p\To\frac{\hbar}{i}\partl{x}$ with the indicated ordering (conjugate
momenta to the right) while still insisting on (\ref{eq:quantization})
as the relation between quantum operator and interior product, we
get an extended definition of the \textbf{interior product} (\ref{eq:bc-interior-product-I},\ref{eq:bc-interior-product-II}):\begin{eqnarray}
\hspace{-0.8cm}\ip_{T^{(t,t',t'')}} & \equiv & \left(\frac{i}{\hbar}\right)^{t'+t''}\hat{T}^{(t,t',t'')}\equiv\label{eq:quantizationII}\\
 & \equiv & T_{m_{1}\ldots m_{t}}\hoch{n_{1}\ldots n_{t'}k_{1}\ldots k_{t''}}\ce^{m_{1}}\cdots\ce^{m_{t}}\frac{\partial^{t'}}{\partial\ce^{n_{1}}\cdots\partial\ce^{n_{t'}}}\frac{\partial^{t''}}{\partial x^{k_{1}}\cdots\partial x^{k_{t''}}}=\label{eq:quantizationIII}\\
\hspace{-0.6cm}\ip_{T^{(t,t',t'')}}\rho^{(r)} & = & T_{m_{1}\ldots m_{t}}\hoch{n_{1}\ldots n_{t'}k_{1}\ldots k_{t''}}\ce^{m_{1}}\cdots\ce^{m_{t}}\left\{ \be_{n_{1}},\left\{ \cdots,\left\{ \be_{n_{t'}},\left\{ p_{k_{1}},\left\{ \cdots,\left\{ p_{k_{t''}},\rho^{(r)}\right\} \right\} \right\} \right\} \right\} \right\} =\qquad\label{eq:i-T}\\
 & = & (t')!\left(\zwek{r}{t'}\right)T_{\mm}\hoch{n_{1}\ldots n_{t'}k_{1}\ldots k_{t''}}\frac{\partial^{t''}}{\partial x^{k_{1}}\cdots\partial x^{k_{t''}}}\rho_{n_{t'}\ldots n_{1}\mm}^{(r)}\label{eq:i-TII}\end{eqnarray}
\rem{\[
\ip_{T^{(t,t',t'')}}=\frac{1}{(t'+t'')!}\left(\zwek{t'+t''}{t'}\right)T\frac{\lpartial^{t'+t''}}{\partial p_{i_{t'+t''}}\ldots\partial p_{i_{t'+1}}\partial\be_{i_{t'}}\ldots\partial\be_{i_{1}}}\frac{\partial^{t'+t''}}{\partial\bs{c}^{i_{1}}\ldots\partial\bs{c}^{i_{t'}}\partial x^{i_{t'+1}}\ldots\partial x^{i_{t'+t''}}}\]
}The operator $\ip_{T}$ will serve us as an embedding of $T$ (a
phase space function, which lies in the extension of the space of
multivector valued forms by the basis element $p_{k}$) into the space
of differential operators acting on forms. Because of the partial
derivatives with respect to $x$, the last line is not a tensor and
$T$ in that sense not a well defined geometric object. Nevertheless
it can be a building block of a geometrically well defined object,
for example in the definition of the \textbf{exterior derivative}
on multivector valued forms which we suggested above. Namely, if we
define%
\footnote{This can of course be seen as a BRST differential, which is well known
to be the sum of the longitudinal exterior derivate plus the Koszul
Tate differential. However, as the constraint surface in our case
corresponds to the configuration space ($p_{k}$ would be the first
class constraint generating the BRST-transformation), it is reasonable
to regard the BRST differential as a natural extension of the exterior
derivative of the configuration space.$\quad\fussend$%
} \rem{Relation zu Zucchini's Beobachtung?}\begin{eqnarray}
\de K^{(k,k')} & \equiv & \left\{ \oo,K^{(k,k')}\right\} =\label{eq:dK}\\
 & = & \partial_{\bs{m}}K_{\mm}\hoch{\nn}-(-)^{k-k'}k'\cdot K_{\mm}\hoch{\nn k}p_{k}\label{eq:dK-coord}\end{eqnarray}
We get via our extended embedding (\ref{eq:i-TII}) the nice relation
\footnote{The exterior derivative on forms has already earlier (\ref{eq:exterior-derivative-via-BRST})
been seen to coincide with the Poisson bracket with $\oo$, which
can be used to demonstrate (\ref{eq:dK-und-Lie}):\begin{eqnarray*}
\left[\de\,,\ip_{K}\right]\rho & = & \de\,(\ip_{K}\rho)-(-)^{\abs{K}}\ip_{K}(\de\rho)=\\
 & = & \left\{ \oo,\ip_{K}\rho\right\} -(-)^{\abs{K}}\ip_{K}\left\{ \oo,\rho\right\} =\\
 & \stackrel{(\ref{eq:bc-interior-product-I})}{=} & \partial_{m_{1}}K_{m_{2}\ldots m_{k+1}}\hoch{n_{1}\ldots n_{k'}}\bs{c}^{m_{1}}\cdots\ce^{m_{k+1}}\Big\{\be_{n_{1}},\left\{ \be_{n_{2}},\left\{ \cdots,\left\{ \be_{n_{k'}},\rho^{(r)}\right\} \right\} \right\} +\\
 &  & +(-)^{k}k'\cdot K_{m_{1}\ldots m_{k}}\hoch{n_{1}\ldots n_{k'}}\bs{c}^{m_{1}}\cdots\ce^{m_{k}}\Big\{\underbrace{\left\{ \oo,\be_{n_{1}}\right\} }_{p_{n_{1}}},\left\{ \be_{n_{2}},\left\{ \cdots,\left\{ \be_{n_{k'}},\rho^{(r)}\right\} \right\} \right\} \Big\}\us{\stackrel{(\ref{eq:i-T})}{=}}{(\ref{eq:dK-coord})}\ip_{\de K}\rho\qquad\fussend\end{eqnarray*}
}\begin{eqnarray}
\ip_{\de K}\rho & = & \left[\de\,,\ip_{K}\right]\rho\stackrel{(\ref{eq:Lie-derivativeI})}{=}-(-)^{k-k'}\Lie_{K}\rho\label{eq:dK-und-Lie}\\
\textrm{with }\quad\Lie_{K}\rho & = & (k')!\left(\zwek{r}{k'-1}\right)K_{\bs{m}\ldots\bs{m}}\hoch{l_{1}\ldots l_{k'}}\partial_{l_{k'}}\rho_{l_{k'-1}\ldots l_{1}\bs{m}\ldots\bs{m}}+\nonumber \\
 &  & -(-)^{k-k'}(k')!\left(\zwek{r}{k'}\right)\partial_{\bs{m}}K_{\bs{m}\ldots\bs{m}}\hoch{l_{1}\ldots l_{k'}}\rho_{l_{k'}\ldots l_{1}\bs{m}\ldots\bs{m}}\end{eqnarray}
 \rem{nur wahr, wenn man auf Formen wirkt!! $[\de,\ip_v]w\neq \Lie_v w$ ...
Lie Ableitung bisher nur im Anhang definiert! Dort vielleicht eine kurze subsection mit $\ip$ und Cartan-Gleichungen, aber ohne $\ip^{(p)}$!?}$\Lie_{K}\rho$ is the natural generalization of the Lie derivative
with respect to vectors acting on forms, which is given similarly
$\Lie_{v}\rho=[\ip_{v},\de\,]\rho$. As $\ip_{K}$ is a higher order
derivative, also $\Lie_{K}$ is a higher order derivative. Nevertheless,
it will be called \textbf{Lie derivative with respect to} $K$ in
this paper. Let us again recall this fact: if $p_{k}$ appears in
a combination like $\de K$, there is a well defined geometric meaning
and $\de K$ is up to a sign nothing else than the Lie derivative
with respect to $K$, when embedded in the space of differential operators
on forms. The commutator with the exterior derivative is a natural
differential in the space of differential operators acting on forms,
and via the embedding it induces the differential $\de$ on $K$.
It should perhaps be stressed that the above definition of $\de K$
corresponds to an extended action of the exterior derivative which
acts also on the basis elements of the tangent space\begin{eqnarray}
\de(\pe_{m}) & = & p_{m}\label{eq:d-auf-partial}\end{eqnarray}
This approach will enable us to give explicit coordinate expressions
for the derived bracket of multivector valued forms even in the general
case where the result is not a tensor: In the space of differential
operators on forms, we have the commutator $[\ip_{K},\ip_{L}]$ and
its derived bracket (\ref{eq:derived-bracketI}) $[\ip_{K},_{\de}\ip_{L}]\equiv[[\ip_{K},\de\,],\ip_{L}]$,
while on the space of multivector valued forms we have the algebraic
bracket $\left[K,L\right]^{\Delta}$ and want to define its derived
bracket up to a sign as $[\de K,L]^{\Delta}$. To this end we also
have to extend the definition (\ref{eq:bc-interior-pproduct-I},\ref{eq:bc-interior-pproductII})
of $\ip^{(p)}$, which appears in the explicit expression of the algebraic
bracket in (\varRef{bcAlgebraicBracketI}) to objects that contain
$p_{k}$. This is done in a way that the old equations for the algebraic
bracket remain formally the same. So let us define%
\footnote{Note that $\sum_{q=0}^{p}\left(\zwek{t'}{q}\right)\left(\zwek{t''}{p-q}\right)=\left(\zwek{t'+t''}{p}\right)$
\rem{and $\sum_{q=0}^{p}\left(\zwek{p}{q}\right)=2^{p}$}$\qquad\fussend$%
}\begin{eqnarray}
\hspace{-.5cm}\lqn{\ip_{T^{(t,t',t'')}}^{(p)}\equiv}\nonumber \\
 &  & \hspace{-.5cm}\equiv\sum_{q=0}^{p}\left(\zwek{t'}{q}\right)\left(\zwek{t''}{p-q}\right)T_{\bs{m}\ldots\bs{m}}\hoch{\bs{n}\ldots\bs{n}i_{1}\ldots i_{q}\,,\, i_{q+1}\ldots i_{p}k_{1}\ldots k_{t''-p+q}}p_{k_{1}}\cdots p_{k_{t''-p+q}}\frac{\partial^{p}}{\partial\bs{c}^{i_{1}}\ldots\partial\bs{c}^{i_{q}}\partial x^{i_{q+1}}\ldots\partial x^{i_{p}}}\qquad\quad\\
 &  & \hspace{-.5cm}=\frac{1}{p!}\sum_{q=0}^{p}\left(\zwek{p}{q}\right)T\frac{\lpartial^{p}}{\partial p_{i_{p}}\ldots\partial p_{i_{q+1}}\partial\be_{i_{q}}\ldots\partial\be_{i_{1}}}\frac{\partial^{p}}{\partial\bs{c}^{i_{1}}\ldots\partial\bs{c}^{i_{q}}\partial x^{i_{q+1}}\ldots\partial x^{i_{p}}}\end{eqnarray}
\rem{\begin{eqnarray*}
\ip_{T^{(t,t',t'')}}^{(p)}\tilde{T}^{(\tilde{t},\tilde{t}',\tilde{t}'')} & = & \sum_{q=0}^{p}(-)^{(t'-q)(\tilde{t}-q)}q!\left(\zwek{\tilde{t}}{q}\right)\left(\zwek{t'}{q}\right)\left(\zwek{t''}{p-q}\right)\times\\
 &  & \times T_{\bs{m}\ldots\bs{m}}\hoch{\bs{n}\ldots\bs{n}i_{1}\ldots i_{q}\,,\, i_{q+1}\ldots i_{p}k\ldots k}\frac{\partial^{p-q}}{\partial x^{i_{q+1}}\ldots\partial x^{i_{p}}}\tilde{T}_{i_{q}\ldots i_{1}\bs{m}\ldots\bs{m}}\hoch{\bs{n}\ldots\bs{n}\,,\, k\ldots k}\end{eqnarray*}
} For $p=t'+t''$ it coincides with the full interior product (\ref{eq:i-TII}):
$\ip_{T^{(t,t',t'')}}^{(t'+t'')}=\ip_{T^{(t,t',t'')}}$. In addition
we have with this definition (after some calculation) $\ip_{\de T}^{(p)}=[\de\,,\ip_{T}^{(p)}]$
and in particular \begin{equation}
\ip_{\de K}^{(p)}=[\de\,,\ip_{K}^{(p)}]\end{equation}
and the equations for the algebraic bracket (\varRef{bcAlgebraicBracket}-\varRef{bcAlgebraicBracketI})
indeed remain formally the same for objects containing $p_{m}$ \begin{eqnarray}
[\ip_{T^{(t,t',t'')}},\ip_{\tilde{T}^{(\tilde{t},\tilde{t}',\tilde{t}'')}}] & \equiv & \ip_{\left[T,\tilde{T}\right]^{\Delta}}\\
\ip_{T}\ip_{\tilde{T}} & = & \sum_{p\geq0}\ip_{\ip_{T}^{(p)}\tilde{T}}\\
{}[T^{(t,t',t'')},\tilde{T}^{(\tilde{t},\tilde{t}',\tilde{t}'')}]^{\Delta} & \equiv & \sum_{p\geq1}\underbrace{\ip_{T}^{(p)}\tilde{T}-(-)^{(t-t')(\tilde{t}-\tilde{t}')}\ip_{\tilde{T}}^{(p)}T}_{\equiv[T,\tilde{T}]_{(p)}^{\Delta}}\\
{}[T,\tilde{T}]_{(1)}^{\Delta} & = & \left\{ T,\tilde{T}\right\} \label{eq:bc-big-br-TtildeT}\end{eqnarray}
 which we can again rewrite in terms of {}``quantum''-operators
(\ref{eq:quantization}) as\begin{eqnarray}
\left[\hat{T}^{(k,k')},\hat{\tilde{T}}^{(l,l')}\right] & = & \sum_{p\geq1}\left(\frac{\hbar}{i}\right)^{p}\widehat{\left[T,\tilde{T}\right]_{(p)}^{\Delta}}=\label{eq:quantum-commutator-T-Ttilde}\\
 & = & \left(\frac{\hbar}{i}\right)\widehat{\left\{ T,\tilde{T}\right\} }+\sum_{p\geq2}\left(\frac{\hbar}{i}\right)^{p}\widehat{\left[T,\tilde{T}\right]_{(p)}^{\Delta}}\end{eqnarray}
It should be stressed that -- although very useful -- $\ip^{(p)}$
is unfortunately NOT a geometric operation any longer in general,
in the sense that $\ip_{\de K}^{(p)}L$ and also $\ip_{L}^{(p)}\de K$
do not have a well defined geometric meaning, although $\de K$ and
$L$ have. $\ip_{\de K}\rho$ and $\ip_{K}^{(p)}L$ are in contrast
well defined. $\ip_{\de K}^{(p)}L$, for example, should rather be
understood as a building block of a coordinate calculation which combines
only in certain combinations (e.g. the bracket $[\,,\,]^{\Delta}$)
to s.th. geometrically meaningful. 

We are now ready to define the \textbf{derived bracket} of the algebraic
bracket for multivector valued forms (see footnote \ref{foot-derived-bracket}
on page \pageref{foot-derived-bracket}) \rem{Footnote wird evtl zu section?}\begin{eqnarray}
\hspace{-1cm}\left[K^{(k,k')}\bs{,}L^{(l,l')}\right] & \equiv & \left[K,_{\de\,}L\right]^{\Delta}\equiv-(-)^{k-k'}\left[\de K,L\right]^{\Delta}=\label{eq:bc-derived-bracketI}\\
 & = & \sum_{p\geq1}-(-)^{k-k'}\ip_{\de K}^{(p)}L+(-)^{(k+1-k')(l-l')+k-k'}\ip_{L}^{(p)}\de K=\label{eq:bc-derived-bracketII}\\
 & = & \sum_{p\geq1}-(-)^{k-k'}\ip_{\de K}^{(p)}L+(-)^{(k-k'+1)(l-l'+1)}(-)^{l-l'}\ip_{\de L}^{(p)}K+(-)^{(k-k')(l-l')+k-k'}\de(\ip_{L}^{(p)}K)\qquad\label{eq:bc-derived-bracketIII}\end{eqnarray}
 The result is geometrical in the sense that after embedding via the
interior product it is a well defined operator acting on forms. This
is the case, because due to our extended definitions we have for \textbf{all}
multivector valued forms the relation \begin{eqnarray}
\left[[\ip_{K},\de],\ip_{L}\right] & = & \ip_{\left[K^{(k,k')}\bs{,}L^{(l,l')}\right]}\end{eqnarray}
 and the lefthand side is certainly a well defined geometric object.\rem{Achtung! Lie-Derivative und interior Product nur geom wohldef, wenn sie auf Form wirken!! betrachte z.B. die Lie-Ableitung von einem Vektor nach einem Vektor.
(\ref{eq:dK-und-Lie}): Lie-derivative of order $p$ via\begin{eqnarray*}
\Lie_{K}^{(p)} & \equiv & [\ip_{K}^{(p)},\de\,]=-(-)^{k-k'}\ip_{\de K}^{(p)}\end{eqnarray*}
}  A considerable effort went into getting a correct coordinate form
for the general derived bracket and for that reason, let us quickly
have a glance at the final result, although it is kind of ugly:%
\footnote{The building blocks are \begin{eqnarray*}
\ip_{\de K}^{(p)}L & = & (-)^{(k'-p)(l-p)}p!\left(\zwek{k'}{p}\right)\left(\zwek{l}{p}\right)\partial_{\bs{m}}K_{\bs{m}\ldots\bs{m}}\hoch{\bs{n}\ldots\bs{n}i_{1}\ldots i_{p}}L_{i_{p}\ldots i_{1}\bs{m}\ldots\bs{m}}\hoch{\bs{n}\ldots\bs{n}}+\\
 &  & -(-)^{k-k'}(-)^{(k'-1-p)(l-p)}(p+1)!\left(\zwek{k'}{p+1}\right)\left(\zwek{l}{p}\right)K_{\mm}\hoch{\nn i_{1}\ldots i_{p}k}L_{i_{p}\ldots i_{1}\bs{m}\ldots\bs{m}}\hoch{\bs{n}\ldots\bs{n}}p_{k}+\\
 &  & -(-)^{k-k'}(-)^{(k'-p)(l-p+1)}p!\left(\zwek{k'}{p}\right)\left(\zwek{l}{p-1}\right)K_{\mm}\hoch{\nn i_{1}\ldots i_{p-1}i_{p}}\partial_{i_{p}}L_{i_{p-1}\ldots i_{1}\bs{m}\ldots\bs{m}}\hoch{\bs{n}\ldots\bs{n}}\end{eqnarray*}
\begin{eqnarray*}
\ip_{L}^{(p)}\de K & = & (-)^{(l'-p)(k+1-p)+p}p!\left(\zwek{k}{p}\right)\left(\zwek{l'}{p}\right)L_{\bs{m}\ldots\bs{m}}\hoch{\bs{n}\ldots\bs{n}k_{1}\ldots k_{p}}\partial_{\bs{m}}K_{k_{p}\ldots k_{1}\bs{m}\ldots\bs{m}}\hoch{\bs{n}\ldots\bs{n}}+\\
 &  & +(-)^{(l'-p)(k+1-p)}p!\left(\zwek{k}{p-1}\right)\left(\zwek{l'}{p}\right)L_{\bs{m}\ldots\bs{m}}\hoch{\bs{n}\ldots\bs{n}k_{1}\ldots k_{p-1}l}\partial_{l}K_{k_{p-1}\ldots k_{1}\bs{m}\ldots\bs{m}}\hoch{\bs{n}\ldots\bs{n}}+\\
 &  & -(-)^{k-k'}(-)^{(l'-p)(k-p)}k'\cdot p!\left(\zwek{k}{p}\right)\left(\zwek{l'}{p}\right)L_{\bs{m}\ldots\bs{m}}\hoch{\bs{n}\ldots\bs{n}k_{1}\ldots k_{p}}K_{k_{p}\ldots k_{1}\bs{m}\ldots\bs{m}}\hoch{\bs{n}\ldots\bs{n}k}p_{k}\qquad\fussend\end{eqnarray*}
} \begin{eqnarray}
\left[K\bs{,}L\right] & = & \sum_{p\geq1}-(-)^{k-k'}(-)^{(k'-p)(l-p)}p!\left(\zwek{l}{p}\right)\left(\zwek{k'}{p}\right)\partial_{\bs{m}}K_{\bs{m}\ldots\bs{m}}\hoch{\bs{n}\ldots\bs{n}l_{1}\ldots l_{p}}L_{l_{p}\ldots l_{1}\bs{m}\ldots\bs{m}}\hoch{\bs{n}\ldots\bs{n}}+\nonumber \\
 &  & +(-)^{k+k'l+k'+p+pl+pk'}p!\left(\zwek{k}{p}\right)\left(\zwek{l'}{p}\right)\partial_{\bs{m}}K_{\bs{m}\ldots\bs{m}k_{p}\ldots k_{1}}\hoch{\bs{n}\ldots\bs{n}}L_{\bs{m}\ldots\bs{m}}\hoch{k_{1}\ldots k_{p}\bs{n}\ldots\bs{n}}+\nonumber \\
 &  & -(-)^{k'l+k'+pl+pk'}p!\left(\zwek{k}{p-1}\right)\left(\zwek{l'}{p}\right)\partial_{l}K_{\bs{m}\ldots\bs{m}k_{p-1}\ldots k_{1}}\hoch{\bs{n}\ldots\bs{n}}L_{\bs{m}\ldots\bs{m}}\hoch{k_{1}\ldots k_{p-1}l\bs{n}\ldots\bs{n}}+\nonumber \\
 &  & +(-)^{(k'-p)(l-p+1)}p!\left(\zwek{l}{p-1}\right)\left(\zwek{k'}{p}\right)K_{\bs{m}\ldots\bs{m}}\hoch{\bs{n}\ldots\bs{n}l_{1}\ldots l_{p-1}k}\partial_{k}L_{l_{p-1}\ldots l_{1}\bs{m}\ldots\bs{m}}\hoch{\bs{n}\ldots\bs{n}}+\nonumber \\
 &  & +(-)^{(k'-1-p)(l-p)}p!(k'-p)\left(\zwek{l}{p}\right)\left(\zwek{k'}{p}\right)K_{\bs{m}\ldots\bs{m}}\hoch{\bs{n}\ldots\bs{n}l_{1}\ldots l_{p}k}L_{l_{p}\ldots l_{1}\bs{m}\ldots\bs{m}}\hoch{\bs{n}\ldots\bs{n}}p_{k}+\nonumber \\
 &  & -(-)^{k'l+l+pk'+lp}k'\cdot p!\left(\zwek{k}{p}\right)\left(\zwek{l'}{p}\right)K_{\bs{m}\ldots\bs{m}k_{p}\ldots k_{1}}\hoch{\bs{n}\ldots\bs{n}k}L_{\bs{m}\ldots\bs{m}}\hoch{k_{1}\ldots k_{p}\bs{n}\ldots\bs{n}}p_{k}\label{eq:bc-derived-bracket-coord}\end{eqnarray}
The result is only a tensor, when both terms with $p_{k}$ on the
righthand side vanish, although the complete expression is in general
geometrically well-defined when considered to be a differential operator
acting on forms via $\ip_{\left[K\bs{,}L\right]}$ as this equals
per definition the well-defined $[[\ip_{K},\de],\ip_{L}]$. The above
coordinate form reduces in the appropriate cases to vector Lie-bracket,
Schouten-bracket, and (up to a total derivative) to the (Fr\"ohlicher)-Nijenhuis-bracket.
If one allows as well sums of a vector and a 1-form, we get the Dorfman
bracket, and also the sum of a vector and a general form gives a result
without $p$. \rem{ In the above equation we have to hide $p_{k}$
in a total derivative in order to have a chance to get a tensorial
bracket! But let us first start again with (\ref{eq:bc-derived-bracketIII})\begin{eqnarray*}
\left[K\bs{,}L\right] & = & \sum_{p=1}^{\infty}-(-)^{k-k'}(-)^{(k'-p)(l-p)}p!\left(\zwek{l}{p}\right)\left(\zwek{k'}{p}\right)\partial_{\bs{m}}K_{\bs{m}\ldots\bs{m}}\hoch{\bs{n}\ldots\bs{n}l_{1}\ldots l_{p}}L_{l_{p}\ldots l_{1}\bs{m}\ldots\bs{m}}\hoch{\bs{n}\ldots\bs{n}}+\\
 &  & +(-)^{(k'-p)(l-p+1)}p!\left(\zwek{l}{p-1}\right)\left(\zwek{k'}{p}\right)K_{\bs{m}\ldots\bs{m}}\hoch{\bs{n}\ldots\bs{n}l_{1}\ldots l_{p-1}k}\partial_{k}L_{l_{p-1}\ldots l_{1}\bs{m}\ldots\bs{m}}\hoch{\bs{n}\ldots\bs{n}}+\\
 &  & +(-)^{(k'-1-p)(l-p)}p!(k'-p)\left(\zwek{l}{p}\right)\left(\zwek{k'}{p}\right)K_{\bs{m}\ldots\bs{m}}\hoch{\bs{n}\ldots\bs{n}l_{1}\ldots l_{p}k}L_{l_{p}\ldots l_{1}\bs{m}\ldots\bs{m}}\hoch{\bs{n}\ldots\bs{n}}p_{k}+\\
 &  & +(-)^{(k-k'+1)(l-l'+1)}(-)^{l-l'}(-)^{(l'-p)(k-p)}p!\left(\zwek{k}{p}\right)\left(\zwek{l'}{p}\right)\partial_{\bs{m}}L_{\bs{m}\ldots\bs{m}}\hoch{\bs{n}\ldots\bs{n}l_{1}\ldots l_{p}}K_{l_{p}\ldots l_{1}\bs{m}\ldots\bs{m}}\hoch{\bs{n}\ldots\bs{n}}+\\
 &  & -(-)^{(k-k'+1)(l-l'+1)}(-)^{(l'-p)(k-p+1)}p!\left(\zwek{k}{p-1}\right)\left(\zwek{l'}{p}\right)L_{\bs{m}\ldots\bs{m}}\hoch{\bs{n}\ldots\bs{n}l_{1}\ldots l_{p-1}k}\partial_{k}K_{l_{p-1}\ldots l_{1}\bs{m}\ldots\bs{m}}\hoch{\bs{n}\ldots\bs{n}}+\\
 &  & -(-)^{(k-k'+1)(l-l'+1)}(-)^{(l'-1-p)(k-p)}p!(l'-p)\left(\zwek{k}{p}\right)\left(\zwek{l'}{p}\right)L_{\bs{m}\ldots\bs{m}}\hoch{\bs{n}\ldots\bs{n}l_{1}\ldots l_{p}k}K_{l_{p}\ldots l_{1}\bs{m}\ldots\bs{m}}\hoch{\bs{n}\ldots\bs{n}}p_{k}+\\
 &  & +(-)^{(k-k')(l-l'+1)}(-)^{(l'-p)(k-p)}p!\left(\zwek{k}{p}\right)\left(\zwek{l'}{p}\right)\de\,\left(L_{\bs{m}\ldots\bs{m}}\hoch{\bs{n}\ldots\bs{n}k_{1}\ldots k_{p}}K_{k_{p}\ldots k_{1}\bs{m}\ldots\bs{m}}\hoch{\bs{n}\ldots\bs{n}}\right)\end{eqnarray*}
} 

Due to our extended definition of the exterior derivative, we can
also define the \textbf{derived bracket of the big bracket} (the Poisson
bracket) via \begin{eqnarray}
\left[K^{(k,k')},_{\de\,}L^{(l,l')}\right]_{(1)}^{\Delta} & \equiv & -(-)^{k-k'}\left[\de K,L\right]_{(1)}^{\Delta}=\label{eq:bc-derived-of-bigbracket}\\
 & = & -(-)^{k-k'}\left\{ \de K,L\right\} \end{eqnarray}
which is just the $p=1$ term of the full derived bracket with the
explicit coordinate expression\begin{eqnarray}
\left[K,_{\de\,}L\right]_{(1)}^{\Delta} & = & -(-)^{k-k'}(-)^{(k'-1)(l-1)}lk'\partial_{\bs{m}}K_{\bs{m}\ldots\bs{m}}\hoch{\bs{n}\ldots\bs{n}l_{1}}L_{l_{1}\bs{m}\ldots\bs{m}}\hoch{\bs{n}\ldots\bs{n}}+\nonumber \\
 &  & -(-)^{k+k'l+l}kl'\partial_{\bs{m}}K_{\bs{m}\ldots\bs{m}k_{1}}\hoch{\bs{n}\ldots\bs{n}}L_{\bs{m}\ldots\bs{m}}\hoch{k_{1}\bs{n}\ldots\bs{n}}+\nonumber \\
 &  & -(-)^{k'l+l}l'\partial_{l}K_{\bs{m}\ldots\bs{m}}\hoch{\bs{n}\ldots\bs{n}}L_{\bs{m}\ldots\bs{m}}\hoch{l\bs{n}\ldots\bs{n}}+\nonumber \\
 &  & +(-)^{(k'-1)l}k'K_{\bs{m}\ldots\bs{m}}\hoch{\bs{n}\ldots\bs{n}k}\partial_{k}L_{\bs{m}\ldots\bs{m}}\hoch{\bs{n}\ldots\bs{n}}+\nonumber \\
 &  & +(-)^{k'(l-1)}(k'-1)lk'K_{\bs{m}\ldots\bs{m}}\hoch{\bs{n}\ldots\bs{n}l_{1}k}L_{l_{1}\bs{m}\ldots\bs{m}}\hoch{\bs{n}\ldots\bs{n}}p_{k}+\nonumber \\
 &  & -(-)^{k'l+k'}k'kl'K_{\bs{m}\ldots\bs{m}k_{1}}\hoch{\bs{n}\ldots\bs{n}k}L_{\bs{m}\ldots\bs{m}}\hoch{k_{1}\bs{n}\ldots\bs{n}}p_{k}\label{eq:bc-derived-of-bigbracket-coord}\end{eqnarray}
\begin{eqnarray}
\left[K\bs{,}L\right] & = & \left[K,_{\de\,}L\right]_{(1)}^{\Delta}-(-)^{k-k'}\sum_{p\geq2}\left[\de K,L\right]_{(p)}^{\Delta}\label{eq:derived-complete-vs-big}\end{eqnarray}
Also this bracket takes a very pleasant coordinate form for generalized
multivectors (see (\ref{eq:derived-of-big-generalized}) on page \pageref{eq:derived-of-big-generalized}).
\rem{ or\begin{eqnarray*}
\left[K\bs{,}_{\de\,}L\right]_{(1)}^{\Delta} & = & -(-)^{k-k'}(-)^{(k'-1)(l-1)}lk'\partial_{\bs{m}}K_{\bs{m}\ldots\bs{m}}\hoch{\bs{n}\ldots\bs{n}j}L_{j\bs{m}\ldots\bs{m}}\hoch{\bs{n}\ldots\bs{n}}+\\
 &  & +(-)^{(k'-1)l}k'K_{\bs{m}\ldots\bs{m}}\hoch{\bs{n}\ldots\bs{n}j}\partial_{j}L_{\bs{m}\ldots\bs{m}}\hoch{\bs{n}\ldots\bs{n}}+\\
 &  & +(-)^{k'(l-1)}lk'(k'-1)K_{\bs{m}\ldots\bs{m}}\hoch{\bs{n}\ldots\bs{n}ji}L_{j\bs{m}\ldots\bs{m}}\hoch{\bs{n}\ldots\bs{n}}p_{i}+\\
 &  & +(-)^{(k-k'+1)(l-l'+1)}(-)^{l-l'}(-)^{(l'-1)(k-1)}kl'\partial_{\bs{m}}L_{\bs{m}\ldots\bs{m}}\hoch{\bs{n}\ldots\bs{n}j}K_{j\bs{m}\ldots\bs{m}}\hoch{\bs{n}\ldots\bs{n}}+\\
 &  & -(-)^{(k-k'+1)(l-l'+1)+(l'-1)k}l'L_{\bs{m}\ldots\bs{m}}\hoch{\bs{n}\ldots\bs{n}j}\partial_{j}K_{\bs{m}\ldots\bs{m}}\hoch{\bs{n}\ldots\bs{n}}+\\
 &  & -(-)^{(k-k'+1)(l-l'+1)}(-)^{l'(k-1)}kl'(l'-1)L_{\bs{m}\ldots\bs{m}}\hoch{\bs{n}\ldots\bs{n}ji}K_{j\bs{m}\ldots\bs{m}}\hoch{\bs{n}\ldots\bs{n}}p_{i}+\\
 &  & +(-)^{(k-k')(l-l'+1)}(-)^{(l'-1)(k-1)}kl'\de\,\left(L_{\bs{m}\ldots\bs{m}}\hoch{\bs{n}\ldots\bs{n}j}K_{j\bs{m}\ldots\bs{m}}\hoch{\bs{n}\ldots\bs{n}}\right)\end{eqnarray*}
} In contrast to the full derived bracket, we have no guarantee for
this derived bracket to be geometrical itself. \rem{coordinate form for generalized vectors?Dorfmann-Schouten-bracket}

\rem{Gegenbeispiel: Consider the case $K=K^{(1,1)}=K_{\bs{m}}\hoch{\bs{n}}$
and $L=v^{(2)}=v^{\bs{m}\bs{m}}$ with \begin{eqnarray*}
\de K & = & \partial_{\bs{m}}K_{\bs{m}}\hoch{\bs{n}}-K_{\bs{m}}\hoch{i}p_{i}\end{eqnarray*}
 \begin{eqnarray*}
\left[K^{(1,1)},_{\de}v^{(2)}\right] & = & \ip_{v^{(2)}}^{(1)}\de K^{(1,1)}+\ip_{v^{(2)}}^{(2)}\de K^{(1,1)}-\ip_{\de K}^{(1)}v\end{eqnarray*}
It turns out that for $\ip_{\ip_{v^{(2)}}^{(1)}\de K^{(1,1)}+\ip_{v^{(2)}}^{(2)}\de K^{(1,1)}-\ip_{\de K}^{(1)}v}\rho^{(r)}$
the Christoffel symbols cancel only in the complete combination, including
$\ip^{(2)}$! (Detailliere Rechnung in Note) } \rem{Remembering
that $\oo=\ce^{k}p_{k}$, which implies $\ip_{\oo}=\ce^{k}\partl{x^{k}}$,
we should note that there are now several ways to express the exterior
derivative\begin{eqnarray*}
\oo & \equiv & \ce^{k}p_{k},\qquad\ip_{Q}=\ce^{k}\partl{x^{k}}=\frac{i}{\hbar}\hat{\oo}\\
\ip_{\oo}\rho & = & \frac{i}{\hbar}\hat{\oo}\rho=\de\rho\qquad\ip_{\oo}K\neq\de K\quad!\\
{}[\ip_{\oo},\ip_{K}] & = & \ip_{[\oo,K]^{\Delta}}=\ip_{\{\oo,K\}}=\ip_{\de K}\\
{}[\hat{\oo},\hat{K}] & = & \left(\frac{\hbar}{i}\right)\widehat{\de K}\\
\left[[\ip_{K},\de\,],\ip_{L}\right] & = & -(-)^{k-k'}\left[\ip_{\de K},\ip_{L}\right]=\left[\left[\ip_{K},\ip_{\oo}\right],\ip_{L}\right]=\ip_{\left[\left[K,\oo\right]^{\Delta},L\right]^{\Delta}}\end{eqnarray*}
} 

Let us eventually note how one can easily adjust the extended exterior
derivative to the twisted case:\begin{eqnarray}
[\de+H\wedge\,,\,\ip_{K}] & \equiv & \ip_{\de_{H}K}\\
\de_{H}K & = & \de K+\left[H,K\right]^{\Delta}=\de K-(-)^{k-k'}\sum_{p\geq1}\ip_{K}^{(p)}H\end{eqnarray}
with $H$ being an odd closed differential form. It should be stressed
that $\de+H\wedge$ is not a differential, but on the operator level
its commutator $[\de+H\wedge,\ldots]$ is a differential and thus
the above defined $\de_{H}$ is a differential as well.

\subsection{Sigma-Models}

\label{sub:Sigma-Models} A sigma model is a field theory whose fields
are embedding functions from a world-volume $\Sigma$ into a target
space $M$, like in string theory. So far there was no sigma-model
explicitly involved into our considerations. One can understand the
previous subsection simply as a convenient way to formulate some geometry.
The phase space introduced there, however, is like the phase space
of a (point particle) sigma model with only one world-volume dimension
-- the time -- which is not showing up in the off-shell phase-space.
Let us now naively consider the same setting like before as a sigma
model with the coordinates $x^{m}$ depending on some worldsheet coordinates%
\footnote{The index $\mu$ will not include the worldvolume time, when considering
the phase space, but it will contain the time in the Lagrangian formalism.
As this should be clear from the context, there will be no notational
distinction. $\qquad\fussend$%
} $\sigma^{\mu}$. The resulting model has a very special field content,
because its anticommuting fields $\ce^{m}(\sigma)$ have the same
index structure as the embedding coordinate $x^{m}(\sigma)$. In one
and two worldvolume-dimensions, $\ce^{m}$ can be regarded as worldvolume-fermions,
and this will be used in the stringy application in \ref{sub:Zabzine}.
In general worldvolume dimensions, $\ce^{m}$ could be seen as ghosts,
leading to a topological theory. In any case the dimension of the
worldvolume will not yet be fixed, as the described mechanism does
not depend on it. 

A multivector valued form on a $C^{\infty}$-manifold $M$ can locally
be regarded as an analytic function of $x^{m},\de x^{m}\equiv\ce^{m}$
and $\pe_{m}\equiv\be_{m}$\begin{eqnarray}
K^{(k,k')}(x,\de x,\pe) & = & K_{m_{1}\ldots m_{k}}\hoch{n_{1}\ldots n_{k'}}(x)\de x^{m_{1}}\wedge\cdots\wedge\de x^{m_{k}}\wedge\pe_{n_{1}}\wedge\cdots\wedge\pe_{n_{k'}}=\label{eq:K-als-analytische-FunktionI}\\
 & \equiv & K_{m_{1}\ldots m_{k}}\hoch{n_{1}\ldots n_{k'}}(x)\ce^{m_{1}}\cdots\ce^{m_{k}}\be_{n_{1}}\cdots\be_{n_{k'}}=K^{(k,k')}(x,\ce,\be)\label{eq:K-als-analytische-FunktionII}\end{eqnarray}
 For sigma models, $x^{m}\To x^{m}(\sigma),p_{m}\To p_{m}(\sigma),\ce^{m}\To\ce^{m}(\sigma)$
and $\be_{m}\To\be_{m}(\sigma)$ become dependent on the worldvolume
variables $\sigma^{\mu}$. They are, however, for every $\sigma$
valid arguments of the function $K$. Frequently only the worldvolume
coordinate $\sigma$ will then be denoted as new argument of $K$,
which has to be understood in the following sense\begin{eqnarray}
\hspace{-1cm}K^{(k,k')}(\sigma)\equiv K^{(k,k')}\left(x(\sigma),\ce(\sigma),\be(\sigma)\right) & = & K_{m_{1}\ldots m_{k}}\hoch{n_{1}\ldots n_{k'}}\left(x(\sigma)\right)\cdot\ce^{m_{1}}(\sigma)\cdots\ce^{m_{k}}(\sigma)\be_{n_{1}}(\sigma)\cdots\be_{n_{k'}}(\sigma)\label{eq:K-of-sigma}\end{eqnarray}
Also functions depending on $p_{m}$, like $\de K(x,\ce,\be,p)$ in
(\ref{eq:dK-coord}), or more general a function $T^{(t,t',t'')}(x,\ce,\be,p)$
as in (\ref{eq:Tcbp}) are denoted in this way\begin{eqnarray}
T^{(t,t',t'')}(\sigma) & \equiv & T^{(t,t',t'')}\left(x(\sigma),\ce(\sigma),\be(\sigma),p(\sigma)\right)\quad(\textrm{see }(\ref{eq:Tcbp}))\label{eq:T-of-sigma}\\
\textrm{e.g. }\de K(\sigma) & \equiv & \de K\left(x(\sigma),\ce(\sigma),\be(\sigma),p(\sigma)\right)\quad(\textrm{see }(\ref{eq:dK-coord}))\label{eq:dK-of-sigma}\\
\textrm{or }\oo(\sigma) & \equiv & \oo\left(\ce(\sigma),p(\sigma)\right)=\ce^{m}(\sigma)p_{m}(\sigma)\quad(\textrm{see }(\ref{eq:BRST-op}))\label{eq:o-of-sigma}\end{eqnarray}
 The expression $\de K(\sigma)$ should \textbf{NOT} be mixed up with
the worldsheet exterior derivative of $K$ which will be denoted by
$\dew K(\sigma)$.%
\footnote{ It is much better to mix it up with a BRST transformation or with
something similar to a worldsheet supersymmetry transformation. We
will come to that later in subsection \ref{sub:Zabzine}. To make
confusion perfect, it should be added that in contrast it is not completely
wrong in subsection \ref{sub:antibracket} to mix up the target space
exterior derivative with the worldsheet exterior derivative...$\qquad\fussend$%
} Every operation of the previous section, like $\ip_{K}^{(p)}L$ or
the algebraic or derived brackets leads again to functions of $x,\ce,\be$
and sometimes $p$. Let us use for all of them the notation as above,
e.g. for the derived bracket of the big bracket (\ref{eq:bc-derived-of-bigbracket},\ref{eq:bc-derived-of-bigbracket-coord})\begin{eqnarray}
\left[K^{(k,k')},_{\de\,}L^{(l,l')}\right]_{(1)}^{\Delta}(\sigma) & \equiv & \left[K^{(k,k')}\bs{,}L^{(l,l')}\right]_{(1)}^{(\Delta)}\left(x(\sigma),\ce(\sigma),\be(\sigma),p(\sigma)\right)\label{eq:bracket-of-sigma}\end{eqnarray}
And even $\de x^{m}=\ce^{m}$ and $\de\be_{m}=p_{m}$ will be seen
as a function (identity) of $\ce^{m}$ or $\be_{m}$, s.th. we denote
\begin{eqnarray}
\de x^{m}(\sigma) & \equiv & \ce^{m}(\sigma)\label{eq:dx-of-sigma}\\
\de\be_{m}(\sigma) & \equiv & p_{m}(\sigma)\label{eq:db-of-sigma}\end{eqnarray}
Although $\de$ acts only in the target space on $x,\be,\ce$ and
$p$, the above obviously suggests to introduce a differential --
say $\es$ -- in the new phase space, which is compatible with the
target space differential in the sense \rem{(\ref{eq:d-auf-partial})}\begin{eqnarray}
\es\left(x^{m}(\sigma)\right) & = & \de x^{m}(\sigma)\equiv\ce^{m}(\sigma)\\
\es\left(\be_{m}(\sigma)\right) & = & \de\be_{m}(\sigma)\equiv p_{m}(\sigma)\end{eqnarray}
 We can generate $\es$ with the Poisson bracket in almost the same
way as $\de$ before in (\ref{eq:BRST-op}):\begin{eqnarray}
\OO & \equiv & \int_{\Sigma}\msigp\quad\oo(\sigma)=\int\msigp\quad\ce^{m}(\sigma)p_{m}(\sigma),\qquad\textrm{\es}\,(\ldots)=\left\{ \OO,\ldots\right\} \label{eq:Omega}\end{eqnarray}
The Poisson bracket between the conjugate fields gets of course an
additional delta function compared to (\ref{eq:Poisson-bracket-bc},\ref{eq:Poisson-bracket-xp}).\begin{eqnarray}
\left\{ p_{m}(\sigma'),x^{n}(\sigma)\right\}  & = & \delta_{m}^{n}\delta^{d_{{\rm w}}-1}(\sigma'-\sigma)\\
\left\{ \be_{m}(\sigma'),\ce^{n}(\sigma)\right\}  & = & \delta_{m}^{n}\delta^{d_{{\rm w}}-1}(\sigma'-\sigma)\end{eqnarray}
 The first important (but rather trivial) observation is then that
for $K(\sigma)$ being a function of $x(\sigma),\ce(\sigma),\be(\sigma)$
as in (\ref{eq:K-of-sigma}) (and not a functional, which could contain
derivatives on or integrations over $\sigma$) we have \begin{eqnarray}
\es\,(K(\sigma)) & = & \left(\ce^{m}(\sigma)\partl{(x^{m}(\sigma))}+p_{m}(\sigma)\partl{(\be_{m}(\sigma))}\right)K\left(x(\sigma),\ce(\sigma),\be(\sigma)\right)=\de K(\sigma)\label{eq:sK-gleich-dK}\end{eqnarray}
The same is true for more general objects of the form of $T$ in (\ref{eq:T-of-sigma}).
Because of this fact the distinction between $\de$ and $\es$ is
not very essential, but in subsection \ref{sub:antibracket} the replacement
of the arguments as in (\ref{eq:T-of-sigma}) will be different and
the distinction very essential in order not to get confused.

The relation between Poisson bracket and big bracket (\ref{eq:bc-big-bracket},\ref{eq:bc-big-br-TtildeT})
gets obviously modified by a delta function\begin{eqnarray}
\left\{ K^{(k,k')}(\sigma'),L^{(l,l')}(\sigma)\right\}  & = & \left[K^{(k,k')},L^{(l,l')}\right]_{(1)}^{\Delta}(\sigma)\,\delta^{d_{{\rm w}}-1}(\sigma'-\sigma)\label{eq:Poisson-big-br}\\
\textrm{or more general }\left\{ T^{(t,t',t'')}(\sigma'),\tilde{T}^{(\tilde{t},\tilde{t}',\tilde{t}'')}(\sigma)\right\}  & = & \left[T^{(t,t',t'')},\tilde{T}^{(\tilde{t},\tilde{t}',\tilde{t}'')}\right]_{(1)}^{\Delta}(\sigma)\,\delta^{d_{{\rm w}}-1}(\sigma'-\sigma)\label{eq:Poisson-big-br-T}\end{eqnarray}
The relation between the derived bracket (using $\es\,$) on the lefthand
side and the derived bracket (using $\de\,$) on the righthand side
is (omitting the overall sign in the definition of the derived bracket)\begin{eqnarray}
\left\{ \es K^{(k,k')}(\sigma'),L^{(l,l')}(\sigma)\right\}  & \stackrel{(\ref{eq:sK-gleich-dK})}{=} & \left\{ \de K^{(k,k')}(\sigma'),L^{(l,l')}(\sigma)\right\} \stackrel{(\ref{eq:Poisson-big-br-T})}{=}\left[\de K^{(k,k')},L^{(l,l')}\right]_{(1)}^{\Delta}(\sigma)\,\delta^{d_{{\rm w}}-1}(\sigma'-\sigma)\qquad\end{eqnarray}
The worldvolume coordinates $\sigma$ remain so far more or less only
spectators. In the subsection \ref{sub:antibracket}, the world-volume
coordinates play a more active part and already in the following subsection
a similar role is taken by an anticommuting extension of the worldsheet. 

Before we proceed, it should be stressed that the replacement of $x,\ce,\be$
and $p$ by $x(\sigma),\ce(\sigma),\be(\sigma)$ and $p(\sigma)$
was just the most naive replacement to do, and it will be a bit extended
in the following section until it can serve as a useful tool in an
application in \ref{sub:Zabzine}. But in principle, one can replace
those variables by any fields with matching index structure and parity
(even composite ones) and study the resulting relations between Poisson
bracket on the one side and geometric bracket on the other side. Also
the differential $\es\,$ can be replaced for example by the twisted
differential or by more general BRST-like transformations. In this
way it should be possible to implement other derived brackets, for
example those built with the Poisson-Lichnerowicz-differential (see
\cite{Kosmann-Schwarzbach:2003en}), in a sigma-model description.
In \ref{sub:antibracket}, a different (but also quite canonical)
replacement is performed and we will see that the different replacement
corresponds to a change of the role of $\sigma$ and an anticommuting
worldvolume coordinate $\tet$ which will be introduced in the following.

\subsection{Natural appearance of derived brackets in Poisson brackets of superfields}

\label{sub:Natural-appearance} In the application to worldsheet theories
in section \ref{sec:Applications-in-string}, there appear superfields,
either in the sense of worldsheet supersymmetry or in the sense of
de-Rham superfields (see e.g. \cite{Cattaneo:1999fm,Zucchini:2004ta}).
Let us view a superfield in general as a method to implement a fermionic
transformation of the fields via a shift in a fermionic parameter
$\tet$ which can be regarded as fermionic extension of the worldvolume.
In our case the fermionic transformation is just the spacetime de-Rham-differential
$\de$, or more precisely $\es\,$, and is not necessarily connected
to worldvolume supersymmetry. In fact, in worldvolumes of dimension
higher than two, supersymmetry requires more than one fermionic parameter
while a single $\tet$ is enough for our purpose to implement $\es$.
In two dimensions, however, this single theta can really be seen as
a worldsheet fermion (see \ref{sub:Zabzine}). But let us neglect
this knowledge for a while, in order to clearly see the mechanism,
which will be a bit hidden again, when applied to the supersymmetric
case in \ref{sub:Zabzine}.

As just said above, we want to implement with superfields the fermionic
transformation $\es$ and not yet a supersymmetry. So let us define
in this section a \textbf{superfield} as a function of the phase space
fields with additional dependence on $\tet$, $Y=Y(x(\sigma),p(\sigma),\ce(\sigma),\be(\sigma),\tet)$,
which obeys %
\footnote{If this seems unfamiliar, compare with the case of worldsheet supersymmetry,
where one introduces a differential operator $\qu\equiv\partial_{\tet}+\tet\partial_{\sigma}$
and the definition of a superfield is, in contrast to here, $\delta_{\feps}Y\stackrel{!}{=}\feps\qu Y$,
where $\delta_{\feps}$ is the supersymmetry transformation of the
component fields (compare \ref{sub:Zabzine}).$\qquad\fussend$%
}\enlargethispage*{2cm}\begin{eqnarray}
\es Y(x(\sigma),p(\sigma),\ce(\sigma),\be(\sigma),\tet) & \stackrel{!}{=} & \partial_{\tet}Y(x(\sigma),p(\sigma),\ce(\sigma),\be(\sigma),\tet)\label{eq:superfield-definition}\\
\textrm{with } &  & \es x^{m}(\sigma)=\ce^{m}(\sigma),\es\be_{m}(\sigma)=p_{m}(\sigma)\quad(\es\tet=0)\end{eqnarray}
 With our given field content it is possible to define two basic conjugate%
\footnote{\label{foot:super-Poisson-bracket}The superfields $\Phi$ and $\Es$
are conjugate with respect to the following \textbf{super-Poisson-bracket}
\begin{eqnarray*}
\left\{ F(\sigma',\tet'),G(\sigma,\tet)\right\}  & \equiv & \int\msigp\backtilde\int d\tilde{\tet}\qquad\big(\delta F(\sigma',\tet')/\delta\Es_{k}(\tilde{\sigma},\tilde{\tet})\funktional{}{\Phi^{k}(\tilde{\sigma},\tilde{\tet})}G(\sigma,\tet)-\delta F(\sigma',\tet')/\delta\Phi^{k}(\tilde{\sigma},\tilde{\tet})\funktional{}{\Es_{k}(\tilde{\sigma},\tilde{\tet})}G(\sigma,\tet)\big)=\\
 & = & \int\msigp\backtilde\int d\tilde{\tet}\qquad\big(\delta F(\sigma',\tet')/\delta\Es_{k}(\tilde{\sigma},\tilde{\tet})\funktional{}{\Phi^{k}(\tilde{\sigma},\tilde{\tet})}G(\sigma,\tet)-(-)^{FG}\delta G(\sigma',\tet')/\delta\Es_{k}(\tilde{\sigma},\tilde{\tet})\funktional{}{\Phi^{k}(\tilde{\sigma},\tilde{\tet})}F(\sigma,\tet)\big)\end{eqnarray*}
which, however, boils down to taking the ordinary graded Poisson bracket
between the component fields (as can be seen in (\ref{eq:conjugate-superfields})).
The \textbf{functional derivatives} from the left and from the right
are defined as usual via\[
\delta_{S}A\equiv\int\msigp\backtilde\int d\tilde{\theta}\quad\delta A/\delta S_{k}(\tilde{\sigma},\tilde{\theta})\cdot\delta S_{k}(\tilde{\sigma},\tilde{\theta})\equiv\int\msigp\backtilde\int d\tilde{\theta}\quad\delta S_{k}(\tilde{\sigma},\tilde{\theta})\cdot\funktl{S_{k}(\tilde{\sigma},\tilde{\theta})}A\]
and similarly for $\Phi$, which leads to \begin{eqnarray*}
\funktl{\Es_{m}(\tilde{\sigma},\tilde{\tet})}\Es_{n}(\sigma,\tet) & = & \delta_{n}^{m}(\tet-\tilde{\tet})\delta^{d_{{\rm w}}-1}(\sigma-\tilde{\sigma})=-\delta\Es_{n}(\sigma,\tet)/\Es_{m}(\tilde{\sigma},\tilde{\tet})\\
\funktl{\Phi^{m}(\tilde{\sigma},\tilde{\tet})}\Phi^{n}(\sigma,\tet) & = & \delta_{m}^{n}(\tilde{\tet}-\tet)\delta^{d_{{\rm w}}-1}(\sigma-\tilde{\sigma})=\delta\Phi^{n}(\sigma,\tet)/\delta\Phi^{m}(\tilde{\sigma},\tilde{\tet})\end{eqnarray*}
The functional derivatives can also be split in those with respect
to the component fields\[
\funktl{\Es_{m}(\tilde{\sigma},\tilde{\tet})}=\funktl{p_{m}(\tilde{\sigma})}-\tilde{\tet}\funktl{\be_{m}(\tilde{\sigma})},\qquad\funktl{\Phi^{m}(\tilde{\sigma},\tilde{\tet})}=\funktl{\ce^{m}(\tilde{\sigma})}+\tilde{\tet}\funktl{x^{m}(\tilde{\sigma})}\qquad\fussend\]
} superfields $\Phi^{m}$ and $\Es_{m}$ \\
which build up a super-phase-space\enlargethispage*{1cm}%
\footnote{For Grassmann variables $\delta(\tet'-\tet)=\tet'-\tet$ in the following
sense\begin{eqnarray*}
\int\de\tet'(\tet'-\tet)F(\tet') & = & \int\de\tet'(\tet'-\tet)\left(F(\tet)+(\tet'-\tet)\partial_{\tet}F(\tet)\right)=\\
 & = & \int\de\tet'\quad\tet'F(\tet)-\tet'\tet\partial_{\tet}F(\tet)-\tet\tet'\partial_{\tet}F(\tet)=\\
 & = & F(\tet)\end{eqnarray*}
We have as usual\begin{eqnarray*}
\tet\delta(\tet'-\tet) & = & \tet(\tet'-\tet)=\tet\tet'=\tet'(\tet'-\tet)=\\
 & = & \tet'\delta(\tet'-\tet)\end{eqnarray*}
Pay attention to the antisymmetry\begin{eqnarray*}
\delta(\tet'-\tet) & = & -\delta(\tet-\tet')\qquad\fussend\end{eqnarray*}
} \begin{eqnarray}
\Phi^{m}(\sigma,\tet) & \equiv & x^{m}(\sigma)+\bs{\theta}\ce^{m}(\sigma)=x^{m}(\sigma)+\bs{\theta}\es x^{m}(\sigma)\\
\Es_{m}(\sigma,\tet) & \equiv & \be_{m}(\sigma)+\tet p_{m}(\sigma)=\be_{m}(\sigma)+\tet\es\be_{m}(\sigma)\\
\left\{ \Es_{m}(\sigma,\tet),\Phi^{n}(\sigma',\tet')\right\}  & = & \left\{ \be_{m}(\sigma),\tet'\ce^{n}(\sigma')\right\} +\tet\left\{ p_{m}(\sigma),x^{n}(\sigma')\right\} =\label{eq:conjugate-superfields}\\
 & = & \underbrace{(\tet-\tet')}_{\equiv\delta(\tet-\tet')}\delta(\sigma-\sigma')\delta_{m}^{n}\end{eqnarray}
$\Phi$ and $\Es$ are obviously superfields in the above sense\begin{eqnarray}
\partial_{\tet}\Phi^{m}(\sigma,\tet) & = & \underbrace{\es x^{m}(\sigma)}_{\ce^{m}(\sigma)}\underbrace{+\tet\es\ce^{m}(\sigma)}_{=0}=\es\Phi^{m}(\sigma,\tet)\label{eq:theta-derivative-eq-exterior-derivative}\\
\partial_{\tet}\Es_{m} & = & \underbrace{\es\be_{m}(\sigma)}_{p_{m}(\sigma)}\underbrace{+\tet\es p_{m}(\sigma)}_{0}=\es\Es_{m}(\sigma,\tet)\label{eq:theta-derivative-eq-exterior-derivativeS}\end{eqnarray}
as well as $\es\Phi(\sigma,\tet)=\ce(\sigma)$ and $\es\Es(\sigma,\tet)=p(\sigma)$
are superfields, and every analytic function of those fields will
be a superfield again. \rem{and $\partial_\sigma\Phi$, aber hier wurscht...}

We will convince ourselves in this subsection that in the Poisson
brackets of general superfields, the derived brackets come with the
complete $\delta$-function (of $\sigma$ and $\tet$) while the corresponding
algebraic brackets come with a derivative of the delta-function. The
introduction of worldsheet coordinates $\sigma$ was not yet really
necessary for this discussion, but it will be useful for the comparison
with the subsequent subsection. Indeed, we do not specify the dimension
$d_{{\rm w}}$ of the worldsheet yet. An argument sigma is representing
several worldsheet coordinates $\sigma^{\mu}$. It should be stressed
again that the differential $\de$ should \textbf{NOT} be mixed up
with the worldsheet exterior derivative $\dew$, which does not show
up in this subsection. 

Similar as in \ref{sub:Sigma-Models}, equations (\ref{eq:K-of-sigma})-(\ref{eq:db-of-sigma}),we
will view all geometric objects as functions of local coordinates
and replace the arguments not by phase space fields but by the just
defined super-phase fields which reduces for $\tet=0$ to the previous
case. \begin{eqnarray}
T^{(t,t',t'')}(\sigma,\tet) & \equiv & T^{(t,t',t'')}\left(\Phi(\sigma,\tet),\es\Phi(\sigma,\tet),\Es(\sigma,\tet),\es\Es(\sigma,\tet)\right)\stackrel{\tet=0}{=}T^{(t,t',t'')}(\sigma)\quad(\textrm{see }(\ref{eq:T-of-sigma}))\label{eq:T-sig-tet}\end{eqnarray}
For example for a multivector valued form we write \begin{eqnarray}
K^{(k,k')}(\sigma,\tet) & \equiv & K^{(k,k')}\big(\Phi^{m}(\sigma,\tet),\underbrace{\es\Phi^{m}(\sigma,\tet)}_{\ce^{m}(\sigma)},\Es_{m}(\sigma,\tet)\big)=\label{eq:K-sig-tet-def}\\
 &  & \hspace{-1cm}=K_{m_{1}\ldots m_{k}}\hoch{n_{1}\ldots n_{k'}}\left(\Phi(\sigma,\tet)\right)\,\underbrace{\es\Phi^{m_{1}}(\sigma,\tet)}_{\ce^{m_{1}}(\sigma)}\ldots\es\Phi^{m_{k}}(\sigma,\tet)\Es_{n_{1}}(\sigma,\tet)\ldots\Es_{n_{k'}}(\sigma,\tet)\us{\stackrel{\tet=0}{=}}{(\ref{eq:K-of-sigma})}K^{(k,k')}(\sigma)\qquad\end{eqnarray}
Likewise for all the other examples of \ref{sub:Sigma-Models}:\begin{eqnarray}
\textrm{e.g. }\de K(\sigma,\tet) & \equiv & \de K\left(\Phi(\sigma,\tet),\es\Phi(\sigma,\tet),\Es(\sigma,\tet),\es\Es(\sigma,\tet)\right)\label{eq:dK-of-sig-tet}\\
\textrm{or }\oo(\sigma,\tet) & \equiv & \oo\left(\es\Phi(\sigma,\tet),\es\Es(\sigma,\tet)\right)=\ce^{m}(\sigma)p_{m}(\sigma)=\oo(\sigma)\label{eq:o-of-sig-tet}\\
\hspace{-1.2cm}\left[K^{(k,k')},_{\de\,}L^{(l,l')}\right]_{(1)}^{\Delta}(\sigma,\tet) & \equiv & \left[K^{(k,k')}\bs{,}L^{(l,l')}\right]_{(1)}^{(\Delta)}\left(\Phi(\sigma,\tet),\es\Phi(\sigma,\tet),\Es(\sigma,\tet),\es\Es(\sigma,\tet)\right)\us{\stackrel{\tet=0}{=}}{(\ref{eq:bracket-of-sigma})}\left[K^{(k,k')}\bs{,}L^{(l,l')}\right]_{(1)}^{(\Delta)}(\sigma)\qquad\quad\\
\de x^{m}(\sigma,\tet) & \equiv & \es\Phi^{m}(\sigma,\tet)=\ce^{m}(\sigma)\\
\de\be_{m}(\sigma,\tet) & \equiv & \es\Es_{m}(\sigma,\tet)=p_{m}(\sigma)\label{eq:db-sig-tet}\end{eqnarray}
For functions of the type $T^{(t,t',t'')}(\sigma,\tet)$ we clearly
have \begin{eqnarray}
\de T^{(t,t',t'')}(\sigma,\tet) & = & \es\left(T^{(t,t',t'')}(\sigma,\tet)\right)\label{eq:dT-gleich-sT}\\
\textrm{in particular }\de K^{(k,k')}(\sigma,\tet) & = & \es\left(K^{(k,k')}(\sigma,\tet)\right)\end{eqnarray}
As all those analytic functions of the basic superfields are superfields
(in the sense of \ref{eq:superfield-definition}) themselves, $\partial_{\tet}$
can be replaced by $\es\,$ in a $\tet$-expansion, so that we get
the important relation \begin{eqnarray}
T^{(t,t',t'')}(\sigma,\tet) & = & T^{(t,t',t'')}(\sigma)+\tet\de T^{(t,t',t'')}(\sigma)\label{eq:wichtig}\\
K^{(k,k')}(\sigma,\tet) & = & K^{(k,k')}(\sigma)+\tet\de K^{(k,k')}(\sigma)\label{eq:wichtigII}\end{eqnarray}
This also implies that $\de T(\sigma,\tet)$ and in particular $\de K(\sigma,\tet)$
do actually not depend on $\tet$:\begin{equation}
\de K^{(k,k')}(\sigma,\tet)=\de K^{(k,k')}(\sigma)\label{eq:wichtigIII}\end{equation}
 Now comes the nice part:

\paragraph{Proposition 1 }

\emph{For all multivector valued forms $K^{(k,k')},L^{(l,l')}$ on
the target space manifold, in a local coordinate patch seen as functions
of $x^{m}$,$\de x^{m}$ and $\pe_{m}$ as in (\ref{eq:multivector-valued-form-K}),
the following equation holds for the corresponding superfields (\ref{eq:K-sig-tet-def})
\begin{equation}
\hspace{-.4cm}\boxed{\{ K^{(k,k')}(\sigma',\tet'),L^{(l,l')}(\sigma,\tet)\}=\delta(\tet'-\tet)\delta(\sigma-\sigma')\cdot\underbrace{[\de K,L]_{(1)}^{\Delta}}_{\lqn{-(-)^{k-k'}\left[K,_{\de}L\right]_{(1)}^{\Delta}}}(\sigma,\tet)+\underbrace{\partial_{\tet}\delta(\tet-\tet')}_{=1}\delta(\sigma-\sigma')[K,L]_{(1)}^{\Delta}(\sigma,\tet)}\!\!\!\label{eq:Proposition1}\end{equation}
 where $[K,L]_{(1)}^{\Delta}$ is the big bracket (\ref{eq:bc-big-bracket})
(Buttin's algebraic bracket, which was previously just the Poisson
bracket, being true now up to a $\delta(\sigma-\sigma')$ only after
setting $\tet=\tet'$) and $\left[K,_{\de}L\right]_{(1)}^{\Delta}$
is the derived bracket of the big bracket (\ref{eq:bc-derived-of-bigbracket}).}\vspace{.5cm}\rem{aehnliche Prop waere fuer die Ersetzung wie im Antifeld-Fall denkbar}

\emph{Proof}$\quad$ Using (\ref{eq:wichtigII}), we can simply plug
$K(\sigma',\tet')=K(\sigma')+\tet'\de K(\sigma')$ and $L(\sigma,\tet)=L(\sigma)+\tet\de L(\sigma)$
into the lefthand side:\begin{eqnarray}
\lqn{\left\{ K(\sigma',\tet'),L(\sigma,\tet)\right\} =}\nonumber \\
 & = & \left\{ K(\sigma'),L(\sigma)\right\} +\tet'\left\{ \de K(\sigma'),L(\sigma)\right\} +(-)^{k-k'}\tet\left\{ K(\sigma'),\de L(\sigma)\right\} +(-)^{k-k'}\tet\tet'\left\{ \de K(\sigma'),\de L(\sigma)\right\} =\qquad\\
 & = & \left\{ K(\sigma'),L(\sigma)\right\} +(\tet'-\tet)\left\{ \de K(\sigma'),L(\sigma)\right\} +\tet\de\left\{ K(\sigma'),L(\sigma)\right\} -\tet\tet'\de\left\{ \de K(\sigma'),L(\sigma)\right\} =\\
 & \stackrel{(\ref{eq:bc-big-bracket})}{=} & \delta(\sigma-\sigma')\left(\left[K,L\right]_{(1)}^{\Delta}(\sigma)+\tet\de\left[K,L\right]_{(1)}^{\Delta}(\sigma)\right)+(\tet'-\tet)\delta(\sigma-\sigma')\left(\left[\de K,L\right]_{(1)}^{\Delta}(\sigma)+\tet\de\left[\de K,L\right]_{(1)}^{\Delta}(\sigma)\right)=\qquad\quad\\
 & \stackrel{(\ref{eq:wichtig})}{=} & \delta(\sigma-\sigma')\left[K,L\right]_{(1)}^{\Delta}(\sigma,\tet)+(\tet'-\tet)\delta(\sigma-\sigma')\left[\de K,L\right]_{(1)}^{\Delta}(\sigma,\tet)\qquad\square\end{eqnarray}

There is yet another way to see that the bracket at the plain delta
functions is the derived bracket of the one at the derivative of the
delta-function, which will be useful later: Denote the coefficients
in front of the delta-functions by $A(K,L)$ and $B(K,L)$: \begin{equation}
\left\{ K(\sigma',\tet'),L(\sigma,\tet)\right\} =A(K,L)\cdot\delta(\tet'-\tet)\delta(\sigma-\sigma')+B(K,L)(\sigma,\tet)\underbrace{\partial_{\tet}\delta(\tet-\tet')}_{=1}\delta(\sigma-\sigma')\end{equation}
In order to hit the delta-functions, it is enough to integrate over
a patch $U(\sigma)$ containing the point parametrized by $\sigma$.
We can thus extract $A$ and $B$ via\allowdisplaybreaks \begin{eqnarray}
A(K,L)(\sigma,\tet) & = & \int\de\tet'\int_{U(\sigma)}\msigp'\left\{ K(\sigma',\tet'),L(\sigma,\tet)\right\} =\\
 & = & \int\de\tet'\int\msigp'\left\{ K(\sigma')+\tet'\de K(\sigma'),L(\sigma,\tet)\right\} =\\
 & = & \int\msigp'\{\underbrace{\de K(\sigma')}_{\stackrel{(\ref{eq:wichtigIII})}{=}\de K(\sigma',\tet)},L(\sigma,\tet)\}\\
B(K,L)(\sigma,\tet) & = & \int\de\tet'\int_{U(\sigma)}\msigp'(\tet'-\tet)\left\{ K(\sigma',\tet'),L(\sigma,\tet)\right\} =\\
 & = & \int\msigp'\left\{ K(\sigma',\tet'),L(\sigma,\tet)\right\} \mid_{\tet'=\tet}\\
\dann A(K,L) & = & B(\de K,L)\end{eqnarray}
It is thus enough to collect in a direct calculation the terms at
the derivative of the delta-function and verify that it leads to the
big bracket.$\qquad\square$

\subsection{Comment on the quantum case}

\label{sub:Canonical-commutator} In (\ref{eq:quantization}) the
embedding via the interior product into the space of operators acting
on forms was interpreted as quantization . In the presence of world-volume
dimensions, the partial derivative as Schroedinger representation
for conjugate momenta is no longer appropriate and one has to switch
to the functional derivative. Remember\begin{eqnarray}
\Phi^{m}(\sigma,\tet) & = & x^{m}(\sigma)+\tet\ce^{m}(\sigma),\qquad\de\Phi^{m}(\sigma,\tet)=\ce^{m}(\sigma)=\de\Phi(\sigma)\\
\Es_{m}(\sigma,\tet) & = & \be_{m}(\sigma)+\tet p_{m}(\sigma),\qquad\de\Es_{m}(\sigma,\tet)=p_{m}(\sigma)=\de\Es(\sigma)\end{eqnarray}
The quantization of the superfields in the Schroedinger representation
(conjugate momenta as super functional derivatives) is consistent
with the quantization of the component fields (see also footnote \ref{foot:super-Poisson-bracket})\begin{eqnarray}
\hat{\Es}_{m}(\sigma,\tet) & \equiv & \frac{\hbar}{i}\funktional{}{\Phi^{m}(\sigma,\tet)}=\frac{\hbar}{i}\funktional{}{\ce^{m}(\sigma)}+\tet\frac{\hbar}{i}\funktional{}{x^{m}(\sigma)}\\
\dann\left[\hat{\Es}_{m}(\sigma,\tet),\hat{\Phi}^{n}(\sigma',\tet')\right] & = & \frac{\hbar}{i}\left(\funktional{}{\ce^{m}(\sigma)}+\tet\funktional{}{x^{m}(\sigma)}\right)\left(x^{n}(\sigma')+\tet'\ce^{n}(\sigma')\right)=\\
 & = & \frac{\hbar}{i}\delta_{m}^{n}\left(\tet-\tet'\right)\delta(\sigma-\sigma')\end{eqnarray}
The quantization of a multivector valued form, containing several
operators $\hat{\Es}$ at the same worldvolume-point, however, leads
to powers of delta functions with the same argument when acting on
some wave functional. This is the usual problem in quantum field theory
and requires a model dependent regularization and renormalization.
We will stay model independent here and therefore will not treat the
quantum case for a present worldvolume coordinate $\sigma$.  Nevertheless
it is instructive to study it for absent $\sigma$, but keeping $\tet$
and considering {}``worldline-superfields'' of the form \begin{eqnarray}
\Phi^{m}(\tet) & = & x^{m}+\tet\ce^{m},\qquad\de\Phi^{m}(\tet)=\ce^{m}\\
\Es_{m}(\tet) & = & \be_{m}+\tet p_{m},\qquad\de\Es_{m}(\tet)=p_{m}\end{eqnarray}
Quantum operator and commutator simplify to\begin{eqnarray}
\hat{\Es}_{m}(\tet) & \equiv & \frac{\hbar}{i}\funktional{}{\Phi^{m}(\tet)}=\frac{\hbar}{i}\partl{\ce^{m}}+\tet\frac{\hbar}{i}\partl{x^{m}}\\
\dann\left[\hat{\Es}_{m}(\tet),\hat{\Phi}^{n}(\tet')\right] & = & \frac{\hbar}{i}\delta_{m}^{n}\left(\tet-\tet'\right)\\
\left[\hat{\Es}_{m}(\tet),\widehat{\de\Phi}^{n}(\tet')\right] & = & \frac{\hbar}{i}\delta_{m}^{n}\end{eqnarray}
\rem{$\de\hat{\Phi}=\widehat{\de\Phi}$ in the sense that $\left[\de\,,\ip_{K}\right]=\ip_{\de K}$.
}In contrast to $\sigma$, products of $\tet$-delta functions are
no problem. 

The important relation $K(\tet)=K+\tet\de K$ (\ref{eq:wichtigII})
can be extended to the quantum case as seen when acting on some $r$-form.
\begin{eqnarray}
\ip_{K^{(k,k')}}\rho^{(r)}(\tet) & \stackrel{(\ref{eq:wichtig})}{=} & \ip_{K}\rho+\tet\de(\ip_{K}\rho)=\\
 & \stackrel{(\ref{eq:dK-und-Lie})}{=} & \ip_{K}\rho+\tet\left(\ip_{\de K}\rho+(-)^{k-k')}\ip_{K}\de\rho\right)=\\
 & = & \ip_{K}(\tet)\left(\rho(\tet)\right)\\
\textrm{with }\ip_{K}(\tet) & \equiv & \ip_{K}+\tet\left[\de\,,\ip_{K}\right]\label{eq:iK-of-tet}\end{eqnarray}
In that sense we have (remember $\hat{K}=\left(\frac{\hbar}{i}\right)^{k'}\ip_{K}$)\begin{eqnarray}
\hat{K}^{(k,k')}(\tet) & = & \hat{K}^{(k,k')}+\tet\widehat{\de K}\label{eq:wichtig-quant}\\
\textrm{with }\widehat{\de K} & \stackrel{(\ref{eq:dK-und-Lie})}{=} & \left[\de\,,\hat{K}\right]\label{eq:dK-und-Lie-quant}\end{eqnarray}
where the explicit form of this quantized multivector valued form
reads \begin{eqnarray}
\hat{K}^{(k,k')}(\tet) & \equiv & \left(\frac{\hbar}{i}\right)^{k'}K_{m_{1}\ldots m_{k}}\hoch{n_{1}\ldots n_{k'}}\left(\Phi(\tet)\right)\,\underbrace{\de\Phi^{m_{1}}(\tet)}_{\ce^{m_{1}}}\ldots\de\Phi^{m_{k}}(\tet)\funktional{}{\Phi^{n_{1}}(\tet)}\ldots\funktional{}{\Phi^{n_{k'}}(\tet)}\label{eq:K-quant}\end{eqnarray}
In the derivation of (\ref{eq:iK-of-tet}), $\ip_{K}$ and $\rho$
both were evaluated at the same $\tet$. Let us eventually consider
the general case: \begin{eqnarray}
\hat{K}^{(k,k')}(\tet')\rho^{(r)}(\tet) & = & \left(\hat{K}+\tet'\widehat{\de K}\right)\left(\rho+\tet\de\rho\right)=\\
 & = & \hat{K}\rho+\tet'\widehat{\de K}\rho+(-)^{k-k'}\tet\hat{K}\de\rho+(-)^{k-k'}\tet\tet'\widehat{\de K}\de\rho=\\
 & = & \hat{K}\rho+\tet\de\left(\widehat{K}\rho\right)+(\tet'-\tet)\left(\widehat{\de K}\rho+\tet\de\left(\widehat{\de K}\rho\right)\right)\end{eqnarray}
The relation between quantum operators acting on forms and the interior
product therefore becomes modified in comparison to (\ref{eq:quantization})
and reads\begin{equation}
\boxed{\hat{K}^{(k,k')}(\tet')\rho^{(r)}(\tet)=\left(\frac{\hbar}{i}\right)^{k'}\Big(\ip_{K}\rho(\tet)+(\tet'-\tet)\underbrace{\ip_{\de K}\rho(\tet)}_{(-)^{k-k'}\Lie_{K}\rho}\Big)}\label{eq:quantization-for-superfields}\end{equation}

\paragraph{Proposition 2 }

\emph{For all multivector valued forms $K^{(k,k')},L^{(l,l')}$ on
the target space manifold, in a local coordinate patch seen as functions
of $x^{m}$,$\de x^{m}$ and $\pe_{m}$ as in (\ref{eq:multivector-valued-form-K}),
the following equations holds for the corresponding quantized worldline-superfields
(\ref{eq:K-quant}) $\hat{K}(\tet)$ and $\hat{L}(\tet)$:}\\
\emph{\Ram{1}{\begin{eqnarray}
[\hat{K}^{(k,k')}(\tet'),\hat{L}^{(l,l')}(\tet)] & = & \sum_{p\geq1}\left(\frac{\hbar}{i}\right)^{p}\Big(\underbrace{\partial_{\tet}\delta(\tet-\tet')}_{=1}\widehat{[K,L]_{(p)}^{\Delta}}(\tet)+\delta(\tet'-\tet)\widehat{[\de K,L]_{(p)}^{\Delta}}(\tet)\Big)\label{eq:Proposition2}\\
{}[\hat{K}^{(k,k')}(\tet'),\hat{L}^{(l,l')}(\tet)]\lqn{\rho(\tilde{\tet})=}\nonumber \\
 &  & \hspace{-2cm}=\left(\frac{\hbar}{i}\right)^{k'+l'}\Big(\ip_{\left[K,L\right]^{\Delta}}\rho^{(r)}(\tilde{\tet})+\delta(\tet-\tilde{\tet})\ip_{\de\left[K,L\right]^{\Delta}}\rho^{(r)}(\tilde{\tet})+\nonumber \\
 &  & \hspace{-2cm}\qquad\qquad\qquad\qquad+\delta(\tet'-\tet)\left(\ip_{\left[\de K,L\right]^{\Delta}}\rho^{(r)}(\tilde{\tet})+\delta(\tet-\tilde{\tet})\ip_{\de\left[\de K,L\right]^{\Delta}}\rho^{(r)}(\tilde{\tet})\right)\Big)\end{eqnarray}
} Again the algebraic bracket (\ref{eq:algebraic-bracketI}) comes
with the derivative of the delta function while the derived bracket
(\ref{eq:bc-derived-bracketI}) comes with the plain delta functions.
But this time the algebraic bracket is not only the big bracket $\left[\,,\,\right]_{(1)}^{\Delta}$,
but the full one.} \vspace{.5cm}

\emph{Proof}$\quad$Let us just plug in (\ref{eq:wichtig-quant})
into the lefthand side:\begin{eqnarray}
[\hat{K}(\tet'),\hat{L}(\tet)] & = & [\hat{K}+\tet'\widehat{\de K}\,,\,\hat{L}+\tet\widehat{\de L}]=\\
 & = & [\hat{K},\hat{L}]+\tet'[\widehat{\de K}\,,\,\hat{L}]+(-)^{k-k'}\tet[\hat{K}\,,\,\widehat{\de L}]-(-)^{k-k'}\tet'\tet[\widehat{\de K}\,,\,\widehat{\de L}]=\\
 & \stackrel{(\ref{eq:dK-und-Lie-quant})}{=} & [\hat{K},\hat{L}]+\tet\left[\de\,,[\hat{K}\,,\,\hat{L}]\right]+(\tet'-\tet)\left([\widehat{\de K}\,,\,\hat{L}]+\tet\left[\de\,,[\widehat{\de K}\,,\,\hat{L}]\right]\right)=\\
 & = & [\hat{K},\hat{L}](\tet)+(\tet'-\tet)[\widehat{\de K}\,,\,\hat{L}]\label{eq:lila-Pause}\end{eqnarray}
Remember now the algebraic bracket (\ref{eq:algebraic-bracket})\begin{eqnarray}
[\ip_{K^{(k,k')}},\ip_{L^{(l,,l')}}] & = & \ip_{[K,L]^{\Delta}}=\sum_{p\geq1}\ip_{[K,L]_{(p)}^{\Delta}}\\
\textrm{with }\left[K,L\right]_{(p)}^{\Delta} & \equiv & \ip_{K}^{(p)}L-(-)^{(k-k')(l-l')}\ip_{L}^{(p)}K\end{eqnarray}
or likewise written in terms of $\hat{K}$ and $\hat{L}$ \bref{eq:quantum-commutator1}\begin{equation}
[\hat{K}^{(k,k')},\hat{L}^{(l,l')}]=\sum_{p\geq1}\left(\frac{\hbar}{i}\right)^{p}\widehat{\left[K,L\right]_{(p)}^{\Delta}}\end{equation}
\eref Due to (\ref{eq:quantum-commutator-T-Ttilde}) we have exactly
the same equation for $[\widehat{\de K}\,,\,\hat{L}]$. Plugging this
back into (\ref{eq:lila-Pause}) completes the proof of (\ref{eq:Proposition2}).
The second equation in the proposition is just a simple rewriting,
when acting on a form, which enables to combine the $p$-th terms
of algebraic and derived bracket to the complete ones.$\qquad\square$

\subsection{Analogy for the antibracket}

\label{sub:antibracket} 
\newcommand{\ix}{\bs{x}^{+}}
 In the previous subsection the target space exterior derivative $\de$
(realized in the $\sigma$-model phase-space by $\es$) was induced
by the the derivative $\partial_{\tet}$ with respect to the anticommuting
coordinate. But thinking of the pullback of forms in the target space
to worldvolume-forms, $\de$ can of course also be induced to some
extend by the derivative with respect to the bosonic worldvolume coordinates
$\sigma^{\mu}$ (including the time, because we are in the Lagrangian
formalism now) or better by the worldvolume exterior derivative $\dew$.
To this end, however, we have to make a different identification of
the basis elements in tangent- and cotangent-space of the target space
with the fields on the worldvolume than before, namely%
\footnote{\label{foot:Strobl}This identification resembles the one in \cite{Alekseev:2004np}
with $\pe_{m}\To p_{m}(z)$ and $\de x^{m}\To\partial x^{m}(z)$,
or $\de x^{m_{1}}\cdots\de x^{m_{p}}\To\epsilon^{\mu_{1}\ldots\mu_{p}}\partial_{\mu_{1}}x^{m_{1}}(\sigma)\cdots\partial_{\mu_{p}}x^{m_{p}}(\sigma)$
in \cite{Bonelli:2005ti}. It is observed in \cite{Alekseev:2004np}
that the Poisson bracket induces the Dorfman bracket between sums
of vectors and 1-forms (in generalized geometry) and in \cite{Bonelli:2005ti}
more generally that the Poisson-bracket for the $p$-brane induces
the corresponding bracket between sums of vectors and $p$-forms (which
is called, Vinogradov bracket in \cite{Bonelli:2005ti}). As $\partial x^{m}$
and $p_{m}$ are commuting phase space variables, higher rank tensors
would automatically be symmetrized (only volume forms, i.e. p-forms
on a p-brane, can be implemented, using the epsilon-tensor). Symmetrized
tensors and brackets inbetween (e.g. the Schouten bracket for symmetric
multivectors) make sense and one could transfer the present analysis
to this setting, but in general a natural exterior derivative is missing.
Therefore the analysis for the above identifications is done in the
antifield-formalism. The appearing derived brackets will also contain
the Dorfman bracket and the corresponding bracket for sums of vectors
and p-forms and in that sense the present approach is a generalization
of the observations above.$\qquad\fussend$%
} \begin{eqnarray}
\de x^{m} & \To & \dew x^{m}(\sigma)=\dew\sigma^{\mu}\partial_{\mu}x^{m}(\sigma),\qquad\qquad\pe_{m}\To\ix_{m}(\sigma)\label{eq:indentification-of-basis-elements}\end{eqnarray}
where $\ix_{m}$ is the antifield of $x^{m}$, i.e. the conjugate
field to $x^{m}$ with respect to the antibracket%
\footnote{The antibracket looks similar to the Poisson-bracket, but their conjugate
fields have opposite parity, which leads to a different symmetry (namely
that of a Lie-bracket of degree +1 (or -1), i.e. the one in a Gerstenhaber
algebra or Schouten-algebra, see footnote \ref{Lie-bracket-of-degree})
\rem{evtl spaeter nimmer footnote}\begin{eqnarray*}
\left(A\bs{,}B\right) & \equiv & \int\msig\backtilde\quad\big(\delta A/\ix_{k}(\tilde{\sigma})\funktional{}{x^{k}(\tilde{\sigma})}B-\delta A/\delta x^{k}(\tilde{\sigma})\funktional{}{\ix_{k}(\tilde{\sigma})}B\big)=\\
 & = & \int\msig\backtilde\quad\big(\delta A/\ix_{k}(\tilde{\sigma})\funktional{}{x^{k}(\tilde{\sigma})}B-(-)^{(A+1)(B+1)}\delta B/\ix_{k}(\tilde{\sigma})\funktional{}{x^{k}(\tilde{\sigma})}A\big)\\
\left(A\bs{,}B\right) & = & -(-)^{(A+1)(B+1)}\left(B\bs{,}A\right)\\
\left(\ix_{m}(\sigma)\bs{,}B\right) & = & \funktional{}{x^{m}(\sigma)}B=-\left(B\bs{,}\ix_{m}(\sigma)\right)\\
\left(x^{m}(\sigma)\bs{,}B\right) & = & -\funktional{}{\ix_{m}(\sigma)}B=(-)^{B}\left(B\bs{,}x^{m}(\sigma)\right)\qquad\fussend\end{eqnarray*}
}. Let us rename \begin{eqnarray}
\tet^{\mu} & \equiv & \dew\sigma^{\mu}\end{eqnarray}
For a target space $r$-form \begin{eqnarray}
\rho^{(r)}(x^{m},\de x^{m}) & \equiv & \rho_{m_{1}\ldots m_{r}}(x)\de x^{m_{1}}\cdots\de x^{m_{r}}\end{eqnarray}
we define (in analogy to (\ref{eq:K-sig-tet-def}), but indicating
that we allow in the beginning only a variation in $\sigma$)\begin{eqnarray}
\rho_{\tet}^{(r)}(\sigma) & \equiv & \rho^{(r)}(x^{m}(\sigma),\dew x^{m}(\sigma))=\rho_{m_{1}\ldots m_{r}}(x(\sigma))\dew x^{m_{1}}(\sigma)\cdots\dew x^{m_{r}}(\sigma)\end{eqnarray}
\textbf{Attention:} this vanishes identically for $r>d_{\textrm{w}}$
(worldvolume dimension). 

The worldvolume exterior derivative then induces the target space
exterior derivative in the following sense\begin{eqnarray}
\dew\rho_{\tet}^{(r)}(\sigma) & = & (\de\rho^{(r)})_{\tet}(\sigma)\end{eqnarray}
Again both sides vanish identically for now $r+1>d_{\textrm{w}}$,
which means that in this way one can calculate with target space fields
of form degree not bigger than the worldvolume dimension. If we want
to have the same relation for $K_{\tet}^{(k,k')}(\sigma)$ (defined
in the analogous way), we have to extend the identification in (\ref{eq:indentification-of-basis-elements})
by \begin{eqnarray}
p_{m} & \To & \dew\ix_{m}(\sigma)\label{eq:identification-of-basis-elements-p}\end{eqnarray}
and get \begin{eqnarray}
\dew K_{\tet}^{(k,k')}(\sigma) & = & (\de K^{(k,k')})_{\tet}(\sigma)\end{eqnarray}
with \begin{eqnarray}
K_{\tet}^{(k,k')}(\sigma) & \equiv & K^{(k,k')}\left(x^{m}(\sigma),\dew x^{m}(\sigma),\ix_{m}(\sigma)\right)\label{eq:sigma-model-realizationI}\\
(\de K^{(k,k')})_{\tet}(\sigma) & \equiv & \de K^{(k,k')}\left(x^{m}(\sigma),\dew x^{m}(\sigma),\ix_{m}(\sigma),\dew\ix_{m}(\sigma)\right)\label{eq:sigma-model-realization-II}\end{eqnarray}
The analysis is thus very similar to that of the previous section.
\rem{hier ist tildeOmega usw versteckt}

\paragraph{Proposition 3a}

\emph{For all multivector valued forms $K^{(k,k')},L^{(l,l')}$ on
the target space manifold, in a local coordinate patch seen as functions
of $x^{m}$,$\de x^{m}$ and $\pe_{m}$, the following equation holds
for the corresponding sigma-model realizations (\ref{eq:sigma-model-realizationI},\ref{eq:sigma-model-realization-II})\begin{equation}
\boxed{(K_{\tet}(\sigma')\bs{,}L_{\tet}(\sigma))=\big(\underbrace{[K,_{\de}L]_{(1)}^{\Delta}}_{\lqn{-(-)^{k-k'}\left[\de K,L\right]_{(1)}^{\Delta}}}\big)_{\tet}(\sigma)\delta^{d_{\textrm{w}}}(\sigma-\sigma')-(-)^{k-k'}\tet^{\mu}\partial_{\mu}\delta^{d_{\textrm{w}}}(\sigma-\sigma')\big(\left[K,L\right]_{(1)}^{\Delta}\big)_{\tet}(\sigma)}\label{eq:PropositionIII}\end{equation}
}

\emph{Proof}$\quad$The proof is very similar to that one of proposition
3b (\ref{eq:PropositionIIIb}) and is therefore omitted at this place.$\quad\square$

\paragraph{Conjugate Superfields}

\newcommand{\Ph}{\bs{\Phi}^{+}}
 With $\tet^{\mu}=\dew\sigma^{\mu}$ we have introduced anticommuting
coordinates and it would be nice to extend the anti-bracket of the
fields $x^{m}$ and $\ix_{m}$ to a super-antibracket of conjugate
superfields. Remember, in the previous subsection we had the superfields
$\Phi^{m}=x^{m}+\tet\ce^{m}$ and its conjugate $\Es_{m}$. There
we had one $\tet$ and two component fields. In general the number
of component fields has to exceed the worldvolume dimension $d_{\textrm{w}}$
(the number of $\tet$'s) by one, s.th. we have to introduce a lot
of new fields to realize conjugate superfields. But before, let us
define the fermionic integration measure $\mu(\tet)$ via \begin{equation}
\int\mu(\tet)f(\tet)=\partl{\tet^{d_{\textrm{w}}}}\cdots\partl{\tet^{1}}f(\tet)=\frac{1}{d_{\textrm{w}}!}\epsilon^{\mu_{1}\ldots\mu_{d_{\textrm{w}}}}\partl{\tet^{\mu_{1}}}\cdots\partl{\tet^{\mu_{d_{\textrm{w}}}}}f(\tet)\end{equation}
The corresponding $d_{\textrm{w}}$-dimensional $\delta$-function
is\begin{eqnarray}
\delta^{d_{\textrm{w}}}(\tet'-\tet) & \equiv & (\tet'^{1}-\tet^{1})\cdots(\tet'^{d_{\textrm{w}}}-\tet^{d_{\textrm{w}}})=\\
 & = & \frac{1}{d_{\textrm{w}}!}\epsilon_{\mu_{1}\ldots\mu_{d_{\textrm{w}}}}(\tet'^{\mu_{1}}-\tet^{\mu_{1}})\cdots(\tet'^{\mu_{d_{\textrm{w}}}}-\tet^{\mu_{d_{\textrm{w}}}})=\\
 & = & \sum_{k=0}^{d_{\textrm{w}}}\frac{1}{k!(d_{\textrm{w}}-k)!}\epsilon_{\mu_{1}\ldots\mu_{d_{\textrm{w}}}}\tet'^{\mu_{1}}\cdots\tet'^{\mu_{k}}\tet^{\mu_{k+1}}\cdots\tet^{\mu_{d_{\textrm{w}}}}\\
\int\mu(\tet')\delta^{d_{\textrm{w}}}(\tet'-\tet)f(\tet') & = & f(\tet)\\
\delta^{d_{\textrm{w}}}(\tet'-\tet) & = & (-)^{d_{\textrm{w}}}\delta^{d_{\textrm{w}}}(\tet-\tet')\end{eqnarray}
For the two conjugate superfields, call them $\Phi^{m}$ and $\Ph_{m}$,
we want to have the canonical super anti bracket\begin{equation}
\left(\Ph_{m}(\sigma',\tet')\bs{,}\Phi^{n}(\sigma,\tet)\right)=\delta_{m}^{n}\delta^{d_{\textrm{w}}}(\sigma'-\sigma)\delta^{d_{\textrm{w}}}(\tet'-\tet)=-\left(\Phi^{n}(\sigma,\tet),\Ph_{m}(\sigma',\tet')\right)\label{eq:super-antibracket}\end{equation}
From the above considerations about the fermionic delta function it
is now clear, how these superfields can be defined (they are known
as \textbf{de Rham superfields}, because of the interpretation of
$\tet^{\mu}$ as $\dew\sigma^{\mu}$; see e.g. \cite{Cattaneo:1999fm,Zucchini:2004ta}):\begin{eqnarray}
\Phi^{m}(\sigma,\tet) & \equiv & x^{m}(\sigma)+\bs{x}_{\mu_{d_{\textrm{w}}}}^{m}(\sigma)\tet^{\mu_{d_{\textrm{w}}}}+x_{\mu_{d_{\textrm{w}}-1}\mu_{d_{\textrm{w}}}}^{m}(\sigma)\tet^{\mu_{d_{\textrm{w}}-1}}\tet^{\mu_{d_{\textrm{w}}}}+\ldots+x_{\mu_{1}\ldots\mu_{d_{\textrm{w}}}}^{m}(\sigma)\tet^{\mu_{1}}\cdots\tet^{\mu_{d_{\textrm{w}}}}\label{eq:deRhamSuperfieldPhi}\\
\Ph_{m}(\sigma',\tet') & \equiv & \frac{1}{d_{\textrm{w}}!}\epsilon_{\mu_{1}\ldots\mu_{d_{\textrm{w}}}}\tet'^{\mu_{1}}\cdots\tet'^{\mu_{d_{\textrm{w}}}}\ix_{m}(\sigma')+\frac{1}{(d_{\textrm{w}}-1)!1!}\epsilon_{\mu_{1}\ldots\mu_{d_{\textrm{w}}}}\tet'^{\mu_{1}}\cdots\tet'^{\mu_{d_{\textrm{w}}-1}}x_{m}^{+}\hoch{\mu_{d_{\textrm{w}}}}(\sigma')+\nonumber \\
 &  & \hspace{-1.5cm}+\frac{1}{(d_{\textrm{w}}-2)!2!}\epsilon_{\mu_{1}\ldots\mu_{d_{\textrm{w}}}}\tet'^{\mu_{1}}\cdots\tet'^{\mu_{d_{\textrm{w}}-2}}\ix_{m}\hoch{\mu_{d_{\textrm{w}}-1}\mu_{d_{\textrm{w}}}}(\sigma')+\ldots+\frac{1}{d_{\textrm{w}}!}\epsilon_{\mu_{1}\ldots\mu_{d_{\textrm{w}}}}\ix_{m}\hoch{\mu_{1}\ldots\mu_{d_{\textrm{w}}}}(\sigma')\qquad\label{eq:deRhamSuperfieldPhipl}\end{eqnarray}
The component fields with the matching number of worldsheet indices
are conjugate to each other, e.g.\begin{eqnarray}
\left(\ix_{m}\hoch{\mu_{1}\mu_{2}}(\sigma')\bs{,}x_{\nu_{1}\nu_{2}}^{n}(\sigma)\right) & = & \delta_{m}^{n}\delta_{\nu_{1}\nu_{2}}^{\mu_{1}\mu_{2}}\delta^{d_{\textrm{w}}}(\sigma-\sigma')\end{eqnarray}
For the notation with boldface symbols for anticommuting variables,
the worldvolume was assumed to be even-dimensional. In this case,
one can analytically continue the coordinate form of multivector-valued
forms of the form \begin{equation}
K^{(k,k')}(x,\de x,\pe)\equiv K_{m_{1}\ldots m_{k}}\hoch{n_{1}\ldots n_{k'}}\de x^{m_{1}}\wedge\cdots\wedge\de x^{m_{k}}\wedge\pe_{n_{1}}\wedge\cdots\wedge\pe_{n_{k'}}\end{equation}
to functions of superfields (in odd worldvolume dimension one would
get a symmetrization of the multivector-indices) and redefine $K(\sigma,\tet)$
of (\ref{eq:K-sig-tet-def}) to \begin{eqnarray}
K^{(k,k')}(\sigma,\tet) & \equiv & K^{(k,k')}\left(\Phi(\sigma,\tet),\dew\Phi(\sigma,\tet),\Ph(\sigma,\tet)\right)=\label{eq:K-sig-tet-def-II}\\
 & = & K_{m_{1}\ldots m_{k}}\hoch{n_{1}\ldots n_{k'}}(\Phi)\dew\Phi^{m_{1}}\cdots\dew\Phi^{m_{k}}\Ph_{n_{1}}\cdots\Ph_{n_{k'}}\end{eqnarray}
All other geometric quantities have to be understood in this new sense
now:\begin{eqnarray}
T^{(t,t',t'')}(\sigma,\tet) & \equiv & T^{(t,t',t'')}\left(\Phi(\sigma,\tet),\dew\Phi(\sigma,\tet),\Ph(\sigma,\tet),\dew\Ph(\sigma,\tet)\right)\quad(\textrm{see }(\ref{eq:Tcbp}))\label{eq:T-sig-tet-II}\end{eqnarray}
 To stay with the examples used in (\ref{eq:T-sig-tet})-(\ref{eq:db-sig-tet}):\begin{eqnarray}
\textrm{e.g. }\de K(\sigma,\tet) & \equiv & \de K\left(\Phi(\sigma,\tet),\dew\Phi(\sigma,\tet),\Ph(\sigma,\tet),\dew\Ph(\sigma,\tet)\right)\qquad(\textrm{compare }(\ref{eq:dK-coord}))\label{eq:dK-of-sig-tet-II}\\
\textrm{or }\oo(\sigma,\tet) & \equiv & \oo\left(\dew\Phi(\sigma,\tet),\dew\Ph(\sigma,\tet)\right)=\dew\Phi^{m}(\sigma,\tet)\dew\Ph_{m}(\sigma,\tet)\quad(\textrm{compare }\oo=\ce^{m}p_{m})\qquad\quad\label{eq:o-of-sig-tet-II}\\
\hspace{-1.2cm}\left[K^{(k,k')},_{\de\,}L^{(l,l')}\right]_{(1)}^{\Delta}(\sigma,\tet) & \equiv & \left[K^{(k,k')}\bs{,}L^{(l,l')}\right]_{(1)}^{(\Delta)}\left(\Phi(\sigma,\tet),\dew\Phi(\sigma,\tet),\Ph(\sigma,\tet),\dew\Ph(\sigma,\tet)\right)\\
\de x^{m}(\sigma,\tet) & \equiv & \dew\Phi^{m}(\sigma,\tet)\\
(\de\pe_{m})(\sigma,\tet)\equiv(\de\be_{m})(\sigma,\tet) & \equiv & \dew\Ph_{m}(\sigma,\tet)\label{eq:db-sig-tet-II}\end{eqnarray}
Note that the former relation $K(\sigma,\tet)=K(\sigma)+\tet\de K(\sigma)$
does NOT hold any longer with those new definitions! Nevertheless
we get a very similar statement as compared to propositions 2 on page
\pageref{eq:Proposition1}:\rem{stattdessen $K(\sigma,\tet)=K(\tet)+\sigma^{\mu}\partial_{\mu}K(\tet)+\ldots$
oder (?) $K(\sigma+\tet)=K(\sigma)+\tet^{\mu}\partial_{\mu}K(\sigma)$}

\paragraph{Proposition 3b}

\emph{For all multivector valued forms $K^{(k,k')},L^{(l,l')}$ on
the target space manifold, in a local coordinate patch seen as functions
of $x^{m}$,$\de x^{m}$ and $\pe_{m}$, the following equation holds
for even worldvolume-dimension $d_{\textrm{w}}$ for the corresponding
superfields (\ref{eq:K-sig-tet-def-II}):\begin{equation}
\boxed{(K(\sigma',\tet')\bs{,}L(\sigma,\tet))=\delta^{d_{\textrm{w}}}(\sigma'-\sigma)\delta^{d_{\textrm{w}}}(\tet'-\tet)\underbrace{\left[K,_{\de}L\right]_{(1)}^{\Delta}}_{\lqn{-(-)^{k-k'}\left[\de K,L\right]_{(1)}^{\Delta}}}(\sigma,\tet)-(-)^{k-k'}\tet^{\mu}\partial_{\mu}\delta^{d_{\textrm{w}}}(\sigma-\sigma')\delta^{d_{\textrm{w}}}(\tet'-\tet)\left[K,L\right]_{(1)}^{\Delta}(\sigma,\tet)}\label{eq:PropositionIIIb}\end{equation}
where $[K,L]_{(1)}^{\Delta}$ is the big bracket (\ref{eq:bc-big-bracket})
and $\left[K,_{\de}L\right]_{(1)}^{\Delta}$ is the derived bracket
of the big bracket (\ref{eq:bc-derived-of-bigbracket}).}

\emph{Note that $\sigma$ and $\tet$ have switched their roles compared
to the previous subsection (\ref{eq:Proposition1}), where the algebraic
bracket came together with the derivative with respect to $\tet$
of the delta-functions, while now it comes along with $\partial_{\mu}$
of the delta-functions.\vspace{.5cm}}

\emph{Proof}$\quad$Let us use again the second idea in the proof
of proposition 2, i.e. first collect the terms with derivatives of
the delta function, only to show that one gets the algebraic bracket,
and after that argue that the term with plain delta functions is its
derived bracket. In doing this, however, we will need to prove an
extension of the above proposition to objects like $\de K$ (or more
general an object $T^{(t,t',t'')}$ as in (\ref{eq:Tcbp})) that contain
the basis element $p_{m}$, which is then replaced by $\dew\Ph_{m}$
as e.g. in (\ref{eq:dK-of-sig-tet-II}).\\
(i) The antibracket between two such objects $T$ and $\tilde{T}$
gets contributions to the derivative of the delta-function only from
the antibrackets between $\dew\Phi^{m}$ and $\Ph_{m}$ and between
$\Phi^{m}$ and $\dew\Ph_{m}$ (compare (\ref{eq:super-antibracket}))\begin{eqnarray}
\left(\Ph_{m}(\sigma',\tet')\bs{,}\dew\Phi^{n}(\sigma,\tet)\right) & = & \delta_{m}^{n}\tet^{\mu}\partial_{\mu}\delta^{d_{\textrm{w}}}(\sigma'-\sigma)\delta^{d_{\textrm{w}}}(\tet'-\tet)\\
\left(\dew\Phi^{n}(\sigma',\tet')\bs{,}\Ph_{m}(\sigma,\tet)\right) & = & \delta_{m}^{n}\tet^{\mu}\partial_{\mu}\delta^{d_{\textrm{w}}}(\sigma'-\sigma)\delta^{d_{\textrm{w}}}(\tet'-\tet)\\
\left(\dew\Ph_{m}(\sigma',\tet')\bs{,}\Phi^{n}(\sigma,\tet)\right) & = & -\delta_{m}^{n}\tet^{\mu}\partial_{\mu}\delta^{d_{\textrm{w}}}(\sigma'-\sigma)\delta^{d_{\textrm{w}}}(\tet'-\tet)\\
\left(\Phi^{n}(\sigma',\tet')\bs{,}\dew\Ph_{m}(\sigma,\tet)\right) & = & -\tet^{\mu}\left(\Phi^{n}(\sigma',\tet')\bs{,}\partial_{\mu}\Ph_{m}(\sigma,\tet)\right)=\delta_{m}^{n}\tet^{\mu}\partial_{\mu}\delta^{d_{\textrm{w}}}(\sigma'-\sigma)\delta^{d_{\textrm{w}}}(\tet'-\tet)\end{eqnarray}
The last case is the only one where we had to take care of an extra
sign stemming from $\tet$ jumping over the graded comma. Comparing
this to (\ref{eq:Poisson-bracket-bc}), where we had \begin{eqnarray}
\left\{ \be_{m},\ce^{n}\right\}  & = & \delta_{m}^{n}\\
\left\{ \ce^{n},\be_{m}\right\}  & = & \delta_{m}^{n}\\
\left\{ p_{m},x^{n}\right\}  & = & \delta_{m}^{n}\\
\left\{ x^{n},p_{m}\right\}  & = & -\delta_{m}^{n}\end{eqnarray}
one recognizes that the only difference is an overall odd factor $\tet^{\mu}\partial_{\mu}\delta^{d_{\textrm{w}}}(\sigma'-\sigma)\delta^{d_{\textrm{w}}}(\tet'-\tet)$
(the delta-function for $\tet$ is an even object for even worldvolume
dimension $d_{\textrm{w}}$) and an additional minus sign for the
lower two lines, but the corresponding indices just get contracted
like for the Poisson bracket. After such a bracket of basis elements
has been calculated (which happens just between the remaining factors
of $T$ (at $\sigma'$) on the left and the remaining factors of $\tilde{T}$
(at $\sigma$) on the right) this overall odd factor has to be pulled
to the very left which gives an additional factor of $(-)^{t-t'}$
(in the notation of (\ref{eq:Tcbp})) plus an additional minus sign
for the upper two lines which compensates the relative minus sign
of before and we get just an overall factor of $-(-)^{t-t'}\tet^{\mu}\partial_{\mu}\delta^{d_{\textrm{w}}}(\sigma'-\sigma)\delta^{d_{\textrm{w}}}(\tet'-\tet)$
in all cases at the very left as compared to the Poisson-bracket.
The remaining terms are still partly at $\sigma$ and partly at $\sigma'$,
but using\begin{equation}
A(\sigma)B(\sigma')\partial_{\mu}\delta(\sigma-\sigma')=A(\sigma)\partial_{\mu}B(\sigma)\delta(\sigma-\sigma')+A(\sigma)B(\sigma)\partial_{\mu}\delta(\sigma-\sigma')\quad\forall A,B\label{eq:hint}\end{equation}
we can take all remaining factors in $T(\sigma',\tet')$ at $\sigma$,
while $\tet'$ is set to $\tet$ anyway by the $\delta$-function.
We have thus verified one of the coefficients of the complete antibracket:\begin{eqnarray}
(T(\sigma',\tet'),\tilde{T}(\sigma,\tet)) & = & -(-)^{t-t'}\tet^{\mu}\partial_{\mu}\delta^{d_{\textrm{w}}}(\sigma-\sigma')\delta^{d_{\textrm{w}}}(\tet'-\tet)\left[T,\tilde{T}\right]_{(1)}^{\Delta}(\sigma,\tet)+\nonumber \\
 &  & +\delta^{d_{\textrm{w}}}(\sigma-\sigma')\delta^{d_{\textrm{w}}}(\tet'-\tet)A(\sigma,\tet)\label{eq:gruenerPunkt}\end{eqnarray}
with $A(\sigma,\tet)$ yet to be determined.\\
(ii) It remains to show that $A(\sigma,\tet)$ is a derived expression
of $\left[T,\tilde{T}\right]_{(1)}^{\Delta}$. A hint to this fact
is already given in (\ref{eq:hint}), but this is not enough, as there
is also a contribution from the $(\Phi^{m},\Ph_{n})$-brackets. In
order to get a precise relation between $A(\sigma,\tet)$ and $\left[T,\tilde{T}\right]_{(1)}^{\Delta}(\sigma,\tet)$,
let us see how one can extract them from the complete antibracket.
In order to hit the delta functions with the integration, it is enough
to integrate over the patch $U(\sigma)$ containing the point which
is parametrized by $\sigma^{\mu}$. The last term in (\ref{eq:gruenerPunkt})
is the only one contributing when integrating over $\sigma'$ and
$\tet$\begin{eqnarray}
A(\sigma,\tet) & = & \int_{U(\sigma)}\de^{d_{\textrm{w}}}\sigma'\int\mu(\tet')\quad(T(\sigma',\tet'),\tilde{T}(\sigma,\tet))\end{eqnarray}
That the first term on the righthand side of (\ref{eq:gruenerPunkt})
does not contribute is not obvious as $U(\sigma)$ might have a boundary.
However, for this term one ends up integrating a $d_{\textrm{w}}$-dimensional
delta-function over a boundary of dimension not higher than $d_{\textrm{w}}-1$,
so that one is left with an at least one-dimensional delta-function
on the boundary which vanishes as the boundary of the open neighbourhood
$U(\sigma)$ of $\sigma$ of course nowhere hits $\sigma$.

Extracting the algebraic bracket $\left[T,\tilde{T}\right]_{(1)}^{\Delta}$
is a bit more tricky. One can do it via\begin{eqnarray}
\hspace{-.7cm}\zwek{{\scriptstyle \textrm{for any fixed}}}{{\scriptstyle \textrm{index }\lambda}}:\,\left[T,\tilde{T}\right]_{(1)}^{\Delta}(\sigma,\tet) & = & -(-)^{t-t'}\int_{U(\sigma)}\de^{d_{\textrm{w}}}\sigma'\int\mu(\tet')\quad\left(\frac{e^{\sigma'^{\lambda}}}{e^{\sigma^{\lambda}}}-1\right)\partl{\tet^{\lambda}}(T(\sigma',\tet')\bs{,}\tilde{T}(\sigma,\tet))\quad\label{eq:rosa-Pause}\end{eqnarray}
The boundary term proportional to $\left(\frac{e^{\sigma'^{\lambda}}}{e^{\sigma^{\lambda}}}-1\right)\delta^{d_{\textrm{w}}}(\sigma-\sigma')$
appearing above on the righthand side after partial integration vanishes
as $\sigma'$ in the prefactor is set to $\sigma$ via the delta function.
\\
The claim is now that $A(\sigma,\tet)=-(-)^{t-t'}\left[\de T,\tilde{T}\right]_{(1)}^{\Delta}(\sigma,\tet)$.
So let us calculate the righthand side via (\ref{eq:rosa-Pause}):\begin{eqnarray}
\left[\de T,\tilde{T}\right]_{(1)}^{\Delta}(\sigma,\tet) & = & -(-)^{t+1-t'}\int_{U(\sigma)}\de^{d_{\textrm{w}}}\sigma'\int\mu(\tet')\quad\left(\frac{e^{\sigma'^{\lambda}}}{e^{\sigma^{\lambda}}}-1\right)\partl{\tet^{\lambda}}(\de T(\sigma',\tet')\bs{,}\tilde{T}(\sigma,\tet))=\\
 & = & -(-)^{t+1-t'}\int\de^{d_{\textrm{w}}}\sigma'\int\mu(\tet')\quad\left(\frac{e^{\sigma'^{\lambda}}}{e^{\sigma^{\lambda}}}-1\right)\partl{\tet^{\lambda}}\tet'^{\mu}\partial_{\mu}'(T(\sigma',\tet')\bs{,}\tilde{T}(\sigma,\tet))\end{eqnarray}
$(T\bs{,}\tilde{T})$ contains in both terms a plain $\delta$-function
for the fermionic variables $\tet$, so that we can replace $\tet'$
by $\tet$. Integration by parts of $\partial_{\mu}'$ (where possible
boundary terms again do not contribute because of the vanishing of
the delta function and its derivative on the boundary) delivers the
desired result\begin{equation}
\left[\de T,\tilde{T}\right]_{(1)}^{\Delta}(\sigma,\tet)=-(-)^{t-t'}\int\de^{d_{\textrm{w}}}\sigma'\int\mu(\tet')\quad(T(\sigma',\tet')\bs{,}\tilde{T}(\sigma,\tet))=-(-)^{t-t'}A(\sigma,\tet)\end{equation}
This completes the proof of proposition 3b. $\qquad\square$

\section{Applications in string theory or 2d CFT}

\label{sec:Applications-in-string}

In the previous section the dimension of the worldvolume was arbitrary
or even dimensional. The appearance of derived brackets (including
e.g. the Dorfman bracket) is thus not a special feature of a 2-dimensional
sigma-model like string theory. There are, however, special features
in string theory. Currents in string theory (which have conformal
weight one) naturally are sums of 1-forms and vectors, if one takes
the identification $\partial_{1}x^{m}(\sigma)\leftrightarrow\de x^{m}$
and $p_{m}(\sigma)\leftrightarrow\pe_{m}$, as in \cite{Alekseev:2004np}
(see footnote \ref{foot:Strobl}), e.g. $\partial x^{m}=\partial_{1}x^{m}-\partial_{0}x^{m}\hat{=}\de x^{m}-\eta^{mn}\pe_{n}$
. This is closely related to the identification in our previous section
in the antifield formalism. In addition, only in two dimensions a
single $\tet$ can be interpreted as a worldsheet Weyl spinor (in
1 dimension it can be seen as a Dirac-spinor, but in higher dimensions
the interpretation of $\tet$ as worldvolume spinor breaks down).
As we ended the last section with the antifield formalism, which therefore
is perhaps still more present, let us start this section in the reversed
order, beginning with the application in the antifield formalism.

\subsection{Poisson sigma-model and Zucchini's {}``Hitchin sigma-model''}

\label{sub:Zucchini} Remember for a moment the Poisson-$\sigma$-model
\cite{Schaller:1994es,Cattaneo:1999fm}. It is a two-dimensional sigma-model
($d_{\textrm{w}}=2$) of the form \begin{equation}
S_{0}=\int_{\Sigma}\,\bs{\eta}_{m}\dew x^{m}+\frac{1}{2}P^{mn}(x)\bs{\eta}_{m}\bs{\eta}_{n}\end{equation}
where $\bs{\eta}_{m}$ is a worldsheet one-form. This model is topological
if and only if the Poisson-structure $P^{mn}(x)$ is integrable, i.e.
the Schouten-bracket of $P$ with itself vanishes\begin{eqnarray}
S_{0}\textrm{ topological} & \iff & \left[P\bs{,}P\right]=0\end{eqnarray}
It gives on the one hand a field theoretic implementation of Kontsevich's
star product \cite{Cattaneo:1999fm} and is on the other hand related
to string theory via a topological limit (big antisymmetric part in
the open string metric), which leads to the relation between string
theory and noncommutative geometry.

The necessary ghost fields for the action can be introduced by extending
$x$ and $\eta$ to de Rham superfields as in (\ref{eq:deRhamSuperfieldPhi},\ref{eq:deRhamSuperfieldPhipl})\begin{eqnarray}
\Phi^{m}(\sigma,\tet) & \equiv & x^{m}(\sigma)+\underbrace{\bs{x}_{\mu}^{m}(\sigma)}_{\epsilon_{\mu\nu}\bs{\eta}^{+\nu n}}\tet^{\mu}+\underbrace{x_{\mu_{1}\mu_{2}}^{m}(\sigma)}_{-\frac{1}{2}\eps_{\mu_{1}\mu_{2}}\beta^{+\, m}}\tet^{\mu_{1}}\tet^{\mu_{2}}\\
\Ph_{m}(\sigma',\tet') & \equiv & \underbrace{\frac{1}{2!}\epsilon_{\mu_{1}\mu_{2}}\ix_{m}\hoch{\mu_{1}\mu_{2}}(\sigma')}_{\equiv\bs{\beta}_{m}(\sigma')}+\tet'^{\mu_{1}}\underbrace{\epsilon_{\mu_{1}\mu_{2}}x_{m}^{+}\hoch{\mu_{2}}(\sigma')}_{\eta_{\mu_{1}m}}+\frac{1}{2}\epsilon_{\mu_{1}\mu_{2}}\tet'^{\mu_{1}}\tet'^{\mu_{2}}\ix_{m}(\sigma')\end{eqnarray}
One can use Hodge-duality to rename some component fields as indicated.
$\bs{\beta}_{m}$ is then the ghost field related to the gauge symmetry.
The action including ghost fields and antifields simply reads\begin{eqnarray}
S & = & \int d^{2}\sigma\,\int\mu(\tet)\quad\Ph_{m}\dew\Phi^{m}+\frac{1}{2}P^{mn}(\Phi)\Ph_{m}\Ph_{n}\end{eqnarray}
The expression under the integral corresponds to the tensor $-\delta_{m}\hoch{n}\de x^{m}\wedge\pe_{n}+\frac{1}{2}P^{mn}\pe_{m}\wedge\pe_{n}$
and the antibracket in the master-equation $(S,S)$ implements the
Schoutenbracket on $P$, which is a well known relation. Therefore
we will concentrate on a second example, which is very similar, but
less known. 

Zucchini suggested in \cite{Zucchini:2004ta} a 2-dimensional sigma-model
which is topological if a generalized complex structure in the target
space is integrable (see subsection \ref{sub:generalized-complex-structure}
on page \pageref{sub:generalized-complex-structure} and \ref{sub:Integrability-of-J}
on page \pageref{sub:Integrability-of-J} to learn more about generalized
complex structures). His model is of the form\begin{eqnarray}
S & = & \int d^{2}\sigma\,\int\mu(\tet)\quad\left(\Ph_{m}\dew\Phi^{m}\,+\,\right)\quad\frac{1}{2}P^{mn}(\Phi)\Ph_{m}\Ph_{n}-\frac{1}{2}Q_{mn}(\Phi)\dew\Phi^{m}\dew\Phi^{n}-J^{n}\tief{m}\dew\Phi^{m}\Ph_{n}\label{eq:Zucchini-Wirkung}\end{eqnarray}
where $P^{mn}$, $Q_{mn}$ and $J^{m}\tief{n}$ are the building blocks
of the generalized complex structure (\ref{eq:J-matrix})\begin{eqnarray}
\mc{J}^{M}\tief{N} & = & \left(\begin{array}{cc}
J^{m}\tief{n} & P^{mn}\\
-Q_{mn} & -J^{n}\tief{m}\end{array}\right)\end{eqnarray}
The first term of (\ref{eq:Zucchini-Wirkung}) can be absorbed by
a field redefinition as already observed in \cite{Zucchini:2005rh}.
Ignoring thus the first term and using our notations of before, $S$
can be rewritten as\begin{equation}
S=\int d^{2}\sigma\,\int\mu(\tet)\quad\frac{1}{2}\mc{J}(\Phi,\dew\Phi,\Ph)\end{equation}
Calculating the master equation explicitely and collecting the terms
which combine to the lengthy tensors for the integrability condition
(see (\ref{eq:integrability-tensor-I})-(\ref{eq:integrability-tensor-IV}))
is quite cumbersome, so we can enjoy using instead proposition 3b
on page \pageref{eq:PropositionIIIb}. For a worldsheet without boundary
its integrated version reads\begin{equation}
\left(\int d^{d_{\textrm{w}}}\sigma'\int\mu(\tet')K(\sigma',\tet'),\int\de^{d_{\textrm{w}}}\sigma\int\mu(\tet)L(\sigma,\tet)\right)=\int d^{d_{\textrm{w}}}\sigma\int\mu(\tet)\left[K,_{\de}L\right]_{(1)}^{\Delta}(\sigma,\tet)\label{eq:PropositionIIIb-integrated}\end{equation}
which leads to the relation\begin{eqnarray}
(S,S) & = & 0\qquad\iff\int d^{2}\sigma\int\mu(\tet)\left[\mc{J},_{\de}\mc{J}\right]_{(1)}^{\Delta}(\sigma,\tet)=0\label{eq:Zucchini-last-equation}\end{eqnarray}
\rem{no condition $J^2=-1$??}The derived bracket of the big bracket
of $\mc{J}$ with itself contains already the Nijenhuis tensor (see
in the appendix in equation (\ref{eq:derived-bracket-for-J}) and
in the discussion around) \begin{eqnarray}
\left[\mc{J},_{\de}\mc{J}\right]_{(1)}^{\Delta} & = & \mc{N}_{M_{1}M_{2}M_{3}}\basis^{M_{1}}\basis^{M_{2}}\basis^{M_{3}}-4\mc{J}^{JI}\mc{J}_{IM}\basis^{M}p_{J}=\\
 & \stackrel{\mc{J}^{2}=-1}{=} & \mc{N}_{M_{1}M_{2}M_{3}}\basis^{M_{1}}\basis^{M_{2}}\basis^{M_{3}}+4\oo\label{eq:relation-between-derived-and-Nij-Tens-in-main-part}\\
\basis^{M} & = & (\de x^{m},\pe_{m}),\qquad p_{J}=(p_{j},0)\\
\oo(\de x,p) & = & \de x^{m}p_{m}\end{eqnarray}
For $\mc{J}^{2}=-1$ the last term is proportional to the generator
$\oo$ (remember (\ref{eq:BRST-op})). In (\ref{eq:Zucchini-last-equation}),
however, it appears with $\de x$ and $p$ replaced by the superfields
as in (\ref{eq:o-of-sig-tet-II}) \begin{eqnarray}
\oo(\sigma,\tet) & = & \dew\Phi^{m}(\sigma,\tet)\dew\Ph_{m}(\sigma,\tet)=-\dew\left(\dew\Phi^{m}(\sigma,\tet)\Ph_{m}(\sigma,\tet)\right)\end{eqnarray}
which is a total worldsheet derivative and therefore drops during
the integration. We are left with the generalized Nijenhuis tensor
as a function of superfields\begin{eqnarray}
\mc{N}(\sigma,\tet) & = & \mc{N}_{M_{1}M_{2}M_{3}}(\Phi)\q{\basis}^{M_{1}}\q{\basis}^{M_{2}}\q{\basis}^{M_{3}}\\
\textrm{with }\q{\basis}^{M} & \equiv & (\dew\Phi^{m},\Ph_{m})\end{eqnarray}
Written in small indices\begin{eqnarray}
\mc{N}(\sigma,\tet) & = & \mc{N}_{m_{1}m_{2}m_{3}}(\Phi)\underbrace{\dew\Phi^{m_{1}}\dew\Phi^{m_{1}}\dew\Phi^{m_{1}}}_{=0}+3\mc{N}^{n}\tief{m_{1}m_{2}}(\Phi)\Ph_{n}\dew\Phi^{m_{1}}\dew\Phi^{m_{2}}+\nonumber \\
 &  & +3\mc{N}_{n}\hoch{m_{1}m_{2}}(\Phi)\dew\Phi^{n}\Ph_{m_{1}}\Ph_{m_{2}}+\mc{N}^{m_{1}m_{2}m_{3}}(\Phi)\Ph_{m}\Ph_{m}\Ph_{m}\end{eqnarray}
 One realizes that the first term vanishes identically (as mentioned
in \cite{Zucchini:2004ta}) and only the remaining three tensors are
required to vanish in order to satisfy (\ref{eq:Zucchini-last-equation}).
\rem{old and wrong: The same model $S=\int d^{d_{\textrm{w}}}\sigma\,\int\mu(\tet)\quad\frac{1}{2}\mc{J}(\Phi,\dew\Phi,\Ph)$
in dimensions higher than 2 should however implement the complete
generalized Nijenhuis tensor. As the worldvolume has to be even-dimensional
for this setting (see proposition 3b on page \pageref{eq:PropositionIIIb}),
we need the worldvolume dimension $d_{\textrm{w }}$ to be at least
four.}\rem{twisted?\\
Quantum case \begin{eqnarray*}
W & = & S+\hbar M_{1}+\hbar^{2}M_{2}+\ldots\\
(W,W) & = & i\hbar\Delta W\\
\Delta & = & (-)^{A}\funktional{}{\Phi^{A}}\funktional{}{\Phi_{A}^{*}}\end{eqnarray*}
}

\subsection{Relation between a second worldsheet supercharge and generalized
complex geometry}

\label{sub:Zabzine}

In \cite{Lindstrom:2004iw} the relation between an extended worldsheet
supersymmetry in string theory and the presence of an integrable generalized
complex structure was explored. Zabzine clarified in \cite{Zabzine:2005qf}
the relation in an model independent way in a Hamiltonian description.
The structures appearing there are almost the same that we have discussed
before although we have to modify the procedure a little bit due to
the interpretation of $\tet$ as a worldsheet spinor.

Consider a sigma-model with 2-dimensional worldvolume (worldsheet)
with manifest $N=1$ supersymmetry on the worldsheet. In the phase
space there is only one $\sigma$-coordinate left. Let us denote the
corresponding superfields, following loosely \cite{Zabzine:2005qf},
by\begin{eqnarray}
\Phi^{m}(\sigma,\tet) & \equiv & x^{m}(\sigma)+\tet\lam^{m}(\sigma)\\
\Es_{m}(\sigma,\tet) & \equiv & \ro_{m}(\sigma)+\tet p_{m}(\sigma)\end{eqnarray}
In comparison to section \ref{sub:Natural-appearance}, there is a
change of notation from $\ce^{m}\To\lam^{m}$ \rem{so lassen?} and
$\be_{m}\To\ro_{m}$ as $\be$ and $\ce$ suggest the interpretation
as ghosts which is not true in this case, where $\lam$ and $\ro$
are worldsheet fermions. Introduce now, following Zabzine, the generator
$\qu$ of the \textbf{manifest SUSY} and the corresponding \textbf{covariant
derivative} $\cov$\begin{eqnarray}
\qu & \equiv & \partial_{\tet}+\tet\partial_{\sigma}\\
\cov & \equiv & \partial_{\tet}-\tet\partial_{\sigma}\end{eqnarray}
with the SUSY algebra\rem{%
\footnote{Further useful relations are \begin{eqnarray*}
\int\de\tet\mc{L} & = & \cov\mc{L}\mid=\partial_{\tet}\mc{L}=\qu\mc{L}\mid\\
\int\de\sigma\int\de\tet\mc{L} & = & \int\de\sigma\,\cov\mc{L}=\int\de\sigma\,\partial_{\tet}\mc{L}=\int\de\sigma\,\qu\mc{L}\qquad\fussend\end{eqnarray*}
}} \begin{eqnarray}
\left[\qu,\qu\right] & = & 2\partial_{\sigma}=-\left[\cov,\cov\right]\\
\left[\qu,\cov\right] & = & 0\end{eqnarray}
$\qu$ is the sum of two nilpotent differential operators, namely
$\partial_{\tet}$ and $\tet\partial_{\sigma}$. Acting on the Superfields
$\Phi^{m}$ and $\Es^{m}$, they induce the differentials $\es$ and
$\tilde{\es}$ on the component fields, which are in turn generated
via the Poisson bracket by phase space functions $\OO$ (the same
as (\ref{eq:Omega})) and $\tilde{\OO}$. \rem{(as a phase space
function similar to (\ref{eq:tildOmega}), but with a completely different
action on multivector valued forms in the present setting, as we will
see below).} \begin{eqnarray}
\OO & \equiv & \int d\sigma\:\lam^{k}p_{k}\label{eq:OmegaII}\\
\tilde{\OO} & = & -\int d\sigma\:\partial_{\sigma}x^{k}\ro_{k}\label{eq:tildOmegaII}\\
\es x^{m}\equiv\left\{ \OO,x^{m}\right\}  & = & \lam^{m}\leftrightarrow\de x^{m},\quad\es\ro_{m}\equiv\left\{ \OO,\ro_{m}\right\} =p_{m}\leftrightarrow\de(\pe_{m}),\\
\tilde{\es}\lam^{m}\equiv\left\{ \tilde{\OO},\lam^{m}\right\}  & = & -\partial_{\sigma}x^{m},\quad\tilde{\es}p_{k}=-\partial_{\sigma}\ro_{k}=\left\{ \tilde{\OO},p_{k}\right\} ,\\
\es\Phi^{m} & = & \partial_{\tet}\Phi^{m},\qquad\es\Es_{m}=\partial_{\tet}\Es_{m}\\
\tilde{\es}\Phi^{m} & = & \tet\partial_{\sigma}\Phi^{m},\qquad\tilde{\es}\Es_{m}=\tet\partial_{\sigma}\Es_{m}\end{eqnarray}
The Poisson-generator for the SUSY transformations of the component
fields induced by%
\footnote{We have \begin{eqnarray*}
\qu\Phi^{m} & = & \lam^{m}+\tet\partial_{\sigma}x^{m},\qquad\qu\Es_{m}=p_{m}+\tet\partial_{\sigma}\ro_{m}\\
\cov\Phi^{m} & = & \lam^{m}(\sigma)-\tet\partial_{\sigma}x^{m},\qquad\cov\Es_{m}=p_{m}-\tet\partial_{\sigma}\ro_{m}\\
\delta_{\feps}x^{m} & = & \feps\lam^{m},\qquad\delta_{\feps}\lam^{m}=-\feps\partial_{\sigma}x^{m}\\
\delta_{\feps}\ro_{m} & = & \feps p_{m},\qquad\delta_{\feps}p_{m}=-\feps\partial_{\sigma}\ro_{m}\qquad\fussend\end{eqnarray*}
} $\qu$ is thus the sum of the generators of $\es$ and $\tilde{\es}$:
\begin{eqnarray}
\Q & = & \OO+\tilde{\OO}=\int d\sigma\:\lam^{k}p_{k}-\partial_{\sigma}x^{k}\ro_{k}=-\int d\sigma\int d\tet\,\qu\Phi^{k}\Es_{k}\label{eq:SUSY-generator}\end{eqnarray}
In (\ref{eq:superfield-definition}) superfields were defined via
$\partial_{\tet}Y=\es Y$ in order to implement the exterior derivative
directly with $\partial_{\tet}$. In that sense $\Phi$, $\Es$, $\de\Phi$,
$\de\Es$ and all analytic functions of them were superfields. In
the context of worldsheet supersymmetry, one prefers of course a supersymmetric
covariant formulation. Let us therefore define in this subsection
proper \textbf{superfields} via \begin{equation}
Y\textrm{ is a superfiled }\quad:\iff\quad\qu Y\stackrel{!}{=}\left\{ \Q,Y\right\} =(\es+\tilde{\es})Y\label{eq:superfield-def-susy}\end{equation}
 \rem{($\es+\tilde{\es}$ is invariant under the change $\es\leftrightarrow\tilde{\es}$
and therefore probably under the change of the two mechanisms (identifications))
}which holds for $\Phi$, $\Es$,$\cov\Phi$, $\cov\Es$, all analytic
functions of them (like our analytically continued multivector valued
forms) and worldsheet spatial derivatives $\partial_{\sigma}$ thereof
(but not for e.g. $\qu\Phi$. This means that although we have $\qu\Phi=(\es+\tilde{\es})\Phi$
this does not hold for a second action, i.e. $\qu^{2}\Phi\neq(\es+\tilde{\es})^{2}\Phi$,
which explains the somewhat confusing fact that the Poisson-generator
$\Q$ has the opposite sign in the algebra than $\qu$\rem{($\qu$
and ($\es+\tilde{\es}$) anticommute!)}\begin{eqnarray}
\left\{ \Q,\Q\right\}  & = & -2P\label{eq:SUSY-algebra}\end{eqnarray}
where we introduced the phase-space generator $P$ for the worldsheet
translation induced by $\partial_{\sigma}$ \begin{eqnarray}
P & \equiv & \int d\sigma\quad\partial_{\sigma}x^{k}p_{k}+\partial_{\sigma}\lam^{k}\ro_{k}=\int d\sigma\int d\tet\quad\partial_{\sigma}\Phi^{k}\Es_{k}\end{eqnarray}
The same phenomenon appears for the differentials $\es$ and $\tilde{\es}$.
The graded commutator of $\partial_{\tet}$ and $\tet\partial_{\sigma}$
is the worldsheet derivative $\left[\partial_{\tet},\tet\partial_{\sigma}\right]=\partial_{\sigma}$,
while the algebra for $\es$ and $\tilde{\es}$ has the opposite sign\begin{eqnarray}
\left[\es,\tilde{\es}\right]Y(\sigma,\tet) & = & -\partial_{\sigma}Y(\sigma,\tet)\\
\es\tilde{\OO}=\left\{ \OO,\tilde{\OO}\right\}  & = & -P=\tilde{\es}\OO\end{eqnarray}
One major statement in \cite{Zabzine:2005qf} is as follows: Making
a general ansatz for a generator of a second, non-manifest supersymmetry,
of the form (some signs are adopted to our conventions)\begin{eqnarray}
\Q_{2} & \equiv & \frac{1}{2}\int d\sigma\int d\tet\quad(P^{mn}(\Phi)\Es_{m}\Es_{n}-Q_{mn}(\Phi)\cov\Phi^{m}\cov\Phi^{n}+2J^{m}\tief{n}(\Phi)\Es_{m}\cov\Phi^{n})\label{eq:Qzwei}\end{eqnarray}
and requiring the same algebra as for $\Q$ in (\ref{eq:SUSY-algebra})\begin{eqnarray}
\left\{ \Q_{2},\Q_{2}\right\}  & = & -2P\label{eq:SUSY-alg-II}\\
\Big(\left\{ \Q,\Q_{2}\right\}  & = & 0\Big)\label{eq:useless-condition}\end{eqnarray}
is equivalent to \begin{eqnarray}
\mc{J}^{M}\tief{N} & \equiv & \left(\begin{array}{cc}
J^{m}\tief{n} & P^{mn}\\
-Q_{mn} & -J^{n}\tief{m}\end{array}\right)\end{eqnarray}
being an integrable generalized complex structure (see in the appendix
\ref{sub:generalized-complex-structure} on page \pageref{sub:generalized-complex-structure}
and \ref{sub:Integrability-of-J} on page \pageref{sub:Integrability-of-J}).
On a worldsheet without boundary, the second condition is actually
superfluous, because it is already implemented via the ansatz: The
expression in the integral is an analytic function of superfields
and therefore a superfield itself. According to (\ref{eq:superfield-def-susy})
we can replace at this point the commutator with $\Q$ with the action
of $\qu$ and get \begin{equation}
\left\{ \Q,\Q_{2}\right\} =\int d\sigma\int d\tet\quad\qu(\ldots)=\int d\sigma\quad\partial_{\sigma}(\ldots)=0\end{equation}
 For the other condition, the actual supersymmetry algebra (\ref{eq:SUSY-alg-II}),
the aim of the present considerations should now be clear. The generalized
complex structure $\mc{J}$ itself is a sum of multivector valued
forms\begin{eqnarray}
\mc{J} & \equiv & \mc{J}^{MN}(x)\basis_{M}\basis_{N}\equiv P^{mn}(x)\pe_{m}\wedge\pe_{n}-Q_{mn}(x)\de x^{m}\de x^{n}+2J^{m}\tief{n}(x)\pe_{m}\wedge\de x^{n}\end{eqnarray}
which can be seen as a function of $x$ and the basis elements \begin{equation}
\mc{J}=\mc{J}(x,\de x,\pe)\end{equation}
In \ref{sub:Natural-appearance} we replaced the arguments of functions
like this with {}``superfields'' $x^{m}\To\Phi^{m}$, $\de x^{m}\to\partial_{\tet}\Phi^{m}$
and $\pe_{m}\To\Es_{m}$. The name superfield might have been misleading,
as $\partial_{\tet}\Phi$ is only a superfield in the sense that it
implements the target-space exterior derivative via $\partial_{\tet}$,
but it is not a superfield in the sense of worldsheet supersymmetry.
In a supersymmetric theory one prefers a supersymmetric covariant
formulation. Working with $\partial_{\tet}\Phi$ as before is therefore
not desirable and we replace $\partial_{\tet}\Phi$ by $\cov\Phi$,
leading directly to $\Q_{2}$ (\ref{eq:Qzwei}) which now can be written
as \begin{eqnarray}
\Q_{2} & = & \frac{1}{2}\int d\sigma\int d\tet\,\mc{J}\left(\Phi(\sigma,\tet),\cov\Phi(\sigma,\tet),\Es(\sigma,\tet)\right)\label{eq:Qzwei-II}\end{eqnarray}
Apart from the change $\partial_{\tet}\Phi\To\cov\Phi$ we expect
from the previous section that the Poisson bracket of $\Q_{2}$ with
itself induces some algebraic and some derived bracket of $\mc{J}$
with itself which then corresponds to the integrability condition
for $\mc{J}$. This is indeed the case, but we first have to study
the changes coming from $\partial_{\tet}\Phi\To\cov\Phi$. In other
words, we need a new formulation of proposition 1 (\ref{eq:Proposition1})
in the case of two-dimensional supersymmetry (Proposition 1 is of
course still valid, but it is not formulated in a supersymmetric covariant
way. It should, however, be applicable to e.g. BRST symmetries ).
Let us redefine the meaning of $K(\sigma,\tet)$ in (\ref{eq:K-sig-tet-def})
for a multivector valued form $K^{(k,k')}$ \begin{eqnarray}
K^{(k,k')}(\sigma,\tet) & \equiv & K^{(k,k')}\big(\Phi^{m}(\sigma,\tet),\cov\Phi^{m}(\sigma,\tet),\Es_{m}(\sigma,\tet)\big)=\label{eq:K-sig-tet-Zabz}\\
 &  & \hspace{-1.6cm}=K_{m_{1}\ldots m_{k}}\hoch{n_{1}\ldots n_{k'}}\left(\Phi(\sigma,\tet)\right)\,\cov\Phi^{m_{1}}(\sigma,\tet)\ldots\cov\Phi^{m_{k}}(\sigma,\tet)\Es_{n_{1}}(\sigma,\tet)\ldots\Es_{n_{k'}}(\sigma,\tet)\us{\stackrel{\tet=0}{=}}{(\ref{eq:K-of-sigma})}K^{(k,k')}(\sigma)\qquad\end{eqnarray}
Likewise for all the other examples in (\ref{eq:T-sig-tet})-(\ref{eq:db-sig-tet}):\begin{eqnarray}
T^{(t,t',t'')}(\sigma,\tet) & \equiv & T^{(t,t',t'')}\left(\Phi(\sigma,\tet),\cov\Phi(\sigma,\tet),\Es(\sigma,\tet),\cov\Es(\sigma,\tet)\right)\stackrel{\tet=0}{=}T^{(t,t',t'')}(\sigma)\quad(\textrm{see }(\ref{eq:T-of-sigma}))\label{eq:T-sig-tet-III}\end{eqnarray}
\begin{eqnarray}
\textrm{e.g. }\de K(\sigma,\tet) & \equiv & \de K\left(\Phi(\sigma,\tet),\cov\Phi(\sigma,\tet),\Es(\sigma,\tet),\cov\Es(\sigma,\tet)\right)\label{eq:dK-of-sig-tet-III}\\
\textrm{or }\oo(\sigma,\tet) & \equiv & \oo\left(\cov\Phi(\sigma,\tet),\cov\Es(\sigma,\tet)\right)\stackrel{(\ref{eq:BRST-op})}{=}\cov\Phi^{m}(\sigma,\tet)\cov\Es_{m}(\sigma,\tet)\us{\stackrel{\tet=0}{=}}{(\ref{eq:o-of-sigma})}\oo(\sigma)\label{eq:o-of-sig-tet-III}\\
\hspace{-1.2cm}[K^{(k,k')},_{\de\,}L^{(l,l')}]_{(1)}^{\Delta}(\sigma,\tet) & \equiv & [K^{(k,k')}\bs{,}L^{(l,l')}]_{(1)}^{(\Delta)}\left(\Phi(\sigma,\tet),\cov\Phi(\sigma,\tet),\Es(\sigma,\tet),\cov\Es(\sigma,\tet)\right)\us{\stackrel{\tet=0}{=}}{(\ref{eq:bracket-of-sigma})}[K^{(k,k')}\bs{,}L^{(l,l')}]_{(1)}^{(\Delta)}(\sigma)\qquad\quad\\
\de x^{m}(\sigma,\tet) & \equiv & \cov\Phi^{m}(\sigma,\tet)=\lam^{m}(\sigma)-\tet\partial_{\sigma}x^{m}(\sigma)\\
\de\pe_{m}(\sigma,\tet) & \equiv & \cov\Es_{m}(\sigma,\tet)=p_{m}(\sigma)-\tet\partial_{\sigma}\ro_{m}(\sigma)\label{eq:db-sig-tet-III}\end{eqnarray}
Expanding $K$ in $\tet$ yields \begin{eqnarray}
K^{(k,k')}(\sigma,\tet) & = & K^{(k,k')}(\sigma)+\tet\left(\bei{\partial_{\tet'}K^{(k,k')}(\sigma,\tet')}{\tet'=0}\right)=\\
 & = & K^{(k,k')}(\sigma)+\tet\left(\bei{\qu\tief{'}K^{(k,k')}(\sigma,\tet')}{\tet'=0}\right)\end{eqnarray}
As $K$ is a superfield, we can replace $\qu$ by $\es+\tilde{\es}$\begin{eqnarray}
K^{(k,k')}(\sigma,\tet) & = & K^{(k,k')}(\sigma)+\tet(\es+\tilde{\es})K^{(k,k')}(\sigma)=\label{eq:wichtig-ZabzI}\\
 & = & K^{(k,k')}(\sigma)+\tet\bei{\left((\de+\ip_{v})K^{(k,k')}\right)(\sigma)}{v^{k}\To-\partial_{\sigma}x^{k}}\label{eq:wichtig-Zabz}\end{eqnarray}
This is the analogue to the non-supersymmetric (\ref{eq:wichtigII})
and delivers the exterior derivative which will lead to the appearance
of the derived bracket. The relation between $\tilde{\es}$ and the
inner product with a vector should perhaps be clarified. Remember
that all multivector forms at $\tet=0$, $K^{(k,k')}(\sigma)$, are
analytic functions of the component fields $x^{m},\lam^{m}$ and $\ro_{m}$
. But among those fields, $\tilde{\es}$ acts only on $\lam^{m}$
and we can express it with partial derivatives (instead of functional
ones) when acting on $K$:\begin{equation}
\tilde{\es}K(\sigma)=-\partial_{\sigma}x^{m}\partl{\lam^{m}}K(x,\lam,\ro)=\bei{\ip_{v}K(\sigma)}{v^{k}=-\partial_{\sigma}x^{k}}\label{eq:stilde-auf-K}\end{equation}
in the Poisson bracket of $\tilde{\es}K$ with another multivector
valued form $L$ at $\tet=0$, nothing acts on $v^{k}=-\partial_{\sigma}x^{k}$
(which would produce a derivative of a delta function), as $L$ does
not contain $p_{k}$. Therefore we have\begin{equation}
\left\{ \tilde{\es}K(\sigma'),L(\sigma)\right\} =\bei{[\ip_{v}K,L](\sigma)}{v^{k}=-\partial_{\sigma}x^{k}}\delta(\sigma-\sigma')\label{eq:stilde-in-bracket}\end{equation}
which we will need below. For superfields we have $Y(\sigma,\tet)=Y(\sigma)+\tet(\es+\tilde{\es})Y(\sigma)$.
Applying the same to $v$ yields\begin{eqnarray}
v^{k}(\sigma)+\tet(\es+\tilde{\es})v^{k}(\sigma) & = & -\partial_{\sigma}x^{k}-\tet(\es+\tilde{\es})\partial_{\sigma}x^{k}(\sigma)=\label{eq:v-sig-tet}\\
 & = & -\partial_{\sigma}x^{k}-\tet\partial_{\sigma}\lambda^{k}(\sigma)=-\partial_{\sigma}\Phi^{k}\label{eq:v-sig-tet-II}\end{eqnarray}
\rem{\begin{eqnarray*}
\tilde{\es}T & = & -\partial_{\sigma}x^{m}\partl{\lam^{m}}T-\partial_{\sigma}\ro_{k}\partl{p_{k}}\end{eqnarray*}
}

\paragraph{Proposition 1b }

\emph{For all multivector valued forms $K^{(k,k')},L^{(l,l')}$ on
the target space manifold, in a local coordinate patch seen as functions
of $x^{m}$,$\de x^{m}$ and $\pe_{m}$, the following equation holds
for the corresponding worldsheet-superfields (\ref{eq:K-sig-tet-Zabz})
}\\
\emph{\Ram{1}{\begin{eqnarray}
\{ K^{(k,k')}(\sigma',\tet'),L^{(l,l')}(\sigma,\tet)\} & = & \cov\left(\delta(\tet-\tet')\delta(\sigma-\sigma')\right)\left[K,L\right]_{(1)}^{\Delta}(\sigma,\tet)+\nonumber \\
 &  & \hspace{-.6cm}+\delta(\tet'-\tet)\delta(\sigma-\sigma')\Big(\underbrace{[\de K,L]_{(1)}^{\Delta}(\sigma,\tet)}_{\lqn{-(-)^{k-k'}\left[K,_{\de}L\right]_{(1)}^{\Delta}}}+\underbrace{[\ip_{v}K,L]_{(1)}^{\Delta}(\sigma,\tet)}_{\quad-(-)^{k-k'}[K,_{\ip_{v}}L]}\Big|_{v^{k}=-\partial_{\sigma}\Phi^{k}}\Big)\qquad\label{eq:Proposition1b}\end{eqnarray}
}}\\
\emph{where  e.g. $[\de K,L]_{(1)}^{\Delta}(\sigma,\tet)\equiv[\de K,L]_{(1)}^{\Delta}\left(\Phi(\sigma,\tet),\cov\Phi(\sigma,\tet),\Es(\sigma,\tet),\cov\Es(\sigma,\tet)\right)$.
}\\
\emph{The integrated version for a worldsheet without boundary reads}\emph{\small \begin{equation}
\boxed{\Big\{\int d\sigma'\int d\tet'K^{(k,k')}(\sigma',\tet'),\int d\sigma\int d\tet\, L^{(l,l')}(\sigma,\tet)\Big\}=(\es+\tilde{\es})\int d\sigma\,\Big([K,_{\de}L]_{(1)}^{\Delta}-(-)^{k-k'}[\ip_{v}K,L]_{(1)}^{\Delta}\Big|_{v^{k}=-\partial_{\sigma}x^{k}}\Big)(\sigma)}\label{eq:Proposition1b-integrated}\end{equation}
\vspace{.2cm}}{\small \par}

\emph{Proof}$\quad$Let us use (\ref{eq:wichtig-ZabzI}) for both
multivector valued fields and plug into the lefthand side of (\ref{eq:Proposition1b})\begin{eqnarray}
\lqn{\left\{ K(\sigma',\tet'),L(\sigma,\tet)\right\} =}\nonumber \\
 & = & \left\{ K(\sigma')+\tet'(\es+\tilde{\es})K(\sigma')\,,\, L(\sigma)+\tet(\es+\tilde{\es})L(\sigma)\right\} =\\
 & = & \left\{ K(\sigma'),L(\sigma)\right\} +\tet'\left\{ (\es+\tilde{\es})K(\sigma'),L(\sigma)\right\} +(-)^{k-k'}\tet\left\{ K(\sigma'),(\es+\tilde{\es})L(\sigma)\right\} +\nonumber \\
 &  & +(-)^{k-k'}\tet\tet'\left\{ (\es+\tilde{\es})K(\sigma'),(\es+\tilde{\es})L(\sigma)\right\} =\\
 & = & \left\{ K(\sigma'),L(\sigma)\right\} +(\tet'-\tet)\left\{ (\es+\tilde{\es})K(\sigma'),L(\sigma)\right\} +\tet(\es+\tilde{\es})\left\{ K(\sigma'),L(\sigma)\right\} +\nonumber \\
 &  & +\tet'\tet(\es+\tilde{\es})\left\{ (\es+\tilde{\es})K(\sigma'),L(\sigma)\right\} -\tet'\tet\left\{ (\es+\tilde{\es})(\es+\tilde{\es})K(\sigma'),L(\sigma)\right\} =\\
 & = & \left(1+\tet(\es+\tilde{\es})\right)\left\{ K(\sigma'),L(\sigma)\right\} +(\tet'-\tet)\left(1+\tet(\es+\tilde{\es})\right)\left\{ (\es+\tilde{\es})K(\sigma'),L(\sigma)\right\} +\nonumber \\
 &  & -\tet'\tet\big\{\underbrace{[\es,\tilde{\es}]}_{-\partial_{\sigma'}}K(\sigma'),L(\sigma)\big\}=\\
 & = & \delta(\sigma-\sigma')\left(1+\tet(\es+\tilde{\es})\right)\left[K,L\right]_{(1)}^{\Delta}(\sigma)+(\tet'-\tet)\left(1+\tet(\es+\tilde{\es})\right)\left\{ (\es+\tilde{\es})K(\sigma'),L(\sigma)\right\} +\nonumber \\
 &  & -(\tet'-\tet)\tet\partial_{\sigma}\delta(\sigma-\sigma')\left[K,L\right]_{(1)}^{\Delta}(\sigma)\end{eqnarray}
Now let us make use of (\ref{eq:stilde-in-bracket}) and (\ref{eq:v-sig-tet-II})
to arrive at\begin{eqnarray}
\lqn{\left\{ K(\sigma',\tet'),L(\sigma,\tet)\right\} =}\nonumber \\
 & = & \cov\left(\delta(\tet-\tet')\delta(\sigma-\sigma')\right)\left[K,L\right]_{(1)}^{\Delta}(\sigma,\tet)+\delta(\tet'-\tet)\delta(\sigma-\sigma')\bei{\left[(\de+\ip_{v})K,L\right]_{(1)}^{\Delta}(\sigma,\tet)}{v^{k}=-\partial_{\sigma}\Phi^{k}}\qquad\end{eqnarray}
which is the first equation of the proposition. Integrating over $\tet'$
and $\sigma'$ results in\begin{eqnarray}
\int d\sigma'\int d\tet'\left\{ K(\sigma',\tet'),L(\sigma,\tet)\right\}  & = & \bei{\left[(\de+\ip_{v})K,L\right]_{(1)}^{\Delta}(\sigma,\tet)}{v^{k}=-\partial_{\sigma}\Phi^{k}}=\\
 & = & \bei{\left[(\de+\ip_{v})K,L\right]_{(1)}^{\Delta}(\sigma)}{v^{k}=-\partial_{\sigma}x^{k}}+\tet(\es+\tilde{\es})\bei{\left[(\de+\ip_{v})K,L\right]_{(1)}^{\Delta}(\sigma)}{v^{k}=-\partial_{\sigma}x^{k}}\end{eqnarray}
A second integration picks out the linear part in $\tet$ and adjusting
the order of the integrations gives the additional sign in (\ref{eq:Proposition1b-integrated}).$\qquad\qquad\qquad\qquad\qquad\qquad\qquad\qquad\qquad\qquad\qquad\qquad\qquad\qquad\qquad\qquad\square$

\subsubsection*{Application to the second supercharge $\Q_{2}$}

We are now ready to apply the proposition in the integrated form (\ref{eq:Proposition1b-integrated})
to the question of the existence of a second worldsheet supersymmetry
$\Q_{2}$. Remember, we want $\{\Q_{2},\Q_{2}\}=-2P$. Due to the
proposition, the lefthand side can be written as\begin{eqnarray}
\{\Q_{2},\Q_{2}\} & = & \frac{1}{4}(\es+\tilde{\es})\int d\sigma\,\Big([\mc{J},_{\de}\mc{J}]_{(1)}^{\Delta}-[\ip_{v}\mc{J},\mc{J}]_{(1)}^{\Delta}\Big|_{v=-\partial_{\sigma}x^{k}\ro_{k}}\Big)(\sigma)\label{eq:gruene-Pause}\end{eqnarray}
For $\mc{J}^{2}=-1$, the second term under the integral simplifies
significantly\begin{eqnarray}
-\frac{1}{4}\int d\sigma[\ip_{v}\mc{J},\mc{J}]_{(1)}^{\Delta}\Big|_{v=-\partial_{\sigma}x^{k}\ro_{k}} & = & -\int d\sigma\, v^{K}\mc{J}_{K}\hoch{L}\mc{J}_{L}\hoch{M}\basis_{M}\Big|_{v=-\partial_{\sigma}x^{k}\ro_{k}}=-\int d\sigma\,\partial_{\sigma}x^{k}\ro_{k}=\tilde{\OO}\end{eqnarray}
 Recalling that \begin{eqnarray}
(\es+\tilde{\es})\tilde{\OO} & = & \es\tilde{\OO}=\tilde{\es}\OO=(\es+\tilde{\es})\OO=-P\\
\textrm{and }\OO & = & \int d\sigma\,\oo(\sigma)\quad(\textrm{see }(\ref{eq:o-of-sigma}))\end{eqnarray}
we can rewrite (\ref{eq:gruene-Pause}) as\begin{eqnarray}
\dann\{\Q_{2},\Q_{2}\} & = & \frac{1}{4}(\es+\tilde{\es})\left(\int d\sigma\,[\mc{J},_{\de}\mc{J}]_{(1)}^{\Delta}+4\OO\right)=\\
 & = & \frac{1}{4}(\es+\tilde{\es})\left(\int d\sigma\,\left([\mc{J},_{\de}\mc{J}]_{(1)}^{\Delta}-4\oo\right)(\sigma)\right)+2\underbrace{\tilde{\es}\OO}_{-P}\end{eqnarray}
The righthand side clearly equals $-2P$ for \begin{eqnarray}
[\mc{J},_{\de}\mc{J}]_{(1)}^{\Delta}-4\oo & = & 0\end{eqnarray}
which is again (according to (\ref{eq:integrability-big-derived}))
just the integrability condition for the generalized almost complex
structure $\mc{J}$.

\rem{hier lungern explizite $(J,J)$-Rechnung und der Quanten-Fall!}

\section{Conclusions}

We have seen two closely related mechanisms in sigma-models with a
special field content which lead to the derived bracket of the target
space algebraic bracket by the target space exterior derivative. This
exterior derivative is implemented in the sigma model in one case
via the derivative with respect to a (worldvolume-) Grassmann coordinate
and in the other case via the derivative with respect to the worldvolume
coordinate itself. In the latter case this derivative has to be contracted
with (worldvolume-) Grassmann coordinates in order to be an odd differential.
This leads to the problem that higher powers of the basis elements
vanish, as soon as the power exceeds the worldvolume dimension as
it happens in Zucchini's application. A big number of Grassmann-variables
is therefore advantageous in that approach. For the other mechanism
one rather prefers to have only one single Grassmann variable as there
is no need for any contraction. There is one worldvolume dimension
more in the Lagrangian formalism and for that reason it was preferable
to apply there the mechanism with worldvolume derivatives and use
the other one in the Hamiltonian formalism. \rem{Ist das wahr, dass beides bei beidem gehen wuerde?}

If one does not consider antisymmetric tensors of higher rank, but
only vectors or one-forms (or forms of worldvolume-dimension), the
partial worldvolume derivative without a Grassmann-coordinate is enough.
There is either no need for antisymmetrization or it can be performed
with the worldvolume epsilon tensor. The nature of the mechanism remains
the same and leads to the observations in \cite{Alekseev:2004np,Bonelli:2005ti}
that the Poisson bracket implements the Dorfman bracket for sums of
vectors and one-forms and the corresponding derived bracket for sums
of vectors and $p$-forms on a $p$-brane \cite{Bonelli:2005ti}.
In that sense, the present article is a generalization of those observations.

There remain a couple of things to do. It should be possible to implement
in the same manner by e.g. a BRST differential other target space
differentials which can depend on some extra-structure and repeat
the same analysis. Symmetric tensors then become more interesting
as well, because they need such an extra-structure anyway for a meaningful
differential. From the string theory point of view, the application
of extended worldsheet supersymmetry corresponds to applications in
the RNS string. But generalized complex geometry contains the tools
to allow RR-fluxes, which are hard to treat in RNS. It would therefore
be nice to find some topological limit in a string theory formalism
which is extendable to RR-fields, like the Berkovits-string \cite{Berkovits:2004px},
leading to a topological sigma model like Zucchini's, in order to
learn more about the correspondence between string theory and generalized
complex geometry.

\rem{noch was ueber Hull und ueber generalized Poisson? und ueber induced differential...}

\subsection*{Acknowledgements }

I would like to thank CEA/Saclay for hospitality during a major part
of work on this subject. I am especially grateful to Ruben Minasian
and also to Pierre Vanhove for making my very pleasant stay in Saclay
possible and to them, Mariana Gra\~na, Claus Jeschek, Frederik Witt
and to my supervisor Max Kreuzer (who initially brought the topic
of generalized complex geometry to my attention) for useful discussions
on the subject. Special thanks go to Yvette Kosmann-Schwarzbach and
Maxim Zabzine for their comments on the first preprint version. The
work was financed in parts by EGIDE in France, by a fellowship from
the {}``Akademisch-soziale Arbeitsgemeinschaft \"Osterreichs'',
by a fellowship to go abroad from the TU Vienna and by the FWF research
project P19051. \rem{ESI-fellowship?} This document was produced
using \LyX{}, which is based on \LaTeX{}.

\appendix

\section{Notation and Conventions}

\label{sec:Conventions}Within the article, a lot of different types
of tensors have to be denoted. The choices and sometimes some logic
behind, will be presented here.

World-volume-coordinates are denoted by $\sigma^{\mu}$, target-space
coordinates by $x^{m}$, target space vector-fields by $a,b,\ldots$
or $v,w,\ldots$, 1-forms by small Greek letters $\alpha,\beta,\ldots$
and generalized $T\oplus T^{*}$-vectors by $\mf{a},\mf{b},\ldots$
or $\mf{v},\mf{w},\ldots$~. For an explicit split in vector and
1-form, the letters from the beginning of the alphabet are better
suited, as there is a better correspondence between Latin and Greek
symbols or one can visually better distinguish between Latin and Greek
symbols. Compare e.g. $\mf{a}=a+\alpha$ and $\mf{v}=v+(?\nu)$.\\
Higher order forms will be in general denoted by $\alpha^{(p)},\beta^{(q)},\ldots$
or $\omega^{(p)},\eta^{(q)},\rho^{(r)},\ldots$. There will be exceptions,
however , for specific forms like the $B$-field $B=B_{mn}\de x^{m}\wedge\de x^{n}$.
Following this logic, we will also denote multivectors (tensors with
antisymmetric upper indices) by small letters, indicating their multivector-degree
in brackets: $a^{(p)},b^{(q)},\ldots$ or $v^{(p)},w^{(q)},\ldots$.
There are again exceptions, e.g. a Poisson structure will often be
denoted by $P=P^{mn}\pe_{m}\wedge\pe_{n}$. The most horrible exception
is the one of the beta-transformation, which is denoted by a large
beta $\Beta^{mn}$ in (\ref{eq:SOnn}), in order to distinguish it
from forms.

Tensors of mixed type will be denoted by capital letters where we
denote in brackets first the number of lower indices and then the
number of upper indices, e.g. $T^{(p,q)}$. Most of the time, we treat
multivector valued forms, e.g. the lower indices as well as the upper
indices are antisymmetrized. The letters denoting form degree and
multivector degree will often be adapted to the letter of the tensor,
e.g. $K^{(k,k')},L^{(l,l')},\ldots$\textbf{}\\
\textbf{Attention:} $k$ and $l$ are also used as dummy indices!
Sometimes (I'm sorry for that) the same letter appears with different
meanings. However, in those situations the dummy indices will carry
indices which might even be one of the degrees $k$ or $k'$, e.g.
$K_{\ldots}\hoch{k_{1}\ldots k_{k'}}L_{k_{k'}\ldots k_{1}\ldots}\hoch{\ldots}$.

Working all the time with graded algebras with a graded symmetric
product (the wedge product), everything in this article has to be
understood as \textbf{graded}. I.e. with commutator we mean the graded
commutator and with the Poisson bracket the graded Poisson bracket.
They will not be denoted differently than the non-graded operations.
Relevant for the sign rules is the \textbf{total degree} which we
define to be form degree minus the multivector degree. In the field
language, it corresponds to the total ghost number which is the pure
ghost number minus the antighost number. It will be denoted by \begin{eqnarray}
\abs{K^{(k,k')}} & = & k-k'\end{eqnarray}
As only degrees appear in the exponent of a minus sign, a simplified
notation is used there\begin{equation}
(-)^{A}\equiv(-1)^{\abs{A}},\quad(-)^{A+B}\equiv(-)^{\abs{A}+\abs{B}},\quad(-)^{AB}\equiv(-)^{\abs{A}\abs{B}}\quad\forall A,B\end{equation}
For the Poisson bracket, the following (less common) sign convention
is chosen: \begin{eqnarray}
\left\{ p_{m},x^{n}\right\}  & = & \delta_{m}^{n}=-\left\{ x^{n},p_{m}\right\} \\
\left\{ b_{m},c^{n}\right\}  & = & \delta_{m}^{n}=-(-)^{bc}\left\{ c^{n},b_{m}\right\} \end{eqnarray}
Derivatives with respect to $x^{m}$ are denoted by $\partiell{}{x^{m}}f\equiv\partial_{m}f\equiv f_{,m}$.
For graded variables left derivatives are denoted by $\partl{\ce}f(\ce)$,
while right derivatives are denoted equivalently by two different
notations\begin{eqnarray}
\partial f(\ce)/\partial\ce & \equiv & f\partr{\ce}\end{eqnarray}
The corresponding notations are used for functional derivatives $\funktl{\ce(\sigma)}$.
With respect to the wedge product, the basis element $\pe_{m}$ is
an odd object ($\pe_{m}\wedge\pe_{n}=-\pe_{n}\wedge\pe_{m}$). The
partial derivative $\partial_{k}$ acting on some coefficient function,
however, is an even operator (it does not change the parity as long
as it is not contracted with a basis element $\de x^{k}$). That is
why we denote the odd basis element $\pe_{m}$ and $\de x^{m}$ as
well as the odd exterior derivative $\de\,$ with boldface symbols.
The interior product itself does not carry a grading in the sense
that $\abs{\ip_{K}\rho}=\abs{K}+\abs{\rho}$, while for the Lie derivative
$\Lie_{K}=\left[\ip_{K},\de\,\right]$ the $\Lie$ carries a grading
in the sense $\abs{\Lie_{K}\rho}=\abs{K}+\abs{\rho}+1$. That is why
the Lie derivative is denoted with a boldface $\Lie$ which is also
very good to distinguish it from generalized multivectors $\mc{K},\mc{L},\ldots$.
The philosophy of writing odd objects in boldface style is also extended
to the combined basis element

\begin{equation}
\basis_{M}\equiv(\pe_{m},\de x^{m}),\quad\basis^{M}\equiv(\de x^{m},\pe_{m})\end{equation}
and to the comma in the derived bracket $\left[\,\bs{,}\,\right]$
in contrast to the commutator $\left[\,,\,\right]$. This should be,
however, just a reminder. It will be obvious for other reasons, which
bracket is meant. But we do \textbf{not} extend this philosophy to
vectors and 1-forms, where it would be consistent (but too much effort)
to write the vectors and basis elements in boldface style and the
coefficients in standard style. We will instead write the vector in
the same style as the coefficient $a=a_{m}\de x^{m}$. 

A square bracket is used as usual to denote the antisymmetrization
of, say $p$, indices (including a normalization factor $\frac{1}{p!}$).
A vertical line is used to exclude some indices from antisymmetrization.
An extreme example would be\begin{equation}
A^{[ab\mid cd|e|fg|hi]}\end{equation}
where $A$ is antisymmetrized only in $a,b,e,h$ and $i$, but not
in $c,d,f$ and $g$. Normally we use only expressions like $A^{[ab\mid cd|efg]}$,
where $a,b,e,f$ and $g$ are antisymmetrized.

\paragraph{Wedge product }

\label{Wedge-product} A significant difference from usual conventions
is that for multivectors, forms and generalized multivectors we include
the normalization of the factor already in the definition of the wedge
product\begin{eqnarray}
\de x^{m_{1}}\cdots\de x^{m_{n}}\equiv\de x^{m_{1}}\wedge\ldots\wedge\de x^{m_{n}} & \equiv & \de x^{[m_{1}}\otimes\ldots\otimes\de x^{m_{n}]}\equiv\sum_{P}\frac{1}{n!}\de x^{m_{P(1)}}\otimes\ldots\otimes\de x^{m_{P(n)}}\\
\pe_{m_{1}}\cdots\pe_{m_{n}}\equiv\pe_{m_{1}}\wedge\cdots\wedge\pe_{m_{n}} & \equiv & \pe_{[m_{1}}\otimes\cdots\otimes\pe_{m_{n}]}\equiv\sum_{P}\frac{1}{n!}\pe_{m_{P(1)}}\otimes\cdots\otimes\pe_{m_{P(n)}}\\
\basis_{M_{1}}\ldots\basis_{M_{n}}\equiv\basis_{M_{1}}\wedge\ldots\wedge\basis_{M_{n}} & \equiv & \basis_{[M_{1}}\otimes\ldots\otimes\basis_{M_{n}]}\equiv\sum_{P}\frac{1}{n!}\basis_{M_{P(1)}}\otimes\ldots\otimes\basis_{M_{P(n)}}\end{eqnarray}
(where we sum over all permutations $P$), such that we omit the usual
factor of $\frac{1}{p!}$ in the coordinate expression of a $p$-form,
or a $p$-vector \begin{eqnarray}
\alpha^{(p)} & \equiv & \alpha_{m_{1}\ldots m_{p}}\de x^{m_{1}}\wedge\cdots\wedge\de x^{m_{p}}\equiv\alpha_{m_{1}\ldots m_{p}}\de x^{m_{1}}\cdots\de x^{m_{p}}\\
v^{(p)} & \equiv & v^{m_{1}\ldots m_{p}}\partial_{m_{1}}\wedge\ldots\wedge\partial_{m_{p}}\end{eqnarray}
Readers who prefer the $\frac{1}{p!}$, can easily reintroduce it
in every equation by replacing e.g. the coefficient functions $v^{m_{1}\ldots m_{p}}\To\frac{1}{p!}v^{m_{1}\ldots m_{p}}$.
The equation for the Schouten bracket ( \ref{eq:Schouten-bracketI}),
for example, would change as follows: \begin{eqnarray}
\left[v^{(p)}\bs{,}w^{(q)}\right]^{m_{1}\ldots m_{p+q-1}} & = & pv^{[m_{1}\ldots m_{p-1}|k}\partial_{k}w^{|m_{p}\ldots m_{p+q-1}]}-qv^{[m_{1}\ldots m_{p}\mid}\tief{,k}w^{k\,\mid m_{p+1}\ldots m_{p+q-1}]}\\
\hspace{-1cm}\To\frac{1}{(p+q-1)!}\left[v^{(p)}\bs{,}w^{(q)}\right]^{m_{1}\ldots m_{p+q-1}} & = & \frac{1}{(p-1)!}\frac{1}{q!}v^{[m_{1}\ldots m_{p-1}|k}\partial_{k}w^{|m_{p}\ldots m_{p+q-1}]}+\nonumber \\
 &  & -\frac{1}{p!}\frac{1}{(q-1)!}v^{[m_{1}\ldots m_{p}\mid}\tief{,k}w^{k\,\mid m_{p+1}\ldots m_{p+q-1}]}\quad\end{eqnarray}

\paragraph{Schematic index notation }

For longer calculations in coordinate form it is useful to introduce
the following notation, where every boldface index is assumed to be
contracted with the corresponding basis element (at the same position
of the index), s.th. the indices are automatically antisymmetrized.
\label{fat-index}\begin{eqnarray}
\omega^{(p)} & = & \omega_{m_{1}\ldots m_{p}}\de x^{m_{1}}\cdots\de x^{m_{p}}\equiv\omega_{\bs{m}\ldots\bs{m}}\\
a^{(p)} & = & a^{n_{1}\ldots n_{p}}\pe_{n_{1}}\wedge\ldots\pe_{n_{p}}\equiv a^{\nn}\\
\mc{K}^{(p)} & = & \mc{K}_{M_{1}\ldots M_{p}}\basis^{M_{1}}\ldots\basis^{M_{p}}\equiv\mc{K}_{\bs{M}\ldots\bs{M}}=\\
 & = & \mc{K}^{M_{1}\ldots M_{p}}\basis_{M_{1}}\ldots\basis_{M_{p}}\equiv\mc{K}^{\bs{M}\ldots\bs{M}}\end{eqnarray}
or for products of tensors e.g. \begin{eqnarray}
\omega_{\bs{m}\ldots\bs{m}}\eta_{\bs{m}\ldots\bs{m}} & \equiv & \omega_{[m_{1}\ldots m_{p}}\eta_{m_{p+1}\ldots m_{p+q}]}\de x^{m_{1}}\cdots\de x^{m_{p+q}}=\\
 & = & \omega_{m_{1}\ldots m_{p}}\eta_{m_{p+1}\ldots m_{p+q}}\de x^{m_{1}}\cdots\de x^{m_{p+q}}=(-)^{pq}\eta_{\bs{m}\ldots\bs{m}}\omega_{\bs{m}\ldots\bs{m}}\end{eqnarray}
A boldface index might be hard to distinguish from an ordinary one,
but this notation is nevertheless easy to recognize, as normally several
coinciding indices appear (which are not summed over as they are at
the same position). Similarly, for multivector valued forms we define%
\footnote{Upper and lower signs are thus treated independently. For calculational
reasons this is not the best way to do. We can interpret every boldface
index on the lefthand side of (\ref{eq:grosser-Vorzeichenkummer})
as a basis element sitting at the position of the index, so that the
order of the basis elements on the lefthand side is first $k\times\de x^{m}$,
$(k'-1)\pe_{m}$, $(l-1)\times\de x^{m}$ and $l'\times\pe_{m}$,
s.th., in order to get the order of the righthand side, we have to
interchange $(k'-1)\pe_{m}$ with $(l-1)\times\de x^{m}$, which gives
a sign factor of $(-)^{(k'-1)(l-1)}$. This is a natural sign factor
which appears all the way in the equations, which could be easily
absorbed into the definition. However, we wanted to keep the sign
factors explicitly in the equations in order to keep the notation
as self-explaining as possible and not confuse the reader too much.$\qquad\fussend$%
}\begin{eqnarray}
K_{\bs{m}\ldots\bs{m}}\hoch{\bs{n}\ldots\bs{n}} & \equiv & K_{m_{1}\ldots m_{k}}\hoch{n_{1}\ldots n_{k'}}\de x^{m_{1}}\wedge\ldots\wedge\de x^{m_{k}}\otimes\pe_{m_{1}}\wedge\ldots\wedge\pe_{m_{k'}}\\
\hspace{-1cm}K_{\bs{m}\ldots\bs{m}}\hoch{\bs{n}\ldots\bs{n}p}L_{p\bs{m}\ldots\bs{m}}\hoch{\bs{n}\ldots\bs{n}} & \equiv & \!\! K_{m_{1}\ldots m_{k}}\hoch{n_{1}\ldots n_{k'-1}p}L_{pm_{1}\ldots m_{l-1}}\hoch{n_{1}\ldots n_{l'}}\de x^{m_{1}}\cdots\de x^{m_{k+l-1}}\!\!\otimes\!\pe_{m_{1}}\cdots\pe_{m_{k'+l'-1}}\quad\label{eq:grosser-Vorzeichenkummer}\end{eqnarray}

}
\Teil{derivedbrackets}{

\section{Review of geometric brackets as derived brackets}

\label{sec:bracket-review} Mathematics in this section is based on
the review article on derived brackets by Kosmann-Schwarzbach \cite{Kosmann-Schwarzbach:2003en}.
The presentation, however, will be somewhat different and in addition
to (or sometimes instead of) the abstract definitions coordinate expressions
will be given.

\subsection{Lie bracket of vector fields, Lie derivative and Schouten bracket}

\label{sub:Lie-and-Schouten} This first subsection is intended to
give a feeling, why the Schouten bracket is a very natural extension
of the Lie bracket of vector fields. It is a good example to become
more familiar with the subject, before we become more general in the
subsequent subsections, but it can be skipped without any harm (note
however the notation introduced before (\ref{eq:first-fat-notation})). 

Consider the ordinary \textbf{Lie-bracket of vector fields} which
turns the tangent space of a manifold into a Lie algebra or the tangent
bundle into a Lie algebroid \rem{stimmt das?}  and which takes in
a local coordinate basis the familiar form\begin{eqnarray}
\left[v\bs{,}w\right]^{m} & = & v^{k}\partial_{k}w^{m}-w^{k}\partial_{k}v^{m}\label{eq:vector-Lie-bracket}\end{eqnarray}
We will convince ourselves in the following that numerous other common
differential brackets are just natural extensions of this bracket
and can be regarded as one and the same bracket. Such a generalized
bracket is e.g. useful to formulate integrability conditions and it
can serve via the Jacobi identity as a powerful tool in otherwise
lengthy calculations\rem{Hier waere das Beispiel mit Nijenhuis angebracht, wenn Leibniz funktionieren wuerde} .
In addition it shows up naturally in some sigma-models as is discussed
in section \ref{sec:sigma-model-induced}. 

Given the Lie-bracket of vector fields, it seems natural to extend
it to higher rank tensor fields by demanding a Leibniz rule on tensor
products of the form $\left[v\bs{,}w_{1}\otimes w_{2}\right]=\left[v\bs{,}w_{1}\right]\otimes w_{2}+w_{1}\otimes\left[v\bs{,}w_{2}\right]$.
Remembering that the Lie-bracket of two vector fields is just the
Lie derivative of one vector field with respect to the other\begin{eqnarray}
\left[v\bs{,}w\right] & = & \Lie_{v}w\end{eqnarray}
the \textbf{Lie derivative} of a general tensor $T=T_{m_{1}\ldots m_{p}}^{n_{1}\ldots n_{q}}\de x^{m_{1}}\otimes\ldots\otimes\de x^{m_{p}}\otimes\pe_{n_{1}}\otimes\cdots\otimes\pe_{n_{q}}$with
respect to a vector field $v$ can be seen as a first extension of
the Lie bracket:\begin{eqnarray}
\left[v\bs{,}T\right] & \equiv & \Lie_{v}T\\
\left[v\bs{,}T\right]_{m_{1}\ldots m_{p}}^{n_{1}\ldots n_{q}} & = & v^{k}\partial_{k}T_{m_{1}\ldots m_{p}}^{n_{1}\ldots n_{q}}-\sum_{i}\partial_{k}v^{n_{i}}T_{m_{1}\ldots m_{p}}^{n_{1}\ldots n_{i-1}k\, n_{i+1}\ldots n_{q}}+\sum_{j}\partial_{m_{j}}v^{k}T_{m_{1}\ldots m_{j-1}k\, m_{j+1}\ldots m_{p}}^{n_{1}\ldots n_{q}}\quad\label{eq:Lie-derivative}\end{eqnarray}
The Lie derivative obeys (as a derivative should) the \textbf{Leibniz
rule} \begin{eqnarray}
\left[v\bs{,}T_{1}\otimes T_{2}\right] & = & \left[v\bs{,}T_{1}\right]\otimes T_{2}+T_{1}\otimes\left[v\bs{,}T_{2}\right]\end{eqnarray}
In fact, giving as input only the Lie derivative of a scalar $\phi$,
namely the directional derivative $\left[v\bs{,}\phi\right]\equiv v^{k}\partial_{k}\phi$,
and the Lie bracket of vector fields (\ref{eq:vector-Lie-bracket}),
the Lie derivative of general tensors (\ref{eq:Lie-derivative}) is
determined by the Leibniz-rule. Insisting on antisymmetry of the bracket,
we have to define \begin{eqnarray}
\left[T\bs{,}v\right] & \equiv & -\left[v\bs{,}T\right]\label{eq:tensor-vector-bracket}\end{eqnarray}
Indeed, it can be checked that the above definitions lead to a valid
Jacobi-identity of the form \rem{Herleitung im 2006-Planer, S.23, beim 23.1., am 6.8.} \begin{eqnarray}
\left[v\bs{,}\left[w\bs{,}T\right]\right] & = & \left[\left[v\bs{,}w\right]\bs{,}T\right]+\left[w\bs{,}\left[v\bs{,}T\right]\right]\quad\textrm{for arbitrary tensors }T\label{eq:Jacobi-for-arbitraryT}\end{eqnarray}
which is perhaps better known in the form \begin{eqnarray}
\left[\Lie_{v},\Lie_{w}\right]T & = & \Lie_{\left[v\bs{,}w\right]}T\label{eq:Jacobi-for-arbitraryTII}\end{eqnarray}
 We have now vectors acting via the bracket on general tensors, but
tensors only acting on vectors via (\ref{eq:tensor-vector-bracket})
. It is thus natural to use Leibniz again to define the action of
tensors on tensors. To make a long story short, this is not possible
for general tensors. It is possible, however, for tensors with only
upper indices which are either antisymmetrized (\textbf{multivectors})
or symmetrized (\textbf{symmetric multivectors}). We will concentrate
in this paper on tensors with antisymmetrized indices (the reason
being the natural given differential for forms which also have antisymmetrized
indices), but the symmetric case makes perfect sense and at some points
we will give short comments. (See e.g. \cite{Dubois-Violette:1994gy}\rem{In  nur im Abstract. \cite{Michor:1987spd}
Trotzdem hilfreich...} for more information on the Schouten bracket
of symmetric tensor fields.)

Given two \textbf{multivector fields} (note that the prefactor $1/p!$
is intentionally missing (see page \pageref{Wedge-product}). \begin{eqnarray}
v^{(p)} & \equiv & v^{m_{1}\ldots m_{p}}\pe_{m_{1}}\wedge\ldots\wedge\pe_{m_{p}},\qquad w^{(q)}\equiv w^{m_{1}\ldots m_{q}}\pe_{m_{1}}\wedge\ldots\wedge\pe_{m_{q}}\end{eqnarray}
their Schouten(-Nijenhuis) bracket, or \textbf{Schouten bracket} for
short, is given in a local coordinate basis by\begin{eqnarray}
\left[v^{(p)}\bs{,}w^{(q)}\right]^{m_{1}\ldots m_{p+q-1}} & = & pv^{[m_{1}\ldots m_{p-1}|k}\partial_{k}w^{|m_{p}\ldots m_{p+q-1}]}-qv^{[m_{1}\ldots m_{p}\mid}\tief{,k}w^{k\,\mid m_{p+1}\ldots m_{p+q-1}]}\qquad\label{eq:Schouten-bracketI}\end{eqnarray}
Realizing that the Lie-derivative (\ref{eq:Lie-derivative}) of a
multivector field $w^{(q)}$ with respect to a vector $v^{(1)}$ is
\begin{eqnarray}
\left[v\bs{,}w^{(q)}\right]^{n_{1}\ldots n_{q}} & = & v^{k}\partial_{k}w^{n_{1}\ldots n_{q}}-q\partial_{k}v^{[n_{1}|}w^{k\,|n_{2}\ldots n_{q}]}\end{eqnarray}
one recognizes that (\ref{eq:Schouten-bracketI}) is a natural extension
of this, obeying a Leibniz rule, which we will write down below in
(\ref{eq:Leibniz-for-Schouten}). However, as the coordinate form
of generalized brackets will become very lengthy at some point, we
will first introduce some \textbf{notation} which is more schematic,
although still exact. Namely we imagine that every \textbf{boldface
index} $\bs{m}$ is an ordinary index $m$ contracted with the corresponding
basis vector $\pe_{m}$ at the position of the index:\begin{equation}
v^{(p)}=v^{m_{1}\ldots m_{p}}\pe_{m_{1}}\wedge\ldots\wedge\pe_{m_{p}}\equiv v^{\mm}\end{equation}
This saves us the writing of the basis vectors as well as the enumeration
or manual antisymmetrization of the indices. As a boldface index might
be hard to distinguish from an ordinary one, we will use this notation
only for several indices, s.th. we get repeated indices $\mm$ which
are easily to recognize (and are not summed over, as they are at the
same vertical position). See in the appendix \ref{sec:Conventions}
on page \pageref{fat-index} for a more detailed explanation. The
Schouten bracket then reads\begin{eqnarray}
\left[v^{(p)}\bs{,}w^{(q)}\right] & = & pv^{\bs{m}\ldots\bs{m}k}\partial_{k}w^{\bs{m}\ldots\bs{m}}-qv^{\bs{m}\ldots\bs{m}}\tief{,k}w^{k\,\bs{m}\ldots\bs{m}}=\label{eq:first-fat-notation}\\
 & = & pv^{\bs{m}\ldots\bs{m}k}\partial_{k}w^{\bs{m}\ldots\bs{m}}-(-)^{p(q-1)}qw^{k\,\bs{m}\ldots\bs{m}}v^{\bs{m}\ldots\bs{m}}\tief{,k}=\\
 & = & pv^{\bs{m}\ldots\bs{m}k}\partial_{k}w^{\bs{m}\ldots\bs{m}}-(-)^{(p-1)(q-1)}qw^{\bs{m}\ldots\bs{m}k}\partial_{k}v^{\bs{m}\ldots\bs{m}}\label{eq:Schouten-bracket-condensed}\end{eqnarray}
In the last line it becomes obvious that the bracket is \textbf{skew-symmetric}
in the sense of a Lie algebra of degree%
\footnote{\label{Lie-bracket-of-degree}A \textbf{Lie bracket} $\left[\:,_{(n)}\:\right]$
\textbf{of degree} $n$ in a graded algebra increases the degree (which
we denote by $\abs{\ldots}$) by $n$\[
\Abs{\left[A,_{(n)}B\right]}=\abs{A}+\abs{B}+n\]
It can be understood as an ordinary graded Lie-bracket, when we redefine
the grading $\norm{\ldots}\equiv\abs{\ldots}+n$, such that the Lie
bracket itself does not carry a grading any longer\begin{eqnarray*}
\Norm{\left[A,_{(n)}B\right]} & = & \Norm{A}+\Norm{B}\end{eqnarray*}
The symmetry properties are thus (\textbf{skew symmetry of degree}
$n$) \begin{eqnarray*}
\left[A,_{(n)}B\right] & = & -(-)^{(\abs{A}+n)(\abs{A}+n)}\left[B,_{(n)}A\right]\end{eqnarray*}
and it obeys the usual graded Jacobi-identity (with shifted degrees)\begin{eqnarray*}
\left[A,_{(n)}\left[B,_{(n)}C\right]\right] & = & \left[\left[A,_{(n)}B\right],_{(n)}C\right]+(-)^{(\abs{A}+n)(\abs{A}+n)}\left[B,_{(n)}\left[A,_{(n)}C\right]\right]\end{eqnarray*}
In addition there might be a Poisson-relation with respect to some
other product which respects the original grading. To be consistent
with both gradings, this relation has to read\begin{eqnarray*}
\left[A,_{(n)}B\cdot C\right] & = & \left[A,_{(n)}B\right]\cdot C+(-)^{(\abs{A}+n)\abs{B}}B\cdot\left[A,_{(n)}C\right]\end{eqnarray*}
This is consistent with $B\cdot C=(-)^{\abs{B}\abs{C}}C\cdot B$ on
the one hand and the skew symmetry of the bracket on the other hand.
One can imagine the grading of the bracket to sit at the position
of the comma.

For the bracket of multivectors we have as degree the vector degree.
Later, when we will have tensors of mixed type (vector and form),
we will use the form degree minus the vector degree as total degree.
Then the Schouten-bracket is of degree +1, which should not confuse
the reader.$\qquad\fussend$%
} \textbf{$-1$}:\begin{eqnarray}
\left[v^{(p)}\bs{,}w^{(q)}\right] & = & -(-)^{(p-1)(q-1)}\left[w^{(q)}\bs{,}v^{(p)}\right]\end{eqnarray}
It obeys the corresponding \textbf{Jacobi identity} \begin{eqnarray}
\left[v_{1}^{(p_{1})}\bs{,}\left[v_{2}^{(p_{2})}\bs{,}v_{3}^{(p_{3})}\right]\right] & = & \left[\left[v_{1}^{(p_{1})}\bs{,}v_{2}^{(p_{2})}\right]\bs{,}v_{3}^{(p_{3})}\right]+(-)^{(p_{1}-1)(p_{2}-1)}\left[v_{2}^{(p_{2})}\bs{,}\left[v_{1}^{(p_{1})}\bs{,}v_{3}^{(p_{3})}\right]\right]\end{eqnarray}
Our starting point was to extend the bracket in a way that it acts
via Leibniz on the wedge product. A Lie algebra which has a second
product on which the bracket acts via Leibniz is known as Poisson
algebra. However, here the bracket has degree $-1$ (it reduces the
multivector degree by one) while the wedge product has no degree (the
degree of the wedge product of multivectors is just the sum of the
degrees). According to footnote \ref{Lie-bracket-of-degree}, we have
to adjust the Leibniz rule. The resulting algebra for Lie brackets
of degree -1 is known as \textbf{Gerstenhaber algebra} or in this
special case \textbf{Schouten algebra} (which is the standard example
for a Gerstenhaber algebra). The \textbf{Leibniz rule} is\begin{eqnarray}
\left[v_{1}^{(p_{1})}\bs{,}\, v_{2}^{(p_{2})}\wedge v_{3}^{(p_{3})}\right] & = & \left[v_{1}^{(p_{1})}\bs{,}\, v_{2}^{(p_{2})}\right]\wedge v_{3}^{(p_{3})}+(-)^{(p_{1}-1)p_{2}}v_{2}^{(p_{2})}\wedge\left[v_{1}^{(p_{1})}\bs{,}\, v_{3}^{(p_{3})}\right]\label{eq:Leibniz-for-Schouten}\end{eqnarray}
The standard example in field theory for a Poisson algebra is the
phase space equipped with the Poisson bracket or the commutator of
operators or matrices.%
\footnote{In fact, working with totally symmetric multivector fields would have
lead to a Poisson algebra instead of a Gerstenhaber algebra. \rem{ist das richtig?}$\qquad\fussend$%
} The Schouten algebra is naturally realized by the \textbf{antibracket}
of the BV antifield formalism (see subsection \ref{sub:antibracket}).

\subsection{Embedding of vectors into the space of differential operators}

\label{sub:Embedding-of-vectors} The Leibniz rule is not the only
concept to generalize the vector Lie bracket to higher rank tensors.
The major difficulty in the definition of brackets between higher
rank tensors is the Jacobi-identity, which should hold for them. It
is therefore extremely useful to have a mechanism which automatically
guarantees the Jacobi identity. A way to get such a mechanism is to
\textbf{embed} the tensors into some space of differential operators,
as for the operators we have the commutator as natural Lie bracket
which might in turn induce some bracket on the tensors we started
with. Vector fields e.g. naturally act on differential forms via the
\textbf{interior product}\begin{eqnarray}
\ip_{v}\omega^{(p)} & \equiv & p\cdot v^{k}\omega_{k\bs{m}\ldots\bs{m}}\label{eq:vector-interior-product}\end{eqnarray}
This can be seen as the embedding of vector fields in the space of
differential operators acting on forms, because the interior product
with respect to a vector is a graded derivative with the grading -1
of the vector (we take as total degree the form degree minus the multivector
degree, which for a vector is just -1)\begin{eqnarray}
\ip_{v}\left(\omega^{(p)}\wedge\eta^{(q)}\right) & = & \ip_{v}\omega^{(p)}\wedge\eta^{(q)}+(-)^{q}\omega^{(p)}\wedge\ip_{v}\eta^{(q)}\end{eqnarray}
Taking the idea of above we can take the commutator of two interior
products. We note, however, that it only induces a trivial (always
vanishing) bracket on the vectorfields\begin{eqnarray}
\left[\ip_{v},\ip_{w}\right] & = & 0=\ip_{0}\label{eq:trivial-alg-bracket}\end{eqnarray}
As the interior product (\ref{eq:vector-interior-product}) does not
include any partial derivative on the vector-coefficient, it was clear
from the beginning that this ansatz does not lead to the Lie bracket
of vector fields or any generalization of it. We have to bring the
exterior derivative into the game, in our notation \begin{equation}
\de\omega^{(p)}=\partial_{\bs{m}}\omega_{\bs{m}\ldots\bs{m}}\end{equation}
 There are two ways to do this

\begin{itemize}
\item \emph{Change the embedding:} Instead of embedding the vectors via
the interior product acting on forms, we can embed them via the Lie-derivative
acting on forms. When acting on forms, the Lie derivative can be written
as the (graded) commutator of interior product and exterior derivative\rem{spaeter
nicht immer injektiv!%
\footnote{This is perhaps a good point to demonstrate how to calculate with
our schematic notation\begin{eqnarray*}
\Lie_{v}\rho^{(r)} & = & \left[\ip_{v},\de\,\right]\rho^{(r)}=\\
 & = & \ip_{v}\de\rho^{(r)}+\de\ip_{v}\rho^{(r)}=\\
 & = & (p+1)\cdot v^{k}\partial_{[k}\omega_{\bs{m}\ldots\bs{m}]}+\partial_{\bs{m}}\left(p\cdot v^{k}\omega_{k\bs{m}\ldots\bs{m}}\right)=\\
 & = & v^{k}\partial_{k}\omega_{\bs{m}\ldots\bs{m}}-p\cdot v^{k}\partial_{\bs{m}}\omega_{k\bs{m}\ldots\bs{m}}+p\cdot\partial_{\bs{m}}v^{k}\omega_{k\bs{m}\ldots\bs{m}}+p\cdot v^{k}\partial_{\bs{m}}\omega_{k\bs{m}\ldots\bs{m}}=\\
 & = & v^{k}\partial_{k}\omega_{\bs{m}\ldots\bs{m}}+p\cdot\partial_{\bs{m}}v^{k}\omega_{k\bs{m}\ldots\bs{m}}\qquad\fussend\end{eqnarray*}
}} \begin{eqnarray}
\Lie_{v} & = & \left[\ip_{v},\de\,\right]\\
\Lie_{v}\omega^{(p)} & = & v^{k}\partial_{k}\omega_{\bs{m}\ldots\bs{m}}+p\cdot\partial_{\bs{m}}v^{k}\omega_{k\,\bs{m}\ldots\bs{m}}\label{eq:vector-Lie-derivative}\end{eqnarray}
Indeed, using the Lie derivative as embedding $v\mapsto\Lie_{v}$,
the commutator of Lie derivatives induces the Lie bracket of vector
fields (a special case of (\ref{eq:Jacobi-for-arbitraryTII}) \begin{eqnarray}
\left[\Lie_{v},\Lie_{w}\right] & = & \Lie_{[v\bs{,}w]}\label{eq:induced-bracket-via-Lie-derivative}\end{eqnarray}

\item \emph{Change the bracket:} In the space of differential operators
acting on forms, the commutator is the most natural Lie bracket. However,
the existence of a nilpotent odd operator acting on our algebra, namely
the commutator with the exterior derivative, enables the construction
of what is called a \textbf{derived bracket}%
\footnote{\label{foot-derived-bracket}Given a bracket $\left[\,,_{(n)}\,\right]$
of degree $n$ (not necessarily a Lie bracket. It can be as well a
\textbf{Loday bracket} where the skew-symmetry property as compared
to footnote \ref{Lie-bracket-of-degree} is missing, but the Jacobi
identity still holds) and a differential $\De$ (derivation of degree
1 and square 0), its \textbf{derived bracket} \cite{Kosmann-Schwarzbach:1996a,Kosmann-Schwarzbach:1996b,Kosmann-Schwarzbach:2003en}
(which is of degree $n+1$) is defined by\[
\left[a,_{(\De)}b\right]=(-)^{n+a+1}\left[\De a,_{(n)}b\right]\]
We put the subscript $(\De\,)$ at the position of the comma, to indicate
that the grading of D is sitting there. The strange sign is just to
make the definition nicer for the most frequent case of an interior
derivation, where $\De a=\left[d,_{(n)}a\right]$ with $d$ some element
of the algebra with degree $\abs{d}=1-n$ and $\left[d,_{(n)}d\right]=0$,
s.th. we have\[
\left[a,_{d}b\right]=\left[\left[a,_{(n)}d\right],_{(n)}b\right]\]
The derived bracket is then again a Loday bracket (of degree $n+1$)
and obeys the corresponding Jacobi-identity (that is always the nontrivial
part). If $a,b$ are elements of a commuting subalgebra ($[a,_{(n)}b]=0$),
the derived bracket even is skew-symmetric and thus a Lie bracket
of degree $n+1$.

In the case at hand we start with a Lie bracket of degree 0 (the commutator)
and take as interior derivation the commutator with the exterior derivative
$\left[\de\,,\ldots\right]$. Note that the exterior derivative itself
is a derivative on forms, but not on the space of differential operators
on forms. Therefore we need the commutator.$\quad\fussend$%
}.\begin{eqnarray}
\left[\ip_{v},_{\de}\ip_{w}\right] & \equiv & \left[\left[\ip_{v},\de\,\right],\ip_{w}\right]\end{eqnarray}
This derived bracket (which is in this case a Lie bracket again, as
we are considering the abelian subalgebra of interior products of
vector fields) indeed induces the Lie bracket of vector fields when
we use the interior product as embedding\begin{eqnarray}
\left[\ip_{v},_{\de}\ip_{w}\right] & = & \ip_{\left[v\bs{,}w\right]}\label{eq:vector-derived-bracket}\end{eqnarray}

\end{itemize}
The above equations plus two additional ones are the well known \textbf{Cartan
formulae} \begin{eqnarray}
\left[\ip_{v},\ip_{w}\right] & = & 0=\left[\de\,,\de\,\right]\\
\Lie_{v} & = & \left[\ip_{v},\de\,\right]\\
\left[\Lie_{v},\de\,\right] & = & 0\\
\left[\Lie_{v},\Lie_{w}\right] & = & \Lie_{\left[v,w\right]}\\
\big[\underbrace{\left[\ip_{v},\de\,\right]}_{\Lie_{v}},\ip_{w}\big]] & = & \ip_{\left[v,w\right]}\end{eqnarray}
(\ref{eq:induced-bracket-via-Lie-derivative}) can be rewritten, using
Jacobi's identity and $\left[\de\,,\de\,\right]=0$, as \begin{eqnarray}
\left[\left[\left[\ip_{v},\de\,\right],\ip_{w}\right],\de\,\right] & = & \left[\ip_{\left[v,w\right]},\de\,\right]\end{eqnarray}
Starting from (\ref{eq:vector-derived-bracket}), one thus arrives
at (\ref{eq:induced-bracket-via-Lie-derivative}) by simply taking
the commutator with $\de\,$. We will therefore concentrate in the
following on the second possibility, using the derived bracket, as
the first one can be deduced from it. Let us just mention that the
generalization in the spirit of the derived bracket (\ref{eq:vector-derived-bracket})
(or more precise its skew-symmetrization) is known as \textbf{Vinogradov
bracket} \cite{Vinogradov:1990,Vinogradov:1992} (see footnote \ref{Vinogradov-bracket}),
while the generalization in the spirit of (\ref{eq:induced-bracket-via-Lie-derivative})
is known as \textbf{Buttin's bracket} \cite{Buttin:1974}.\label{ite:Buttins-bracket}

\subsection{Derived bracket for multivector valued forms}

\label{sub:multivector-form-brackets} \rem{section similar zum Text!}
Let us now consider a much more general case, namely the space of
multivector valued forms, i.e. tensors which are antisymmetric in
the upper as well as in the lower indices. With the Schouten bracket
we have a bracket for multivectors, which are antisymmetric in all
(upper) indices. There exists as well a bracket for vector valued
forms, namely tensors with one upper index and arbitrary many antisymmetrized
lower indices. This bracket (which we have not yet discussed) is the
(Fr\"ohlicher-) Nijenhuis bracket (see (\ref{eq:Nijenhuis-bracket-coord})),
which shows up in the integrability condition for almost complex structures.
Multivector valued forms have arbitrary many antisymmetrized upper
and arbitrary antisymmetrized lower indices and thus contain both
cases. The antisymmetrization appears quite naturally in field theory
(we give only a few remarks about completely symmetric indices, which
appear as well, but which will not be subject of this paper). It makes
also sense to define brackets on sums of tensors of different type
(e.g. the Dorfman bracket for generalized complex geometry). Those
brackets are then simply given by linearity.

So let us consider two vector valued forms (we denote the number of
lower indices and the number of upper indices in this order via superscripts)%
\footnote{One can certainly map a tensor $K_{m}\hoch{n}\de x^{m}\otimes\pe_{n}$
to one where the basis elements are antisymmetrized $K_{m}\hoch{n}\de x^{m}\wedge\pe_{n}\stackrel{\textrm{see page }\pageref{Wedge-product}}{\equiv}\frac{1}{2}K_{m}\hoch{n}\de x^{m}\otimes\pe_{n}-\frac{1}{2}K_{m}\hoch{n}\pe_{n}\otimes\de x^{m}$
and vice versa. In the field theory applications we will always get
a complete antisymmetrization. This mapping is the reason why we take
care for the horizontal positions of the indices. It should just indicate
the order of the basis elements which was chosen for the mapping.$\quad\fussend$%
}\begin{eqnarray}
K^{(k,k')} & \equiv & K_{\bs{m}\ldots\bs{m}}\hoch{\bs{n}\ldots\bs{n}}\equiv K_{m_{1}\ldots m_{k}}\hoch{n_{1}\ldots n_{k'}}\de x^{m_{1}}\cdots\de x^{m_{k}}\otimes\pe_{n_{1}}\cdots\pe_{n_{k'}}\\
L^{(l,l')} & \equiv & L_{\underbrace{{\scriptstyle \bs{m}\ldots\bs{m}}}_{l}}\hoch{\bs{n}\ldots\bs{n}}_{\underbrace{}_{l'}}\end{eqnarray}
Note the use of the schematic index notation, which we used for upper
indices already in subsection \ref{sub:Lie-and-Schouten} and which
is explained in the appendix \ref{sec:Conventions} on page \pageref{fat-index}.
Following the ideas of above, we want to embed those vector valued
forms in some space of differential operators. As we have upper as
well as lower indices now, it is less clear why we should choose the
space of operators acting on forms and not on some other tensors for
the embedding. However, the space of forms is the only one where we
have a natural exterior derivative without using any extra structure%
\footnote{One can define an exterior derivative -- the \textbf{Lichnerowicz-Poisson
differential} -- on the space of multivectors as well (via the Schouten
bracket), but for this we need an integrable Poisson structure: $\de_{P}N^{(q)}\equiv\left[P^{(2)}\bs{,}N^{(q)}\right]$,
with $\left[P^{(2)}\bs{,}P^{(2)}\right]=0\qquad\fussend$%
}. Therefore we will define again a natural embedding into the space
of differential operators acting on forms as a generalization of the
interior product. Namely, we will act with a multivector valued form
$K$ on a form $\rho$ by just contracting all upper indices with
form-indices and antisymmetrizing the remaining lower indices s.th.
we get again a form as result. The formal definition goes in two steps.
First one defines the interior product with multivectors. For a decomposable
multivector $v^{(p)}=v_{1}\wedge\ldots\wedge v_{p}$ set\begin{eqnarray}
\ip_{v_{1}\wedge\ldots\wedge v_{p}}\rho^{(r)} & \equiv & \ip_{v_{1}}\cdots\ip_{v_{p}}\rho^{(r)}\label{eq:multivec-inter-prod}\end{eqnarray}
This fixes the interior product for a generic multivector uniquely
(contracting all indices with form-indices). The next step is to define
for a multivector valued form $K^{(k,k')}=\eta^{(k)}\wedge v^{(k')}$
which is decomposable in a form and a multivector, that it acts on
a form by first acting with the multivector as above and then wedging
the result with the form \begin{eqnarray}
\ip_{\eta^{(k)}\wedge v^{(k')}}\rho & \equiv & \eta^{(k)}\wedge\ip_{v^{(k)}}\rho=(-)^{k'k}\ip_{v^{(k')}\wedge\eta^{(k)}}\rho\end{eqnarray}
 It is kind of a normal ordering that $\ip_{v^{(k')}}$ acts first:\begin{equation}
\ip_{\eta}\ip_{v}=\ip_{\eta^{(k)}\wedge v^{(k')}}=(-)^{kk'}\ip_{v^{(k')}\wedge\eta^{(k)}}\neq\ip_{v}\ip_{\eta}\end{equation}
  For a generic multivector valued form, the above definitions fix
the following coordinate form of the \textbf{interior product}%
\footnote{\label{foot:interior-product}The name 'interior product' is misleading
in the sense that the operation is (for decomposable tensors) a composition
of interior and exterior wedge product. It will, however, in the generalizations
of Cartan's formulae play the role of the interior product. We will
therefore stick to this name. We can also see it as a short name for
'interior product of maximal order' in the sense that all upper indices
are contracted as opposed to an interior 'product of order $p$',
where we contract only $p$ upper indices. 'Order' is in the sense
of the order of a derivative. While $\ip_{v}$ is a derivative for
any vector $v$, the general interior product acts like a higher order
derivative.$\qquad\fussend$%
} with a multivector valued form\begin{eqnarray}
\ip_{K^{(k,k')}}\rho^{(r)} & \equiv & (k')!\left(\zwek{r}{k'}\right)K_{\bs{m}\ldots\bs{m}}\hoch{l_{1}\ldots l_{k'}}\rho_{\underbrace{{\scriptstyle l_{k'}\ldots l_{1}\bs{m}\ldots\bs{m}}}_{r}}\label{eq:interior-productI}\end{eqnarray}
So we are just contracting all the upper indices of $K$ with an appropriate
number of indices of the form and are wedging the remaining lower
indices. The origin of the combinatorial prefactor is perhaps more
transparent in the phase space formulation (\ref{eq:bc-interior-product-II})
in subsection \ref{sub:bc-phase-space}. For multivectors $v^{(p)}$
and $w^{(q)}$ the operator product of $\ip_{v^{(p)}}$ and $\ip_{w^{(q)}}$
induces, due to (\ref{eq:multivec-inter-prod}) simply the wedge product
of the multivectors\begin{equation}
\ip_{v^{(p)}}\ip_{w^{(q)}}=\ip_{v^{(p)}\wedge w^{(q)}}\label{eq:multivector-product-of-int-pr}\end{equation}
But for general multivector-valued forms we have instead%
\footnote{\label{foot:noncomm-prod}The product of interior products in (\ref{eq:product-of-interior-products})
induces a noncommutative product for the multivector-valued forms,
whose commutator is the algebraic bracket, namely\begin{eqnarray*}
K*L & \equiv & \sum_{p\geq0}\ip_{K}^{(p)}L\\
\left[K,L\right]^{\Delta} & = & K*L-(-)^{(k-k')(l-l')}L*K\qquad\fussend\end{eqnarray*}
}\begin{eqnarray}
\ip_{K^{(k,k')}}\ip_{L^{(l,l')}} & = & \sum_{p=0}^{k'}\ip_{\ip_{K}^{(p)}L}=\ip_{K\wedge L}+\sum_{p=1}^{k'}\ip_{\ip_{K}^{(p)}L}\label{eq:product-of-interior-products}\end{eqnarray}
with \begin{eqnarray}
\ip_{K^{(k,k')}}^{(p)}L^{(l,l')} & \equiv & (-)^{(k'-p)(l-p)}p!\left(\zwek{k'}{p}\right)\left(\zwek{l}{p}\right)K_{\bs{m}\ldots\bs{m}}\hoch{\bs{n}\ldots\bs{n}l_{1}\ldots l_{p}}L_{l_{p}\ldots l_{1}\bs{m}\ldots\bs{m}}\hoch{\bs{n}\ldots\bs{n}}\label{eq:interior-productIII}\end{eqnarray}
\rem{(see (\ref{eq:bc-product-of-interior-products}) and the comments
above). }For $p=k'$, $\ip_{K}^{(p)}$ reduces to the interior product
(\ref{eq:interior-productI}). Both are in general not a derivative
any longer. $\ip^{(p)}$ is, however, a $p$-th order derivative,
as contracting $p$ indices means taking the $p$-th derivative with
respect to $p$ basis elements (see \ref{eq:bc-interior-pproduct-I}
in subsection \ref{sub:bc-phase-space}). Our embedding $\ip_{K^{(k,k')}}$
in (\ref{eq:interior-productI}) is therefore a $k'$-th order derivative.
For $p=0$ on the other hand, $\ip_{K}^{(p)}$ is just a wedge product
with $K$ \rem{For convenience, we also give a last further generalization,
which we will only rarely use\begin{eqnarray}
\ip_{K^{(k,k')}}^{(p,q)}L^{(l,l')} & \equiv & (-)^{q(l+l'+k+k')+(k'-p)(l-p)}p!q!\left(\zwek{k}{q}\right)\left(\zwek{k'}{p}\right)\left(\zwek{l}{p}\right)\left(\zwek{l'}{q}\right)\times\nonumber \\
 &  & \times K_{l_{q}\ldots l_{1}\bs{m}\ldots\bs{m}}\hoch{\bs{n}\ldots\bs{n}l_{1}\ldots l_{p}}L_{l_{p}\ldots l_{1}\bs{m}\ldots\bs{m}}\hoch{\bs{n}\ldots\bs{n}l_{1}\ldots l_{q}}\label{eq:interior-productIV}\end{eqnarray}
$\ip_{K^{(k,k')}}^{(p,q)}$ in (\ref{eq:interior-productIV}) is a
$(p+q)$-th order derivative. Let us also define a generalized \textbf{wedge
product} via \rem{wie genau definier ich jetzt die fetten Indizes?siehe auch 2. Gl oben}\begin{eqnarray*}
\hspace{-1cm}K^{(k,k')}\wedge L^{(l,l')} & \equiv & (-)^{k'l}K_{\bs{m}\ldots\bs{m}}\hoch{\bs{n}\ldots\bs{n}}L_{\bs{m}\ldots\bs{m}}\hoch{\bs{n}\ldots\bs{n}}\equiv\\
 & \equiv & (-)^{k'l}K_{m_{1}\ldots m_{k}}\hoch{n_{1}\ldots n_{k}}L_{m_{k+1}\ldots m_{k+l}}\hoch{n_{k'+1}\ldots n_{k'+l'}}\de x^{m_{1}}\cdots\de x^{m_{k+l}}\otimes\pe_{n_{1}}\cdots\pe_{n_{k'+l'}}=\qquad\\
 & = & \ip_{K}^{(0)}L\end{eqnarray*}
} 

While for vectors the commutator of two interior products (\ref{eq:trivial-alg-bracket})
did only induce a trivial bracket on vectors, which is the same for
multivectors due to (\ref{eq:multivector-product-of-int-pr}), this
is different for multivector-valued forms. \begin{eqnarray}
\hspace{-1cm}\left[\ip_{K^{(k,k')}},\ip_{L^{(l,,l')}}\right] & = & \ip_{\left[K,L\right]^{\Delta}}\label{eq:algebraic-bracket}\\
\left[K,L\right]^{\Delta} & \equiv & \sum_{p\geq1}\underbrace{\ip_{K}^{(p)}L-(-)^{(k-k')(l-l')}\ip_{L}^{(p)}K}_{\equiv[K,L]_{(p)}^{\Delta}}=\label{eq:algebraic-bracketI}\\
 & = & \sum_{p\geq1}\;(-)^{(k'-p)(l-p)}p!\left(\zwek{k'}{p}\right)\left(\zwek{l}{p}\right)K_{\bs{m}\ldots\bs{m}}\hoch{\bs{n}\ldots\bs{n}l_{1}\ldots l_{p}}L_{l_{p}\ldots l_{1}\bs{m}\ldots\bs{m}}\hoch{\bs{n}\ldots\bs{n}}+\nonumber \\
 &  & -(-)^{(k-k')(l-l')}(-)^{(l'-p)(k-p)}p!\left(\zwek{l'}{p}\right)\left(\zwek{k}{p}\right)L_{\bs{m}\ldots\bs{m}}\hoch{\bs{n}\ldots\bs{n}l_{1}\ldots l_{p}}K_{l_{p}\ldots l_{1}\bs{m}\ldots\bs{m}}\hoch{\bs{n}\ldots\bs{n}}\qquad\label{eq:algebraic-bracket-coord}\end{eqnarray}
where we introduced an \textbf{algebraic bracket} $\left[K,L\right]^{\Delta}$
in the second line, which is is due to Buttin \cite{Buttin:1974},
and which is a generalization of the Nijenhuis-Richardson bracket
for vector-valued forms (\ref{eq:Richardson-Nijenhuis-bracket-coord}).
As it was induced via the embedding from the graded commutator, it
has the same properties, i.e. it is graded antisymmetric and obeys
the graded Jacobi identity. Actually, the term with lowest $p$, so
$[K,L]_{(p=1)}^{\Delta}$, is itself an algebraic bracket, which appears
in subsection \ref{sub:Algebraic-brackets} as canonical Poisson bracket\rem{Kosmann-Schwarzbach: Poisson-bracket on $T^*$?}.
It is known under the name \textbf{Buttin's algebraic bracket} (\cite{Buttin:1974},
denoted in \cite{Kosmann-Schwarzbach:2003en} by $\left[\,,\,\right]_{B}^{0}$)
or as \textbf{big bracket\begin{eqnarray}
[K,L]_{(1)}^{\Delta} & = & \ip_{K}^{(1)}L-(-)^{(k-k')(l-l')}\ip_{L}^{(1)}K=\label{eq:bigbracket}\\
 & = & (-)^{(k'-1)(l-1)}k'l\cdot K_{\bs{m}\ldots\bs{m}}\hoch{\bs{n}\ldots\bs{n}l_{1}}L_{l_{1}\bs{m}\ldots\bs{m}}\hoch{\bs{n}\ldots\bs{n}}+\nonumber \\
 &  & -(-)^{(k-k')(l-l')}(-)^{(l'-1)(k-1)}l'k\cdot L_{\bs{m}\ldots\bs{m}}\hoch{\bs{n}\ldots\bs{n}l_{1}}K_{l_{1}\bs{m}\ldots\bs{m}}\hoch{\bs{n}\ldots\bs{n}}\label{eq:bigbracket-coord}\end{eqnarray}
} But as for the vector fields in subsection \ref{sub:Embedding-of-vectors},
we are rather interested in the derived bracket of $\left[K,L\right]^{\Delta}$,
or at the bracket induced via an embedding based on the Lie derivative.
An obvious generalization of the Lie derivative is the commutator
$\left[\ip_{K},\de\,\right]$, which will be a derivative of the same
order as $\ip_{K}$ and therefore is not a derivative in the sense
that it obeys the Leibniz rule. Although it is common to use this
generalization, I am not aware of an appropriate name for it. Let
us just call it the \textbf{Lie derivative} \textbf{with respect to
$K$} (being a derivative of order $k'$) \textbf{}\rem{Achtung, hier hab ich 'I' verwendet}
\begin{eqnarray}
\Lie_{K^{(k,k')}} & \equiv & \left[\ip_{K^{(k,k')}},\de\,\right]\label{eq:Lie-derivativeI}\\
\Lie_{K^{(k,k')}}\rho & = & (k')!\left(\zwek{r+1}{k'}\right)K_{\bs{m}\ldots\bs{m}}\hoch{l_{1}\ldots l_{k'}}\partial_{[l_{k'}}\rho_{l_{k'-1}\ldots l_{1}\bs{m}\ldots\bs{m}]}+\nonumber \\
 &  & -(-)^{k-k'}(k')!\left(\zwek{r}{k'}\right)\partial_{\bs{m}}\left(K_{\bs{m}\ldots\bs{m}}\hoch{l_{1}\ldots l_{k'}}\rho_{l_{k'}\ldots l_{1}\bs{m}\ldots\bs{m}}\right)=\\
 & = & (k')!\left(\zwek{r}{k'-1}\right)K_{\bs{m}\ldots\bs{m}}\hoch{l_{1}\ldots l_{k'}}\partial_{l_{k'}}\rho_{l_{k'-1}\ldots l_{1}\bs{m}\ldots\bs{m}}+\nonumber \\
 &  & -(-)^{k-k'}(k')!\left(\zwek{r}{k'}\right)\partial_{\bs{m}}K_{\bs{m}\ldots\bs{m}}\hoch{l_{1}\ldots l_{k'}}\rho_{l_{k'}\ldots l_{1}\bs{m}\ldots\bs{m}}\label{eq:Lie-derivativeII}\end{eqnarray}
The Lie derivative above is an ingredient to calculate the \textbf{derived
bracket} (remember footnote \ref{foot-derived-bracket} on page \pageref{foot-derived-bracket})
which is given by%
\footnote{\label{Vinogradov-bracket} The \textbf{Vinogradov bracket} \cite{Vinogradov:1992,Vinogradov:1990}
(see also \cite{Kosmann-Schwarzbach:2003en}) is a bracket in the
space of all graded endomorphisms in the space of differential forms
$\Omega^{\bullet}(M)$\begin{eqnarray*}
\left[a\bs{,}b\right]_{V} & = & \frac{1}{2}\left(\left[\left[a,d\right],b\right]-(-)^{b}\left[a,\left[b,d\right]\right]\right)\quad\forall a,b\in\Omega^{\bullet}(M)\end{eqnarray*}
It is the skew symmetrization of a derived bracket. The embedding
of the multivector valued forms into the endomorphisms $\Omega^{\bullet}(M)$
via the interior product which we consider is neither closed under
the Vinogradov bracket nor under the derived bracket in the general
case.$\quad\fussend$%
} \begin{eqnarray}
\left[\ip_{K},_{\de}\ip_{L}\right] & \equiv & \left[\left[\ip_{K},\de\,\right],\ip_{L}\right]\equiv\ip_{\left[K\bs{,}L\right]}\quad\textrm{if possible}\label{eq:derived-bracketI}\end{eqnarray}
One should distinguish the derived bracket on the level of operators
on the left from the derived bracket on the tensors $\left[K\bs{,}L\right]$
on the right. Only in special cases the result of the commutator on
the left can be written as the interior product of another tensorial
object which then can be considered as the derived bracket with respect
to the algebraic bracket $\left[\,,\,\right]^{\Delta}$. Therefore
one normally does not find an explicit general expression for this
derived bracket in literature. In \ref{sub:Extended-exterior-derivative},
however, the meaning of exterior derivative and interior product are
extended in order to be able to write down an explicit general coordinate
expression (\ref{eq:bc-derived-bracket-coord}) which reduces in the
mentioned special cases to the well known results (see e.g. \ref{sub:Nijenhuis-bracket}).
\rem{other derived brackets with other differentials!}

Closely related to the derived bracket in (\ref{eq:derived-bracketI})
of above is \textbf{Buttin's differential bracket}, given by \begin{eqnarray}
\left[\Lie_{K},\Lie_{L}\right] & \equiv & \Lie_{\left[K\bs{,}L\right]_{B}}\quad\textrm{if possible}\label{eq:Buttins-bracket}\end{eqnarray}
Because of $\left[\de\,,\de\,\right]=0$ and $\Lie_{K}=\left[\ip_{K},\de\right]$
we have (using Jacobi)\begin{eqnarray}
\left[\Lie_{K},\Lie_{L}\right] & = & \left[\left[\ip_{K},_{\de}\ip_{L}\right],\,\de\,\right]=\left[\left[\ip_{K},_{\de}\ip_{L}\right],\,\de\,\right]\stackrel{!}{=}[\ip_{\left[K\bs{,}L\right]_{B}},\,\de\,]\label{eq:relation-Buttin-derived}\end{eqnarray}
Comparing with (\ref{eq:derived-bracketI}) s.th. in cases where $\left[K\bs{,}L\right]$
exists, the brackets have to coincide up to a closed term, or locally
a total derivative\begin{eqnarray}
\ip_{\left[K\bs{,}L\right]} & = & \ip_{\left[K\bs{,}L\right]_{B}}+\left[\de\,,\ldots\right]\end{eqnarray}
Using again the extended definition of exterior derivative and interior
product of \ref{sub:Extended-exterior-derivative}, this relation
can be rewritten as\begin{eqnarray}
\left[K\bs{,}L\right] & = & \left[K\bs{,}L\right]_{B}+\de\left(\ldots\right)\label{eq:relation-Buttin-derived-on-Tensor-level}\end{eqnarray}
The Nijenhuis bracket (\ref{eq:Nij-derived-coord}) is the major example
for this relation. \rem{Hier schlummert ein grosser ausgeschnittener Part mit erweiterter auesserer Ableitung und $L^(p)$, Koordinatenform der derived bracket usw...!} \rem{und hier schlummert gleich noch eine Subsection ueber Sums of multivector valued forms}

\subsection{Examples}

\label{sub:Examples}

\subsubsection{Schouten(-Nijenhuis) bracket}

\label{sub:Schouten-bracketII} Let us shortly review the Schouten
bracket under the new aspects. For multivectors $v^{(p)},w^{(q)}$
the algebraic bracket vanishes\begin{eqnarray}
\left[\ip_{v^{(p)}},\ip_{w^{(q)}}\right] & = & 0\end{eqnarray}
 The \textbf{Schouten bracket} $\left[v^{(p)}\bs{,}w^{(q)}\right]$
coincides with the derived bracket as well as with Buttin's differential
bracket, i.e. we have \begin{eqnarray}
\left[\left[\ip_{v^{(p)}},\de\,\right],\ip_{w^{(q)}}\right] & = & \ip_{\left[v^{(p)}\bs{,}w^{(q)}\right]}\\
\left[\Lie_{v^{(p)}},\Lie_{w^{(q)}}\right] & = & \Lie_{\left[v^{(p)}\bs{,}w^{(q)}\right]}\end{eqnarray}
Its coordinate form -- given already before in (\ref{eq:Schouten-bracket-condensed})
-- is\begin{eqnarray}
\left[v^{(p)}\bs{,}w^{(q)}\right] & = & pv^{\bs{m}\ldots\bs{m}k}\partial_{k}w^{\bs{m}\ldots\bs{m}}-(-)^{(p-1)(q-1)}qw^{\bs{m}\ldots\bs{m}k}\partial_{k}v^{\bs{m}\ldots\bs{m}}\end{eqnarray}

The vector Lie bracket is a special case of the Schouten bracket as
well as of the Nijenhuis bracket.

\subsubsection{(Fr\"ohlicher-)Nijenhuis bracket and its relation to the Richardson-Nijenhuis
bracket}

\label{sub:Nijenhuis-bracket} Consider vector valued forms, i.e.
tensors of the form\begin{eqnarray}
K^{(k,1)} & \equiv & K_{m_{1}\ldots m_{k}}\hoch{n}\de x^{m_{1}}\wedge\cdots\wedge\de x^{m_{k}}\wedge\pe_{n}\cong K_{m_{1}\ldots m_{k}}\hoch{n}\de x^{m_{1}}\wedge\cdots\wedge\de x^{m_{k}}\otimes\pe_{n}\end{eqnarray}
The algebraic bracket of two such tensors, defined via the graded
commutator (note that $\abs{\ip_{K}}=\abs{K}=k-1$)\begin{eqnarray}
\left[\ip_{K},\ip_{L}\right] & = & \ip_{\left[K,L\right]^{\Delta}}\end{eqnarray}
consists only of the first term in the expansion, because we have
only one upper index to contract.\begin{eqnarray}
\left[K^{(k,1)},L^{(l,1)}\right]^{\Delta} & = & \left[K^{(k,1)},L^{(l,1)}\right]_{(1)}^{\Delta}=\ip_{K}^{(1)}L-(-)^{(k-1)(l-1)}\ip_{L}^{(1)}K=\\
 & = & \hspace{-1cm}\stackrel{(\ref{eq:bigbracket-coord})}{=}l\, K_{\bs{m}\ldots\bs{m}}\hoch{j}L_{j\bs{m}\ldots\bs{m}}\hoch{\bs{n}}-(-)^{(k-1)(l-1)}k\, L_{\bs{m}\ldots\bs{m}}\hoch{j}K_{j\bs{m}\ldots\bs{m}}\hoch{\bs{n}}\label{eq:Richardson-Nijenhuis-bracket-coord}\end{eqnarray}
It is thus just the big bracket or Buttin's algebraic bracket but
in this case it is known as \textbf{Richardson-Nijenhuis-bracket}. 

The Lie derivative of a form with respect to $K$ (in the sense of
(\ref{eq:Lie-derivativeI})) is because of $k'=1$ really a (first
order) derivative and takes the form\begin{eqnarray}
\Lie_{K^{(k,1)}} & \equiv & \left[\ip_{K^{(k,1)}},\de\,\right]\\
\Lie_{K^{(k,1)}}\rho^{(r)} & = & K_{\bs{m}\ldots\bs{m}}\hoch{l}\partial_{l}\rho_{\bs{m}\ldots\bs{m}}+(-)^{k}r\partial_{\bs{m}}K_{\bs{m}\ldots\bs{m}}\hoch{l}\rho_{l\bs{m}\ldots\bs{m}}\end{eqnarray}
The \textbf{(Froehlicher-)Nijenhuis} bracket is defined as the unique
tensor $\left[K\bs{,}L\right]_{N}$, s.th.\begin{equation}
[\Lie_{K}\bs{,}\Lie_{L}]=\Lie_{[K\bs{,}L]_{N}}\end{equation}
It is therefore an example of Buttin's differential bracket. Its explicit
coordinate form reads\begin{eqnarray}
\left[K\bs{,}L\right]_{N} & \equiv & K_{\bs{m}\ldots\bs{m}}\hoch{j}\partial_{j}L_{\bs{m}\ldots\bs{m}}\hoch{\bs{n}}+(-)^{k}l\partial_{\bs{m}}K_{\bs{m}\ldots\bs{m}}\hoch{j}L_{j\bs{m}\ldots\bs{m}}\hoch{\bs{n}}+\nonumber \\
 &  & -(-)^{kl}L_{\bs{m}\ldots\bs{m}}\hoch{j}\partial_{j}K_{\bs{m}\ldots\bs{m}}\hoch{\bs{n}}-(-)^{kl}(-)^{l}k\partial_{\bs{m}}L_{\bs{m}\ldots\bs{m}}\hoch{j}K_{j\bs{m}\ldots\bs{m}}\hoch{\bs{n}}\label{eq:Nijenhuis-bracket-coord}\\
 & = & "\Lie_{K}L-(-)^{kl}\Lie_{L}K"\end{eqnarray}

A different point of view on the Nijenhuis bracket is via the \textbf{derived
bracket} on the level of the differential operators acting on forms:
\begin{equation}
\left[\ip_{K},_{\de\,}\ip_{L}\right]\equiv\left[\left[\ip_{K},\de\right],\ip_{L}\right]\end{equation}
It induces the Nijenhuis-bracket only up to a total derivative (the
Lie-derivative-term) \begin{eqnarray}
\left[\ip_{K},_{\de\,}\ip_{L}\right] & \equiv & \ip_{\left[K\bs{,}L\right]_{N}}-(-)^{k(l-1)}\Lie_{\ip_{L}K}\label{eq:zwei-drueber}\end{eqnarray}
Using the extended definition of the exterior derivative in the sense
of (\ref{eq:d-auf-partial}) and of the interior product (\ref{eq:i-TII}),
one can write the Lie derivative as an interior product (see \ref{eq:dK-und-Lie})
$\Lie_{\ip_{L}K}=-(-)^{l+k}\ip_{\de\,(\ip_{L}K)}$ and $\left[\left[\ip_{K},\de\right],\ip_{L}\right]=(-)^{k}\left[\ip_{\de K},\ip_{L}\right]=(-)^{k}\ip_{\left[\de K,L\right]^{\Delta}}$,
so that we can rewrite (\ref{eq:zwei-drueber}) as\begin{eqnarray}
\left[K\bs{,}L\right] & \equiv & \left[K\bs{,}L\right]_{N}+(-)^{(k-1)l}\de\,(\ip_{L}K)\\
\textrm{with }\left[K\bs{,}L\right] & \equiv & (-)^{k}\left[\de K,L\right]^{\Delta}\end{eqnarray}
In that sense, $\left[K\bs{,}L\right]$ is the derived bracket of
the Richardson Nijenhuis bracket while the Nijenhuis bracket differs
by a total derivative. The explicit coordinate form can be read off
from (\ref{eq:bc-derived-bracketIII},\ref{eq:bc-derived-bracket-coord})
(with only the $p=1$ term surviving) \begin{eqnarray}
\left[K\bs{,}L\right] & = & (-)^{k}\ip_{\de K}^{(1)}L+(-)^{kl}(-)^{l}\ip_{\de L}^{(1)}K+(-)^{(k-1)l}\de(\ip_{L}^{(p)}K)=\\
 & = & K_{\bs{m}\ldots\bs{m}}\hoch{j}\partial_{j}L_{\bs{m}\ldots\bs{m}}\hoch{\bs{n}}+(-)^{k}l\partial_{\bs{m}}K_{\bs{m}\ldots\bs{m}}\hoch{j}L_{j\bs{m}\ldots\bs{m}}\hoch{\bs{n}}+\nonumber \\
 &  & -(-)^{kl}L_{\bs{m}\ldots\bs{m}}\hoch{j}\partial_{j}K_{\bs{m}\ldots\bs{m}}\hoch{\bs{n}}-(-)^{kl}(-)^{l}k\partial_{\bs{m}}L_{\bs{m}\ldots\bs{m}}\hoch{j}K_{j\bs{m}\ldots\bs{m}}\hoch{\bs{n}}+\nonumber \\
 &  & +(-)^{(k-1)l}\de\big(\underbrace{kL_{\bs{m}\ldots\bs{m}}\hoch{j}K_{j\bs{m}\ldots\bs{m}}\hoch{\bs{n}}}_{\ip_{L}K}\big)\label{eq:Nij-derived-coord}\end{eqnarray}
\rem{von der anderen Koordinatenform bekommt man direkt ganz genau dasselbe!}
where the last part is non-tensorial due to the appearance of the
basis element $p_{i}$ (see subsection \ref{sub:Extended-exterior-derivative}):\begin{eqnarray}
\de\left(\ip_{L}K\right)=\de\big(kL_{\bs{m}\ldots\bs{m}}\hoch{j}K_{j\bs{m}\ldots\bs{m}}\hoch{\bs{n}}\big) & = & k\partial_{\bs{m}}\left(L_{\bs{m}\ldots\bs{m}}\hoch{j}K_{j\bs{m}\ldots\bs{m}}\hoch{\bs{n}}\right)-(-)^{l+k}L_{\bs{m}\ldots\bs{m}}\hoch{j}K_{j\bs{m}\ldots\bs{m}}\hoch{i}p_{i}\end{eqnarray}
The remaining part coincides with the coordinate form of the \textbf{Nijenhuis
bracket} as given in (\ref{eq:Nijenhuis-bracket-coord}).

One can nicely summarize the algebra of graded derivations on forms
as \begin{eqnarray}
\lqn{\left[\Lie_{K_{1}^{(k_{1})}}+\ip_{L_{1}^{(l_{1})}}\,,\,\Lie_{K_{2}^{(k_{2})}}+\ip_{L_{2}^{(l_{2})}}\right]=}\nonumber \\
 & = & \Lie_{[K_{1}\bs{,}K_{2}]_{N}+\ip_{L_{1}}K_{2}-(-)^{(l_{2}-1)k_{1}}\,\ip_{L_{2}}K_{1}}+\ip_{\left[K_{1}\bs{,}L_{2}\right]_{N}-(-)^{(l_{1}-1)k_{2}}\left[K_{2}\bs{,}L_{1}\right]_{N}+\left[L_{1},L_{2}\right]^{\Delta}}\label{eq:Max-algebra}\end{eqnarray}
\rem{hier schlummern generalized complex examples. hauptsaechlich \begin{eqnarray*}
\lqn{(-)^{\textsc{k}-1}\left[\de\mc{K},\mc{L}\right]_{(p)}^{\Delta}=}\\
 & = & p!\left(\zwek{\textsc{k}}{p}\right)\left(\zwek{\textsc{l}}{p-1}\right)\cdot\mc{K}_{\bs{M}\ldots\bs{M}}\hoch{i_{1}\ldots i_{p-1}i_{p}}\partial_{i_{p}}\mc{L}_{i_{p-1}\ldots i_{1}\bs{M}\ldots\bs{M}}+\\
 &  & -(-)^{(\textsc{k}+1)(\textsc{l}+1)}p!\left(\zwek{\textsc{l}}{p}\right)\left(\zwek{\textsc{k}}{p-1}\right)\mc{L}_{\bs{M}\ldots\bs{M}}\hoch{i_{1}\ldots i_{p}}\partial_{i_{p}}\mc{K}_{i_{p-1}\ldots i_{1}\bs{M}\ldots\bs{M}}+\\
 &  & +(-)^{\textsc{k}-1}p!\left(\zwek{\textsc{k}}{p}\right)\left(\zwek{\textsc{l}}{p}\right)\left(\partial_{\bs{M}}\mc{K}_{\bs{M}\ldots\bs{M}}\hoch{i_{1}\ldots i_{p}}\mc{L}_{i_{p}\ldots i_{1}\bs{M}\ldots\bs{M}}-(-)^{p}\partial_{\bs{M}}\mc{K}_{\bs{M}\ldots\bs{M}i_{p}\ldots i_{1}}\mc{L}^{i_{1}\ldots i_{p}}\tief{\bs{M}\ldots\bs{M}}\right)+\\
 &  & +(p+1)!\left(\zwek{\textsc{k}}{p+1}\right)\left(\zwek{\textsc{l}}{p}\right)\left(\mc{K}_{\bs{M}\ldots\bs{M}}\hoch{i_{1}\ldots i_{p}L}\mc{L}_{i_{p}\ldots i_{1}\bs{M}\ldots\bs{M}}-(-)^{p}\mc{K}_{\bs{M}\ldots\bs{M}i_{p}\ldots i_{1}}\hoch{L}\mc{L}^{i_{1}\ldots i_{p}}\tief{\bs{M}\ldots\bs{M}}\right)p_{L}\end{eqnarray*}
For $p=1$, we get the derived bracket of the big bracket which is
most pleasant, because it can be written in terms of capital indices:\begin{eqnarray*}
(-)^{\textsc{k}-1}\left[\de\mc{K},\mc{L}\right]_{(1)}^{\Delta} & = & \textsc{k}\cdot\mc{K}_{\bs{M}\ldots\bs{M}}\hoch{i_{1}}\partial_{i_{1}}\mc{L}_{\bs{M}\ldots\bs{M}}-(-)^{(\textsc{k}+1)(\textsc{l}+1)}\textsc{l}\cdot\mc{L}_{\bs{M}\ldots\bs{M}}\hoch{i_{1}}\partial_{i_{1}}\mc{K}_{\bs{M}\ldots\bs{M}}+\\
 &  & +(-)^{\textsc{k}-1}\textsc{kl}\partial_{\bs{M}}\mc{K}_{\bs{M}\ldots\bs{M}}\hoch{I}\mc{L}_{I\bs{M}\ldots\bs{M}}+\textsc{k}\left(\textsc{k}-1\right)\textsc{l}\mc{K}_{\bs{M}\ldots\bs{M}}\hoch{IJ}\mc{L}_{I\bs{M}\ldots\bs{M}}p_{J}\end{eqnarray*}
}

}
\Teil{GCG}{

\section{Some aspects of generalized (complex) geometry}

\label{sec:Generalized-complex-geometry}

For introductions into Hitchin's \cite{Hitchin:2004ut} generalized
complex geometry (GCG) see e.g. Zabzine's review \cite{Zabzine:2006uz}
or Gualtieri's thesis \cite{Gualtieri:007}. For a survey of compactification
with fluxes and its relation to GCG see Gra$\tilde{\textrm{n}}$a's
review \cite{Grana:2005jc}.

\subsection{Basics}

In \textbf{generalized geometry} one is looking at structures (e.g.
a complex structure) on the direct sum of tangent and cotangent bundle
$T\oplus T^{*}$. Let us call a section of this bundle a \textbf{generalized
vector} (field) or synonymously \textbf{generalized 1-form}, \textbf{}which
is the sum of a vector field and a 1-form\begin{eqnarray}
\mf{a} & = & a+\alpha=\\
 & = & a^{m}\pe_{m}+\alpha_{m}\de x^{m}\end{eqnarray}
Using the \textbf{combined basis elements} \begin{eqnarray}
\basis_{M} & \equiv & (\pe_{m},\de x^{m})\label{eq:combined-basis}\end{eqnarray}
a generalized vector $\mf{a}$ can be written as \begin{eqnarray}
\mf{a} & = & \mf{a}^{M}\basis_{M}\label{eq:generalized-vector}\\
\mf{a}^{M} & = & (a^{m},\alpha_{m})\end{eqnarray}
There is a \textbf{canonical metric} $\mc{G}$ on $T\oplus T^{*}$\begin{eqnarray}
\erw{\mf{a},\mf{b}} & \equiv & \mf{\alpha}(b)+\mf{\beta}(a)=\\
 & = & \alpha_{m}b^{m}+\beta_{m}a^{m}\equiv\\
 & \equiv & \mf{a}^{M}\mc{G}_{MN}\mf{b}^{N}\end{eqnarray}
with \begin{eqnarray}
\mc{G}_{MN} & \equiv & \left(\begin{array}{cc}
0 & \delta_{m}^{n}\\
\delta_{n}^{m} & 0\end{array}\right)\label{eq:canonical-metric}\end{eqnarray}
which has \textbf{signature} (d,-d) (if d is the dimension of the
base manifold). The above definition differs by a factor of 2 from
the most common one. We prefer, however, to have an inverse metric
of the same form\begin{eqnarray}
\mc{G}^{MN} & \equiv & \left(\mc{G}^{-1}\right)^{MN}=\left(\begin{array}{cc}
0 & \delta_{n}^{m}\\
\delta_{m}^{n} & 0\end{array}\right)\label{eq:inverse-canonical-metric}\end{eqnarray}
As it is constant, we can always pull it through partial derivatives.
Using this metric to lower and raise indices just interchanges vector
and form component. We can equally rewrite $\mf{a}$ in (\ref{eq:generalized-vector})
with a basis with upper capital indices and the vector coefficients
with lower indices\begin{eqnarray}
\basis^{M} & \equiv & \left(\de x^{m},\pe_{m}\right)\label{eq:combined-basisII}\\
\mf{a} & = & \mf{a}_{M}\basis^{M}\\
\mf{a}_{M} & = & (\alpha_{m},a^{m})\end{eqnarray}
Note that in the present paper there is no existence of any metric
on the tangent bundle assumed. Therefore we cannot raise or lower
small indices. In cases where 1-form and vector have a similar symbol,
the position of the small index therefore uniquely determines which
is which (e.g. $\omega_{m}$ and $w^{m}$). 

In addition to the canonical metric $\mc{G}_{MN}$ there is also a
\textbf{canonical antisymmetric 2-form} $\mc{B}$, s.th. $\alpha(b)-\beta(a)=\mf{a}^{M}\mc{B}_{MN}\mf{b}^{N}$
with coordinate form \begin{eqnarray}
\mc{B}_{MN} & \equiv & \left(\begin{array}{cc}
0 & -\delta_{m}^{n}\\
\delta_{n}^{m} & 0\end{array}\right)\label{eq:canonical-B-tensor}\end{eqnarray}
Raising the indices with $\mc{G}^{MN}$ yields \begin{eqnarray}
\mc{B}^{M}\tief{N} & = & \left(\begin{array}{cc}
\delta_{n}^{m} & 0\\
0 & -\delta_{m}^{n}\end{array}\right)=-B_{N}\hoch{M}\\
\mc{B}^{MN} & = & \left(\begin{array}{cc}
0 & \delta_{n}^{m}\\
-\delta_{m}^{n} & 0\end{array}\right)\label{eq:upper-canonical-B-tensor}\end{eqnarray}
We can thus use $\mc{B}$ and $\mc{G}$ to construct \textbf{projection
operators} $\mc{P_{T}}$ and $\mc{P_{T^{\!*}}}$ to tangent and cotangent
space\begin{eqnarray}
\mc{P}_{\mc{T}}\hoch{M}\tief{N} & \equiv & \frac{1}{2}\left(\delta^{M}\tief{N}+B^{M}\tief{N}\right)=\left(\begin{array}{cc}
\delta_{n}^{m} & 0\\
0 & 0\end{array}\right)\\
\mc{P}_{\mc{T}^{*}}\hoch{M}\tief{N} & \equiv & \frac{1}{2}\left(\delta^{M}\tief{N}-B^{M}\tief{N}\right)=\left(\begin{array}{cc}
0 & 0\\
0 & \delta_{m}^{n}\end{array}\right)\\
\mc{P}_{\mc{T}}\mf{a} & = & a,\qquad\mc{P}_{\mc{T}^{*}}\mf{a}=\alpha\end{eqnarray}

\subsection{Generalized almost complex structure}

\label{sub:generalized-complex-structure}A \textbf{generalized almost
complex structure} is a linear map from $T\oplus T^{*}$ to itself
which squares to minus the identity-map, i.e. in components\begin{eqnarray}
\mc{J}^{M}\tief{K}\mc{J}^{K}\tief{N} & = & -\delta_{N}^{M}\label{eq:Jsquare-is-one}\end{eqnarray}
It is called a \textbf{generalized complex structure} if it is integrable
(see subsection \ref{sub:Integrability-of-J}). It should be \textbf{compatible}
with our canonical metric $\mc{G}$ which means that it should behave
like multiplication with $i$ in a Hermitian scalar product of a complex
vector space%
\footnote{\label{scalar-product} In a complex vector space with Hermitian scalar
product $\erw{a,b}=\overline{\erw{b,a}}$ we have $\erw{a,ib}=-\erw{ia,b}$.$\qquad\fussend$
}\begin{eqnarray}
\erw{\mf{v},\mc{J}\mf{w}} & = & -\erw{\mc{J}\mf{v},\mf{w}}\iff(\mc{G}\mc{J})^{T}=-\mc{G}\mc{J}\iff\mc{J}_{MN}=-\mc{J}_{NM}\label{eq:Jis-antisym}\end{eqnarray}
This property is also known as \textbf{antihermiticity} of $\mc{J}$.
Because of (\ref{eq:Jis-antisym}), $\mc{J}$ can be written as \begin{eqnarray}
\mc{J}^{M}\tief{N} & = & \left(\begin{array}{cc}
J^{m}\tief{n} & P^{mn}\\
-Q_{mn} & -J^{n}\tief{m}\end{array}\right)\qquad\mc{J}_{MN}=\left(\begin{array}{cc}
-Q_{mn} & -J^{n}\tief{m}\\
J^{m}\tief{n} & P^{mn}\end{array}\right)\label{eq:J-matrix}\end{eqnarray}
where $P^{mn}$ and $Q_{mn}$ are antisymmetric matrices, and (\ref{eq:Jsquare-is-one})
translates into \begin{eqnarray}
J^{2}-PQ & = & -\one\label{eq:alg-PQJ-cond}\\
JP-PJ^{T} & = & 0\label{eq:alg-JP-cond}\\
-QJ+J^{T}Q & = & 0\label{eq:alg-JQ-cond}\end{eqnarray}
Here it becomes obvious that the generalized complex structure contains
the case of an ordinary almost complex structure $J$ with $J^{2}=-1$
for $Q=P=0$ as well as the case of an almost symplectic structure
of a non-degenerate 2-form $Q$ with existing inverse $PQ=\one$ for
$J=0$. In addition to those algebraic constraints, the integrability
of the generalized almost complex structure gives further differential
conditions (see subsection \ref{sub:Integrability-of-J}) which boil
down in the two special cases to the integrability of the ordinary
complex structure or to the integrability of the symplectic structure.

Because of $\mc{J}^{2}=-\one$, $\mc{J}$ has eigenvalues $\pm i$.
The corresponding eigenvectors span the space of \textbf{generalized
holomorphic vectors} $L$ or generalized antiholomorphic vectors $\bar{L}$
respectively. This provides a natural splitting of the complexified
bundle \begin{equation}
(T\oplus T^{*})\otimes\mathbb{C}=L\oplus\bar{L}\end{equation}
The \textbf{projector} $\Pi$ to the space of eigenvalue $+i$ (namely
$L$) can be be written as \begin{eqnarray}
\Pi & \equiv & \frac{1}{2}\left(\one-i\mc{J}\right)\end{eqnarray}
while the projector to $\bar{L}$ is just the complex conjugate $\bar{\Pi}=\frac{1}{2}\left(\one+i\mc{J}\right)=G^{-1}\Pi^{T}G$.
Indeed, for any generalized vector field $\mf{v}$ we have \begin{eqnarray}
\mc{J}\Pi\mf{v} & = & i\Pi\mf{v}\end{eqnarray}
$L$ and $\bar{L}$ are what one calls \textbf{maximally isotropic
subspaces}, i.e. spaces which are \emph{isotropic} \begin{eqnarray}
\erw{\mf{v},\mf{w}} & = & 0\quad\forall\mf{v},\mf{w}\in L\end{eqnarray}
(this is because $\Pi^{T}G\Pi=\mc{G}\bar{\Pi}\Pi=0$) and which have
half the dimension of the complete bundle. As the canonical metric
$\langle\cdots\rangle$ is nondegenerate, this is the maximal possible
dimension for isotropic subbundles.

\subsection{Dorfman and Courant bracket }

Something which seems to be a bit unnatural in this whole business
in the beginning is the introduction of the Courant bracket, which
is the antisymmetrization of the so-called Dorfman-bracket. The \textbf{Dorfman
bracket} in turn is the natural generalization of the Lie bracket
from the point of view of derived brackets (\ref{eq:derived-bracketI})%
\footnote{\label{twisted-Dorfman} The twisted Dorfman bracket is defined similarly
via \begin{eqnarray*}
\left[\left[\ip_{\mf{a}},\de+H\wedge\,\right],\ip_{\mf{b}}\right] & \equiv & \ip_{\left[\mf{a}\bs{,}\mf{b}\right]_{H}}\end{eqnarray*}
Remembering that $H\wedge=\ip_{H}$ and using $[\ip_{a},\ip_{H}]=\ip_{[a,H]^{\Delta}}=\ip_{\ip_{a}^{(1)}H}$,
we get\begin{eqnarray*}
\left[\mf{a}\bs{,}\mf{b}\right]_{H} & \equiv & \left[a\bs{,}b\right]-\ip_{b}\ip_{a}H\qquad\fussend\end{eqnarray*}
}\rem{Achtung! sind sicher irgendwelche Trivialitaeten in folgender
Fussnote!%
\footnote{If we take Buttin's generalization in the sense that $\left[\Lie_{\mf{a}},\Lie_{\mf{b}}\right]=\Lie_{\left[\mf{a}\bs{,}\mf{b}\right]_{\textrm{B}}}$
, we get the same expression as in (\ref{eq:Dorfman-bracket}), but
without the total derivative at the end. In fact, this bracket obeys
the Jacobi identity and is antisymmetric at the same time which makes
it it quite attractive. It is not known to me whether there are objections
to use Buttin's bracket to define integrability in generalized geometry.
It would certainly lead to a different geometry.$\qquad\fussend$%
}}\begin{eqnarray}
\left[\left[\ip_{\mf{a}},\de\,\right],\ip_{\mf{b}}\right] & = & \ip_{\left[\mf{a}\bs{,}\mf{b}\right]}\\
\textrm{where }\left[\mf{a}\bs{,}\mf{b}\right] & \equiv & \left[a\bs{,}b\right]+\Lie_{a}\beta-\Lie_{b}\alpha+\de\,(\ip_{b}\alpha)=\label{eq:Dorfman-bracket}\\
 & = & \left[a\bs{,}b\right]+\Lie_{a}\beta-\ip_{b}(\de\alpha)=\label{eq:Dorfman-bracketII}\\
 & = & \Lie_{a}\mf{b}-\ip_{b}(\de\alpha)\label{eq:Dorfman-bracketIII}\end{eqnarray}
To get a homogeneous coordinate expression, we define\begin{eqnarray}
\partial_{M} & \equiv & \left(\partial_{m},0\right)\quad\dann\partial^{M}=\left(0,\partial_{m}\right)\end{eqnarray}
\newpage\noindent The Dorfman bracket can then be written as%
\footnote{\label{foot:dual-coord}It is perhaps interesting to note that this
notation of the partial derivative with capital index suggests the
extension to a derivative with respect to some dual coordinate \[
\partial^{m}\equiv\partial_{\tilde{x}_{m}}\]
We could understand this as coordinates of a dual manifold whose tangent
space coincides in some sense with the cotangent space of the original
space and vice versa. This might be connected to Hull's doubled geometry
\cite{Dabholkar:2005ve,Hull:2006qs,Hull:2006va,Hull:2004in}.

To see that such an ad-hoc extension of the Dorfman bracket is not
completely unfounded, note that there is a more general notion of
a Dorfman bracket (or Courant bracket) in the context of Lie-bialgebroids
(for a definition see e.g. \cite[p.32,20]{Gualtieri:007}). There
we have two Lie algebroids $L$ and $L^{*}$ which are dual with respect
to some inner product and which both carry some Lie bracket. (For
$T$ and $T^{*}$, only $T$ carries a Lie bracket in the beginning.
For a non-trivial Lie bracket of forms on $T^{*}$ we need some extra
structure like e.g. a Poisson structure which would lead to the Koszul
bracket on forms.) The Lie bracket on $L$ induces a differential
$\de$ on $L^{*}$ and the Lie bracket on $L^{*}$ induces a differential
$\de^{*}$ on $L$. The definition for the Dorfman bracket on the
Lie bialgebroid $L\oplus L^{*}$ is then \begin{eqnarray*}
\left[\mf{a}\bs{,}\mf{b}\right] & \equiv & \left[a\bs{,}b\right]+\Lie_{a}\beta-\Lie_{b}\alpha+\de\,(\ip_{b}\alpha)+\\
 &  & +\left[\alpha\bs{,}\beta\right]+\Lie_{\alpha}b-\Lie_{\beta}a+\de^{*}(\ip_{\beta}a)\end{eqnarray*}
The first line is the part we are used to from our usual Dorfman bracket
on $T\oplus T^{*}$, while second line is the corresponding part coming
from the nontrivial structure on $L^{*}$. Taking now $L=T$, $L^{*}=T^{*}$
and assuming that $[\alpha\bs{,}\beta]$ and $\Lie_{\alpha}$ and
$\de^{*}$ are a Lie bracket, Lie derivative and exterior derivative
built in the ordinary way, but with the new partial derivative w.r.t.
the dual coordinates $\partial^{m}$, the coordinate form of the Dorfman
bracket remains exactly the one of (\ref{eq:Dorfman-bracket-coord},\ref{eq:Dorfman-bracket-coordII}),
but with $\partial_{M}=(\partial_{m},0)$ replaced by $\partial_{M}=(\partial_{m},\partial^{m}).\qquad\fussend$\rem{relating $\pe^m$ and $\de x^m$: integrable structures: like GCS}%
}\begin{eqnarray}
\left[\mf{a}\bs{,}\mf{b}\right]^{M} & = & \mf{a}^{K}\partial_{K}\mf{b}^{M}+\left(\partial^{M}\mf{a}_{K}-\partial_{K}\mf{a}^{M}\right)\mf{b}^{K}\label{eq:Dorfman-bracket-coord}\\
\textrm{or }\left[\mf{a}\bs{,}\mf{b}\right]_{M} & = & \mf{a}^{K}\partial_{K}\mf{b}_{M}+2\partial_{[M}\mf{a}_{K]}\mf{b}^{K}\label{eq:Dorfman-bracket-coordII}\end{eqnarray}
Apart from the term in the middle $\partial^{M}\mf{a}_{K}$, (\ref{eq:Dorfman-bracket-coord})
looks formally the same as the Lie bracket of vector fields (\ref{eq:vector-Lie-bracket}).
The Dorfman bracket is in general not antisymmetric but it obeys a
\textbf{Jacobi-identity} (Leibniz from the left) of the form\begin{eqnarray}
\left[\mf{a}\bs{,}\left[\mf{b}\bs{,}\mf{c}\right]\right] & = & \left[\left[\mf{a}\bs{,}\mf{b}\right]\bs{,}\mf{c}\right]+\left[\mf{b}\bs{,}\left[\mf{a}\bs{,}\mf{c}\right]\right]\label{eq:Jacobi-for-Dorfman}\end{eqnarray}
Although the Dorfman bracket is all we need, most of the literature
on generalized complex geometry so far works with its antisymmetrization,
which is called \textbf{Courant bracket} \begin{eqnarray}
\left[\mf{a}\bs{,}\mf{b}\right]_{-} & \equiv & \left[a\bs{,}b\right]+\Lie_{a}\beta-\Lie_{b}\alpha+\frac{1}{2}\de\,(\ip_{b}\alpha-\ip_{a}\beta)\label{eq:Courant-bracket}\\
\left[\mf{a}\bs{,}\mf{b}\right]_{-M} & = & \mf{a}^{K}\partial_{K}\mf{b}_{M}-\partial_{K}\mf{a}_{M}\mf{b}^{K}+\frac{1}{2}\left(\partial_{M}\mf{a}_{K}\mf{b}^{K}-\mf{a}^{K}\partial_{M}\mf{b}_{K}\right)\label{eq:Courant-bracket-coord}\end{eqnarray}
and which does not obey any Jacobi identity. As it is much simpler
to go from Dorfman to Courant, than the other way round, we will only
work with the Dorfman bracket. On any isotropic subspace ($\ip_{b}\alpha+\ip_{a}\beta=0$)
the two coincide anyway, i.e. they become a Lie bracket, obeying Jacobi
and being antisymmetric. 

We call a transformation a \textbf{symmetry of the bracket} when the
bracket of two vectors transforms in the same way as the vectors\begin{eqnarray}
\left[(\mf{b}+\delta\mf{b})\bs{,}(\mf{c}+\delta\mf{c})\right] & = & \left[\mf{b}\bs{,}\mf{c}\right]+\delta\left[\mf{b}\bs{,}\mf{c}\right]\\
\delta\left[\mf{b}\bs{,}\mf{c}\right] & = & \left[\delta\mf{b}\bs{,}\mf{c}\right]+\left[\mf{b}\bs{,}\delta\mf{c}\right]+\left[\delta\mf{b}\bs{,}\delta\mf{c}\right]\end{eqnarray}
I.e. infinitesimal symmetry transformations (where the last term drops)
have to obey a product rule. Similar as for the Lie-bracket of vector
fields, infinitesimal transformations are generated by the bracket
itself. Let us call the corresponding derivative, in analogy to the
Lie derivative, the \textbf{Dorfman derivative} of a generalized vector
with respect to a generalized vector. \begin{equation}
\delta\mf{b}=\Dorf_{\mf{a}}\mf{b}\equiv\left[\mf{a},\mf{b}\right]\end{equation}
 These transformations are therefore, due to the Jacobi-identity (\ref{eq:Jacobi-for-Dorfman})
always symmetries of the bracket. From (\ref{eq:Dorfman-bracketIII})
we can see that the Dorfman derivative consists of a usual Lie derivative
and second part which acts only on the vector part of $\mf{b}$ by
contracting it with the exact 2-form $\de\alpha$\begin{eqnarray}
\Dorf_{a}\mf{b} & = & \Lie_{a}\mf{b}\label{eq:Dorfman-transformationI}\\
\Dorf_{\alpha}b & = & -\ip_{b}(\de\alpha)=b^{m}(\partial_{n}\alpha_{m}-\partial_{m}\alpha_{n})\de x^{n}\label{eq:Dorfman-transformationII}\end{eqnarray}
In fact, it is enough for the 2-form to be closed, in order to get
a symmetry. If we replace $-\de\alpha$ by a \emph{closed 2-form}
$B$, the transformation is known as $B$\textbf{-transform\begin{eqnarray}
\delta_{B}b & = & \ip_{b}B\end{eqnarray}
}

Finally, we should note that the $B$-transform is part of the $O(d,d)$-transformations,
i.e. the transformations which leave the canonical metric invariant.
As usual for orthogonal groups the infinitesimal generators are antisymmetric
when the second index is pulled down with the corresponding metric.
The generators of an $O(d,d)$-transformation can therefore be written
as \cite[p.6]{Gualtieri:007}\begin{eqnarray}
\Omega_{MN} & = & \left(\begin{array}{cc}
B_{mn} & -A_{m}\hoch{n}\\
A_{n}\hoch{m} & \textrm{\Beta}^{mn}\end{array}\right)\\
\Omega^{M}\tief{N} & = & \left(\begin{array}{cc}
A_{n}\hoch{m} & \Beta^{mn}\\
B_{mn} & -A_{m}\hoch{n}\end{array}\right)\label{eq:SOnn}\end{eqnarray}
In addition to the $B$-transform, acting with $\Omega$ on a generalized
vector induces the so-called \textbf{beta-transform} on the 1-form
component%
\footnote{The letter $\beta$ for the beta-transformations does not really fit
into the philosophy of the present notations, where we use small Greek
letters for 1-forms (or sometimes p-forms) only, but not for multivectors.
As the transformation is, however, commonly known as beta-transformation,
we use a large $\sBeta$, in order to distinguish it from the one-forms
$\beta$, which are floating around.$\quad\fussend$%
} as well as $Gl(d)$-transformations of vector and 1-form component
via $A$. For constant tensors, the Lie-derivative is just a $Gl(d)$
transformation. Therefore both symmetries of the Dorfman bracket are
symmetries of the canonical metric $\mc{G}$ as well. For this reason
the canonical metric is invariant under the \textbf{Dorfman derivative}
$\Dorf_{\mf{v}}$with respect to a generalized vector $\mf{v}$, which
we define on generalized rank $p$ tensors using (\ref{eq:Dorfman-bracket-coord})
in a way that it acts via Leibniz on tensor products (like the Lie
derivative) and as a directional derivative on scalars\begin{eqnarray}
(\Dorf_{\mf{v}}\mc{T})^{M_{1}\ldots M_{p}} & \equiv & \mf{v}^{K}\partial_{K}\mc{T}^{M_{1}\ldots M_{p}}+\sum_{i}(\partial^{M_{i}}\mf{v}_{K}-\partial_{K}\mf{v}^{M_{i}})T^{M_{1}\ldots M_{i-1}KM_{i+1}\ldots M_{p}}\label{eq:Dorfman-derivative}\\
\Dorf_{\mf{v}}(\mc{A}\otimes\mc{B}) & = & \Dorf_{v}\mc{A}\otimes\mc{B}+\mc{A}\otimes\Dorf_{v}\mc{B}\\
\Dorf_{\mf{v}}(\phi) & = & \mf{v}^{K}\partial_{K}\phi=v^{k}\partial_{k}\phi\end{eqnarray}
Acting on the canonical metric, one recovers the fact, that the Dorfman
derivative contains the isometries of the metric \begin{eqnarray}
\Dorf_{\mf{v}}\mc{G} & = & 2(\partial^{M_{1}}\mf{v}_{K}-\partial_{K}\mf{v}^{M_{1}})\mc{G}^{KM_{2}}=0\end{eqnarray}
Comparing the role of Lie-derivative and Dorfman-derivative, the $B$-transform
should be understood as an extension of diffeomorphisms. In string
theory it shows up in the Buscher-rules for T-duality (\cite{Buscher:1987sk,Buscher:1987qj})
\rem{ist das wahr?} and can perhaps be better understood geometrically
via Hull's \rem{and Dhabolka's?}  doubled geometry \cite{Dabholkar:2005ve,Hull:2006qs,Hull:2006va}
(compare to footnote \ref{foot:dual-coord}). The beta-transform is
not a symmetry of the Dorfman bracket as it stands. However, if we
introduce dual coordinates as suggested in footnote \ref{foot:dual-coord},
the beta-transform would show up in the symmetry-transformations of
the extended Dorfman bracket generated by itself.%
\footnote{Taking the Dorfman bracket of footnote \ref{foot:dual-coord}, we
get as Dorfman derivative of a generalized vector $\mf{c}$ instead
of (\ref{eq:Dorfman-transformationI},\ref{eq:Dorfman-transformationII})
the extended transformation \begin{eqnarray*}
\Dorf_{a}\mf{c} & \equiv & \Lie_{a}\mf{c}-\ip_{\gamma}(\de^{*}a)\\
\Dorf_{\alpha}\mf{c} & \equiv & -(\ip_{c}\de\alpha)+\Lie_{\alpha}\mf{c}\end{eqnarray*}
I.e. the first line is extended by a beta-transformation of $\gamma$
with $\sBeta=-\de^{*}a$ and the $B$-transform of $\alpha$ ($B=-\de\alpha$)
in the second line is extended by a Lie derivative with respect to
$\alpha$.$\qquad\fussend$%
}

On an isotropic subspace $L$ (e.g. the generalized holomorphic subspace)
Courant- and Dorfman-bracket coincide and have the properties of a
Lie bracket. It is therefore possible to define a Schouten bracket
on generalized multivectors on $\bigwedge^{\bullet}L$ which have
e.g. only generalized holomorphic indices (compare \cite[p.21]{Gualtieri:007}).
If we use again the notation with repeated boldface indices \begin{equation}
\mc{A}^{(p)}\equiv\mc{A}_{\bs{M}\ldots\bs{M}}\equiv\mc{A}_{M_{1}\ldots M_{p}}\basis^{M_{1}}\cdots\basis^{M_{2}}\end{equation}
we get as coordinate form for this \textbf{Dorfman-Schouten} \textbf{bracket} \rem{Schouten-bracket on $\Lambda^\bullet L$}\begin{eqnarray}
\left[\mc{A}^{(p)}\bs{,}\mc{B}^{(q)}\right] & = & p\mc{A}^{\bs{M}\ldots\bs{M}K}\partial_{K}\mc{B}^{\bs{M}\ldots\bs{M}}+q\left(p\partial^{\bs{M}}\mc{A}_{K}\hoch{\bs{M}\ldots\bs{M}}-\partial_{K}\mc{A}^{\bs{M}\ldots\bs{M}}\right)\mc{B}^{K\bs{M}\ldots\bs{M}}\label{eq:Dorfman-Schouten-bracket}\end{eqnarray}
In the first term in the bracket on the righthand side, the $\partial^{\bs{M}}$
can as well be shifted with a minus sign to $\mc{B}$, because in
$\bigwedge^{\bullet}L$ we have only isotropic indices in the sense
that\begin{eqnarray}
\mc{A}^{\bs{M}\ldots\bs{M}}\tief{K}\mc{B}^{K\bs{M}\ldots\bs{M}} & = & 0\label{eq:isotropic-multivectors}\end{eqnarray}
For this reason, the Dorfman-Schouten bracket has really the required
skew-symmetry of a Schouten-bracket\begin{eqnarray}
\left[\mc{A}^{(p)}\bs{,}\mc{B}^{(q)}\right] & = & -(-)^{(q+1)(p+1)}\left[\mc{B}^{(q)}\bs{,}\mc{A}^{(p)}\right]\end{eqnarray}
On $\bigwedge^{\bullet}L$ this bracket coincides with the derived
bracket of the big bracket, as the extra term with $p_{M}$ in (\ref{eq:derived-of-big-generalized})
vanishes because of (\ref{eq:isotropic-multivectors}).

\subsection{Integrability}

\label{sub:Integrability-of-J} Integrability for an ordinary complex
structure means that there exist in any chart $\dim_{M}/2$ holomorphic
vector fields (with respect to the almost complex structure) which
can be integrated to holomorphic coordinates $z^{a}$ in this chart
of the manifold and make it a complex manifold. Those vector fields
are then just $\partial/\partial z^{a}$. Those coordinate differentials
have vanishing Lie bracket among each other (partial derivatives commute).
In turn, every set of vectors with vanishing Lie bracket can be integrated
to coordinates. The existence of such a set of integrable holomorphic
vector fields is guaranteed when the holomorphic subbundle is closed
under the Lie bracket, i.e. the Lie bracket of two holomorphic vector
fields is again a holomorphic vector field. \rem{Ist das wahr?} 

As the Dorfman bracket restricted to the generalized holomorphic subbundle
$L\subset(T\oplus T^{*})\otimes\mathbb{C}$ has the properties of
a Lie bracket, we can demand exactly the same for generalized holomorphic
vectors as above for holomorphic ones. \rem{The interpretation will
then be that there is some complex manifold \rem{die gleiche, oder eine andere??? Hull's geometry, oder doch ganz normale?} 
whose complex coordinates define the generalized holomorphic vector
fields. Anyway,} The condition for the generalized complex structure
to be integrable is thus that the generalized holomorphic subbundle
$L$ is closed under the Dorfman bracket, i.e. in terms of the projectors\begin{eqnarray}
\bar{\Pi}\left[\Pi\mf{v}\bs{,}\Pi\mf{w}\right] & = & 0\label{eq:generalized-integrability}\\
\iff\left[\mf{v}\bs{,}\mf{w}\right]-\left[\mc{J}\mf{v}\bs{,}\mc{J}\mf{w}\right]+\mc{J}\left[\mc{J}\mf{v}\bs{,}\mf{w}\right]+\mc{J}\left[\mf{v}\bs{,}\mc{J}\mf{w}\right] & = & 0\label{eq:generalized-integrabilityII}\end{eqnarray}
In the following two sub-subsections we will show that this is equivalent
to the vanishing of a \textbf{generalized Nijenhuis-tensor} \cite[p.25]{Gualtieri:007}
of the coordinate form%
\footnote{\label{foot:holomorphic-indices}This looks formally like the generalized
Schouten bracket (e.g. \cite[p.21]{Gualtieri:007}) on $\bigwedge^{\bullet}L$
(with L being the generalized holomorphic bundle)\rem{ (\ref{eq:Dorfman-Schouten-bracket})}
of $\mc{J}$ with itself (see also the statement below (\ref{eq:derived-of-big-generalized})),
but it is not, as $\mc{J}$ has neither holomorphic nor antiholomorphic
indices\begin{eqnarray*}
\Pi\mc{J} & = & i\Pi\neq\mc{J}\\
\bar{\Pi}\mc{J} & = & -i\Pi\neq\mc{J}\end{eqnarray*}
In fact, we get zero if we contract both indices with the holomorphic
projector \rem{holomorphic upper indices are antiholomorphic lower indices!}\begin{eqnarray*}
\Pi^{N}\tief{L}\Pi^{M}\tief{K}\mc{J}^{KL} & = & \Pi\mc{J}\Pi^{T}=i\Pi\bar{\Pi}=0\end{eqnarray*}
The same happens for two antiholomorphic projectors. But we can project
one index with an holomorphic projector and the other one with an
antiholomorphic one. This yields\begin{eqnarray*}
\bar{\Pi}^{N}\tief{L}\Pi^{M}\tief{K}\mc{J}^{KL} & = & \Pi\mc{J}\Pi=i\Pi\end{eqnarray*}
Up to a constant prefactor the bracket of $\Pi$ with $\Pi$ coincides
with the bracket of $\mc{J}$ with $\mc{J}$. And like for the ordinary
complex structure, where we have the Nijenhuis bracket of the complex
structure with itself, which has one index in $T$ and the second
in $T^{*}$, we could here take $\Pi$ with one index in $L$ and
the other in $\bar{L}$ and regard the bracket as generalized Nijenhuis
bracket of $\Pi$ with itself.\rem{besser rausnehmen?}$\quad\fussend$%
}${}^{,}$%
\footnote{If instead the twisted Dorfman bracket (see footnote \ref{twisted-Dorfman})
is used, one gets the integrability condition for a twisted generalized
complex structure with a twisted generalized Nijenhuis tensor. Consider
the closed three form $H=H_{M_{1}M_{2}M_{3}}\basis^{M_{1}}\basis^{M_{2}}\basis^{M_{3}}$
with $H_{m_{1}m_{2}m_{3}}$ the only nonvanishing components. The
twisted generalized Nijenhuis tensor then reads\rem{siehe Notizblock 2006, S.73 und S.60}
\begin{eqnarray*}
\mc{N}_{M_{1}M_{2}M_{3}}^{H} & = & \mc{N}_{M_{1}M_{2}M_{3}}+6H_{M_{1}M_{2}M_{3}}-18\mc{J}_{M_{1}}\hoch{K}H_{KM_{2}L}\mc{J}^{L}\tief{M_{3}}\end{eqnarray*}
Like (\ref{eq:integrability-tensor-I})-(\ref{eq:integrability-tensor-IV})
this twisted generalized Nijenhuis tensor as well matches with the
tensors given in \cite{Zucchini:2004ta} if one redefines $H_{mnk}\to\frac{1}{3!}H_{mnk}$.$\qquad\fussend$%
} \rem{Referenz zu Gualtieri! Nur Koordinatenform ist neu....}\begin{equation}
\boxed{\frac{1}{4}\mc{N}^{M_{1}M_{2}M_{3}}\equiv\mc{J}^{[M_{1}|K}\partial_{K}\mc{J}^{|M_{2}M_{3}]}+\mc{J}^{[M_{1}|K}\mc{J}_{K}\hoch{|M_{2},M_{3}]}\stackrel{!}{=}0}\label{eq:generalized-integrabilityIII}\end{equation}
Recalling that \begin{eqnarray}
\mc{J}^{MN} & = & \left(\begin{array}{cc}
P^{mn} & J^{m}\tief{n}\\
-J^{n}\tief{m} & -Q_{mn}\end{array}\right),\qquad\mc{J}_{M}\hoch{N}=\left(\begin{array}{cc}
-J^{n}\tief{m} & -Q_{mn}\\
P^{mn} & J^{m}\tief{n}\end{array}\right),\qquad\partial^{M}=(0,\partial_{m})\label{eq:J-tensor-II}\end{eqnarray}
we can rewrite this condition in ordinary tensor components, just
to compare it with the conditions given in literature (for the antisymmetrization
of the capital indices we take into account that in the last term
of (\ref{eq:generalized-integrabilityIII}) the indices $M_{1}$ and
$M_{2}$ are automatically antisymmetrized because of $\mc{J}^{2}=-1$):\begin{eqnarray}
\frac{1}{4}\mc{N}^{m_{1}m_{2}m_{3}} & = & P^{[m_{1}|k}\partial_{k}P^{|m_{2}m_{3}]}\stackrel{!}{=}0\label{eq:integrability-tensor-I}\\
\frac{1}{4}\mc{N}_{n}\hoch{m_{1}m_{2}} & = & \frac{1}{3}\left(-J^{k}\tief{n}\partial_{k}P^{[m_{1}m_{2}]}+2P^{[m_{1}|k}\partial_{k}J^{|m_{2}]}\tief{n}-P^{[m_{1}|k}J^{|m_{2}]}\tief{k,n}+J^{[m_{1}|}\tief{k}P^{k|m_{2}]}\tief{,n}\right)\stackrel{!}{=}0\qquad\quad\\
\frac{1}{4}\mc{N}^{n}\tief{m_{1}m_{2}} & = & \frac{1}{3}\left(-P^{nk}\partial_{k}Q_{[m_{1}m_{2}]}+2J^{k}\tief{[m_{1}|}\partial_{k}J^{n}\tief{|m_{2}]}+2J^{n}\tief{k}J^{k}\tief{[m_{1},m_{2}]}-2P^{nk}Q_{k[m_{1},m_{2}]}\right)\stackrel{!}{=}0\\
\frac{1}{4}\mc{N}_{m_{1}m_{2}m_{3}} & = & J^{k}\tief{[m_{1}|}\partial_{k}Q_{|m_{2}m_{3}]}+J^{k}\tief{[m_{1}|}Q_{k|m_{2},m_{3}]}-Q_{[m_{1}|k}J^{k}\tief{|m_{2},m_{3}]}\stackrel{!}{=}0\label{eq:integrability-tensor-IV}\end{eqnarray}
If we compare those expressions with the tensors $A,B,C$ and $D$
given in (2.16) of \cite[p.7]{Zucchini:2004ta}, we recognize (replacing
$Q$ by $-Q$) that our first line is just $\frac{1}{3}A$, the second
line is $-\frac{1}{3}B$ (using (\ref{eq:alg-JP-cond})), the third
$\frac{1}{3}C$ and the fourth line is $-\frac{1}{3}D$. There, in
turn, it is claimed that the expressions are equivalent to those originally
given in (3.16)-(3.19) of \cite[p.7]{Lindstrom:2004iw}.

\subsubsection{Coordinate based way to derive the generalized Nijenhuis-tensor}

\label{sub:Coordinate-based-way}\rem{dasselbe fuer twisted case machen! nur 3-Form, oder auch hoeher?}
In this sub-subsection we will see that calculations with capital-index
notation is rather convenient. So we simply calculate (\ref{eq:generalized-integrabilityII})
brute force by using the explicit coordinate formula for the Dorfman-bracket\bref{eq:Dorfman-bracket-coord}
\begin{equation}
\left[\mf{v}\bs{,}\mf{w}\right]^{M}=\mf{v}^{K}\partial_{K}\mf{w}^{M}+\left(\partial^{M}\mf{v}_{K}-\partial_{K}\mf{v}^{M}\right)\mf{w}^{K}\end{equation}
\eref  The brackets of interest are: \begin{eqnarray}
\left[\mf{v}\bs{,}\mc{J}\mf{w}\right]^{N} & = & \mf{v}^{K}\partial_{K}\mc{J}^{N}\tief{L}\mf{w}^{L}+\mc{J}^{N}\tief{L}\mf{v}^{K}\partial_{K}\mf{w}^{L}+\left(\partial^{N}\mf{v}_{K}-\partial_{K}\mf{v}^{N}\right)(\mc{J}\mf{w})^{K}\\
(\mc{J}\left[\mf{v}\bs{,}\mc{J}\mf{w}\right])^{M} & = & \underline{\mf{v}^{K}\mc{J}^{M}\tief{N}\partial_{K}\mc{J}^{N}\tief{L}\mf{w}^{L}}-\mf{v}^{K}\partial_{K}\mf{w}^{M}+\mc{J}^{M}\tief{N}\left(\partial^{N}\mf{v}_{K}-\partial_{K}\mf{v}^{N}\right)(\mc{J}\mf{w})^{K}\\
\left[\mc{J}\mf{v}\bs{,}\mf{w}\right]^{N} & = & \mc{J}^{K}\tief{L}\mf{v}^{L}\partial_{K}\mf{w}^{N}+\left(\partial^{N}\mc{J}_{KL}-\partial_{K}\mc{J}^{N}\tief{L}\right)\mf{v}^{L}\mf{w}^{K}+\left(\mc{J}_{K}\hoch{L}\partial^{N}\mf{v}_{L}-\mc{J}^{N}\tief{L}\partial_{K}\mf{v}^{L}\right)\mf{w}^{K}\\
(\mc{J}\left[\mc{J}\mf{v}\bs{,}\mf{w}\right])^{M} & = & \mc{J}^{M}\tief{N}(\mc{J}\mf{v})^{K}\partial_{K}\mf{w}^{N}+\underline{\mc{J}^{M}\tief{N}\left(\partial^{N}\mc{J}_{KL}-\partial_{K}\mc{J}^{N}\tief{L}\right)\mf{v}^{L}\mf{w}^{K}}+\nonumber \\
 &  & -(\mc{J}\mf{w})^{L}\mc{J}^{M}\tief{N}\partial^{N}\mf{v}_{L}+\partial_{K}\mf{v}^{M}\mf{w}^{K}\\
\left[\mc{J}\mf{v}\bs{,}\mc{J}\mf{w}\right]^{M} & = & \mc{J}^{K}\tief{N}\mf{v}^{N}\partial_{K}\mc{J}^{M}\tief{L}\mf{w}^{L}+\mc{J}^{K}\tief{N}\mf{v}^{N}\mc{J}^{M}\tief{L}\partial_{K}\mf{w}^{L}+\nonumber \\
 &  & \left(\partial^{M}\mc{J}_{KN}\mf{v}^{N}-\partial_{K}\mc{J}^{M}\tief{N}\mf{v}^{N}\right)\mc{J}^{K}\tief{L}\mf{w}^{L}+\left(\mc{J}_{KN}\partial^{M}\mf{v}^{N}-\mc{J}^{M}\tief{N}\partial_{K}\mf{v}^{N}\right)\mc{J}^{K}\tief{L}\mf{w}^{L}=\\
 & = & (\mc{J}\mf{v})^{K}\mc{J}^{M}\tief{L}\partial_{K}\mf{w}^{L}-\mc{J}^{M}\tief{N}\partial_{K}\mf{v}^{N}(\mc{J}\mf{w})^{K}+\nonumber \\
 &  & +\underline{\left(\mc{J}^{K}\tief{L}\partial^{M}\mc{J}_{KN}+2\mc{J}^{K}\tief{[N|}\partial_{K}\mc{J}^{M}\tief{|L]}\right)\mf{v}^{N}\mf{w}^{L}}+\partial^{M}\mf{v}_{L}\mf{w}^{L}\end{eqnarray}
The underlined terms sum up in the complete expression to the generalized
Nijenhuis tensor, while the rest cancels\begin{eqnarray}
0 & \stackrel{!}{=} & \left[\mf{v}\bs{,}\mf{w}\right]^{M}-\left[\mc{J}\mf{v}\bs{,}\mc{J}\mf{w}\right]^{M}+(\mc{J}\left[\mc{J}\mf{v}\bs{,}\mf{w}\right])^{M}+(\mc{J}\left[\mf{v}\bs{,}\mc{J}\mf{w}\right])^{M}=\\
 & = & \left(2\mc{J}^{M}\tief{K}\partial_{[N}\mc{J}^{K}\tief{L]}-\mc{J}^{K}\tief{L}\partial^{M}\mc{J}_{KN}+\mc{J}^{MK}\partial_{K}\mc{J}_{LN}-2\mc{J}^{K}\tief{[N|}\partial_{K}\mc{J}^{M}\tief{|L]}\right)\mf{v}^{N}\mf{w}^{L}=\\
 & = & \mf{v}_{N}\left(3\mc{J}^{[M|}\tief{K}\mc{J}^{K|L,N]}+3\mc{J}^{[N|K}\partial_{K}\mc{J}^{|ML]}\right)\mf{w}_{L}=\\
 & = & \frac{3}{4}\mf{v}_{N}\mc{N}^{NML}\mf{w}_{L}\end{eqnarray}

\subsubsection{Derivation via derived brackets}

\label{sub:Derivation-via-derived-bracket}Eventually we want to see
directly how the generalized Nijenhuis tensor is connected to derived
brackets. We will use our insight from the subsections \ref{sub:Algebraic-brackets}
and \ref{sub:Extended-exterior-derivative}. Remember, our basis $\basis^{M}=(\de x^{m},\pe_{m})$
was identified with the conjugate (ghost-)variables $\basis^{M}\equiv(\ce^{m},\be_{m})$.
One can define generalized multi-vector fields of the form\begin{eqnarray}
\mc{K}^{(\textsc{k})} & \equiv & \mc{K}_{\bs{M}\ldots\bs{M}}\equiv\mc{K}_{M_{1}\ldots M_{\textsc{k}}}\basis^{M_{1}}\cdots\basis^{M_{\textsc{k}}}\end{eqnarray}
They are in fact just sums of multivector valued forms:\begin{equation}
\mc{K}_{\bs{M}\ldots\bs{M}}=\sum_{k=0}^{\textsc{k}}\left(\zwek{\textsc{k}}{k}\right)\mc{K}_{\underbrace{{\scriptstyle \mm}}_{k}}\underbrace{\hoch{\nn}}_{\textsc{k}-k}\equiv\sum_{k=0}^{\textsc{k}}K^{(k,\textsc{k}-k)}\end{equation}
The big bracket, or Buttin's algebraic bracket is then just the canonical
Poisson bracket \begin{eqnarray}
\left[\mc{K},\mc{L}\right]_{(1)}^{\Delta} & \equiv & \textsc{kl}\mc{K}_{\bs{M}\ldots\bs{M}}\hoch{I}\mc{L}_{I\bs{M}\ldots\bs{M}}=\left\{ \mc{K},\mc{L}\right\} \label{eq:multvec-bigbrack}\\
\left\{ \basis_{M},\basis_{N}\right\}  & = & \mc{G}_{MN}\label{eq:canon-Poisson}\end{eqnarray}
The coordinate expression for its derived bracket (compare to (\ref{eq:bc-derived-of-bigbracket},\ref{eq:bc-derived-of-bigbracket-coord}))
reads \begin{eqnarray}
(-)^{\textsc{k}-1}\left[\de\mc{K}^{(\textsc{k})},\mc{L}^{(\textsc{L})}\right]_{(1)}^{\Delta} & = & \textsc{k}\cdot\mc{K}_{\bs{M}\ldots\bs{M}}\hoch{I}\partial_{I}\mc{L}_{\bs{M}\ldots\bs{M}}-(-)^{(\textsc{k}+1)(\textsc{l}+1)}\textsc{l}\cdot\mc{L}_{\bs{M}\ldots\bs{M}}\hoch{I}\partial_{I}\mc{K}_{\bs{M}\ldots\bs{M}}+\nonumber \\
 &  & +(-)^{\textsc{k}-1}\textsc{kl}\partial_{\bs{M}}\mc{K}_{\bs{M}\ldots\bs{M}}\hoch{I}\mc{L}_{I\bs{M}\ldots\bs{M}}+\textsc{k}\left(\textsc{k}-1\right)\textsc{l}\mc{K}_{\bs{M}\ldots\bs{M}}\hoch{IJ}\mc{L}_{I\bs{M}\ldots\bs{M}}p_{J}\label{eq:derived-of-big-generalized}\end{eqnarray}
with $p_{J}\equiv(p_{j},0)$ and $\partial_{I}\equiv(\partial_{i},0)$.
In the case were both $\mc{K}$ and $\mc{L}$ only have generalized
holomorphic indices, the $p$-term drops and this expression should
coincide with the Schouten-bracket on $\bigwedge^{\bullet}L$ for
the holomorphic Lie-algebroid $L$ (see e.g. \cite[p.21]{Gualtieri:007}
and footnote \ref{foot:holomorphic-indices}). For two rank-two objects,
like the generalized complex structure $\mc{J}$, this reduces to
\begin{eqnarray}
\left[\mc{K},_{\de\,}\mc{L}\right]_{(1)}^{\Delta} & = & 2\cdot\mc{K}_{\bs{M}}\hoch{I}\partial_{I}\mc{L}_{\bs{M}\bs{M}}+2\cdot\mc{L}_{\bs{M}}\hoch{I}\partial_{I}\mc{K}_{\bs{M}\bs{M}}-4\partial_{\bs{M}}\mc{K}_{\bs{M}}\hoch{I}\mc{L}_{I\bs{M}}+4\mc{K}^{IJ}\mc{L}_{I\bs{M}}p_{J}\end{eqnarray}
which reads for two coinciding tensors $\mc{J}$\begin{eqnarray}
\left[\mc{J},_{\de\,}\mc{J}\right]_{(1)}^{\Delta} & = & 4\cdot\mc{J}_{\bs{M}}\hoch{I}\partial_{I}\mc{J}_{\bs{M}\bs{M}}-4\partial_{\bs{M}}\mc{J}_{\bs{M}}\hoch{I}\mc{J}_{I\bs{M}}-4\mc{J}^{JI}\mc{J}_{I\bs{M}}p_{J}=\label{eq:derived-bracket-for-J}\\
 & \us{\stackrel{(\ref{eq:generalized-integrabilityIII})}{=}}{\mc{J}^{2}=-1} & \mc{N}_{\bs{M}\ldots\bs{M}}+4\underbrace{p_{M}\basis^{M}}_{\lqn{=\oo\textrm{ (\ref{eq:BRST-op})}}}\label{eq:relation-between-Nij-und-der-big}\end{eqnarray}
 where $\oo=\de x^{k}p_{k}=-\de(\de x^{k}\wedge\pe_{k})$. We will
verify this relation between the generalized Nijenhuis tensor and
the derived bracket in the following calculation, where we calculate
$\mc{N}$ using the big bracket (\ref{eq:multvec-bigbrack}) all the
time. This bracket is like a matrix multiplication if one of the objects
has only one index. We will use this fact frequently for the multiplication
of $\mc{J}$ with a vector\begin{eqnarray}
\mc{J}\mf{v} & \equiv & \mc{J}^{M}\tief{N}\mf{v}^{N}\basis_{M}=\frac{1}{2}\left\{ \mc{J},\mf{v}\right\} \\
\dann\left\{ \mc{J},\left\{ \mc{J},\mf{v}\right\} \right\}  & = & 4\mc{J}^{2}\mf{v}=-4\mf{v}=\left\{ \left\{ \mf{v},\mc{J}\right\} ,\mc{J}\right\} \\
\left\{ \left\{ \mf{v},\mc{J}\right\} ,\left\{ \mc{J},\mf{w}\right\} \right\}  & = & -4\mf{v}^{K}\mf{w}_{K}=-4\left\{ \mf{v},\mf{w}\right\} \end{eqnarray}
If both objects are of higher rank, however, antisymmetrization of
the remaining indices modifies the result. We thus have to be careful
with the following examples \rem{the first does not enter the calculations!}\begin{eqnarray}
\left\{ \mc{J},\mc{J}\right\}  & = & 4\mc{J}_{\bs{M}}\hoch{K}\mc{J}_{K\bs{M}}=-4\mc{G}_{\bs{M}\bs{M}}=0\quad(!\textrm{ because of antisymmetrization)}\label{eq:doesnt-enter}\\
\left\{ \mc{J},\left\{ \mc{J},\de\mf{v}\right\} \right\}  & = & \mc{J}_{\bs{M}}\hoch{K}\mc{J}_{[K|}\hoch{L}(\de\mf{v})_{L|\bs{M}]}\neq-4\de\mf{v}\end{eqnarray}
As mentioned earlier, the Dorfman bracket (\ref{eq:Dorfman-bracket})
used in our integrability condition is just the derived bracket of
the algebraic bracket. I.e. we have \begin{eqnarray}
\left[\mf{v}\bs{,}\mf{w}\right] & = & \left[\de\mf{v},\mf{w}\right]^{\Delta}=\label{eq:Dorf-ist-der}\\
 & = & \left[\de\mf{v},\mf{w}\right]_{(1)}^{\Delta}+\underbrace{\sum_{p\geq2}\left[\de\mf{v},\mf{w}\right]_{(p)}^{\Delta}}_{=0}=\\
 & = & \left\{ \de\mf{v},\mf{w}\right\} \label{eq:Dorf-ist-der-of-big}\end{eqnarray}
where the differential $\de$ has to be understood in the extended
sense of (\ref{eq:exterior-derivative-via-BRST},\ref{eq:dK}), namely
as Poisson-bracket with the BRST-like generator \begin{eqnarray}
\oo & = & \basis^{M}p_{M}=\ce^{m}p_{m}\stackrel{\textrm{locally }}{=}\de(x^{m}p_{m})=-\de(\ce^{m}\be_{m})\\
p_{M} & \equiv & (p_{m},0)\\
\de\mf{v} & \equiv & \left\{ \oo,\mf{v}\right\} =\partial_{\bs{M}}v_{\bs{M}}+\mf{v}^{K}p_{K}\end{eqnarray}
where $p_{m}$ is the conjugate variable to $x^{m}$. We can now rewrite
the integrability condition (\ref{eq:generalized-integrabilityII})
as \begin{eqnarray}
\left\{ \de\mf{v},\mf{w}\right\} -\frac{1}{4}\left\{ \de\left\{ \mc{J},\mf{v}\right\} ,\left\{ \mc{J},\mf{w}\right\} \right\} +\frac{1}{4}\left\{ \mc{J},\left\{ \de\left\{ \mc{J},\mf{v}\right\} ,\mf{w}\right\} \right\} +\frac{1}{4}\left\{ \mc{J},\left\{ \de\mf{v},\left\{ \mc{J},\mf{w}\right\} \right\} \right\}  & \stackrel{!}{=} & 0\label{eq:generalized-integrability-in-brackets}\end{eqnarray}
 Remember that the Poisson bracket is a graded one, and $\mf{v},\mf{w}$
and $\de$ are odd, while $\mc{J}$ is even. \rem{hier schlummert ein Einschub ueber das graded equal sign} 

Let us now start with applying Jacobi to the second term of (\ref{eq:generalized-integrability-in-brackets})
\begin{eqnarray}
-\frac{1}{4}\left\{ \de\left\{ \mc{J},\mf{v}\right\} ,\left\{ \mc{J},\mf{w}\right\} \right\}  & = & -\frac{1}{4}\left\{ \left\{ \de\left\{ \mc{J},\mf{v}\right\} ,\mc{J}\right\} ,\mf{w}\right\} -\frac{1}{4}\left\{ \mc{J},\left\{ \de\left\{ \mc{J},\mf{v}\right\} ,\mf{w}\right\} \right\} \end{eqnarray}
so that we get\begin{eqnarray}
0 & \stackrel{!}{=} & \left\{ \de\mf{v},\mf{w}\right\} -\frac{1}{4}\left\{ \left\{ \de\left\{ \mc{J},\mf{v}\right\} ,\mc{J}\right\} ,\mf{w}\right\} +\frac{1}{4}\left\{ \mc{J},\left\{ \de\mf{v},\left\{ \mc{J},\mf{w}\right\} \right\} \right\} =\\
 & = & \left\{ \de\mf{v},\mf{w}\right\} -\frac{1}{4}\left\{ \left\{ \left\{ \de\mc{J},\mf{v}\right\} ,\mc{J}\right\} ,\mf{w}\right\} -\frac{1}{4}\left\{ \left\{ \left\{ \mc{J},\de\mf{v}\right\} ,\mc{J}\right\} ,\mf{w}\right\} +\frac{1}{4}\left\{ \mc{J},\left\{ \de\mf{v},\left\{ \mc{J},\mf{w}\right\} \right\} \right\} =\\
 & = & \left\{ \de\mf{v},\mf{w}\right\} -\frac{1}{4}\left\{ \left\{ \left\{ \mf{v},\de\mc{J}\right\} ,\mc{J}\right\} ,\mf{w}\right\} +\frac{1}{4}\left\{ \left\{ \left\{ \de\mf{v},\mc{J}\right\} ,\mc{J}\right\} ,\mf{w}\right\} +\frac{1}{4}\left\{ \mc{J},\left\{ \de\mf{v},\left\{ \mc{J},\mf{w}\right\} \right\} \right\} \label{eq:kurzePause}\end{eqnarray}
It would be nice to separate $\mf{w}$ completely by moving it for
the last term into the last bracket like in the first three terms.
We thus consider only the last term for a moment and calculate it
in two different ways (first using Jacobi for second and third bracket
and after that using Jacobi for first and second bracket):\begin{eqnarray}
\frac{1}{4}\left\{ \mc{J},\left\{ \de\mf{v},\left\{ \mc{J},\mf{w}\right\} \right\} \right\}  & \stackrel{1.}{=} & \frac{1}{4}\left\{ \mc{J},\left\{ \left\{ \de\mf{v},\mc{J}\right\} ,\mf{w}\right\} \right\} +\frac{1}{4}\left\{ \mc{J},\left\{ \mc{J},\left\{ \de\mf{v},\mf{w}\right\} \right\} \right\} =\\
 & = & \frac{1}{4}\left\{ \mc{J},\left\{ \left\{ \de\mf{v},\mc{J}\right\} ,\mf{w}\right\} \right\} -\left\{ \de\mf{v},\mf{w}\right\} \\
 & \stackrel{2.}{=} & \frac{1}{4}\left\{ \left\{ \mc{J},\de\mf{v}\right\} ,\left\{ \mc{J},\mf{w}\right\} \right\} +\frac{1}{4}\left\{ \de\mf{v},\left\{ \mc{J},\left\{ \mc{J},\mf{w}\right\} \right\} \right\} =\\
 & = & \frac{1}{4}\left\{ \mc{J},\left\{ \left\{ \mc{J},\de\mf{v}\right\} ,\mf{w}\right\} \right\} +\frac{1}{4}\left\{ \left\{ \left\{ \mc{J},\de\mf{v}\right\} ,\mc{J}\right\} ,\mf{w}\right\} -\left\{ \de\mf{v},\mf{w}\right\} =\\
 & = & -\frac{1}{4}\left\{ \mc{J},\left\{ \left\{ \de\mf{v},\mc{J}\right\} ,\mf{w}\right\} \right\} +\left\{ \de\mf{v},\mf{w}\right\} -2\left\{ \de\mf{v},\mf{w}\right\} +\frac{1}{4}\left\{ \left\{ \left\{ \mc{J},\de\mf{v}\right\} ,\mc{J}\right\} ,\mf{w}\right\} \end{eqnarray}
Comparing both calculations yields\begin{eqnarray}
\frac{1}{4}\left\{ \mc{J},\left\{ \de\mf{v},\left\{ \mc{J},\mf{w}\right\} \right\} \right\}  & = & -\frac{1}{8}\left\{ \left\{ \mc{J},\left\{ \mc{J},\de\mf{v}\right\} \right\} ,\mf{w}\right\} -\left\{ \de\mf{v},\mf{w}\right\} \end{eqnarray}
We can plug this back in (\ref{eq:kurzePause}) and leave away the
outer bracket with $\mf{w}$:\begin{eqnarray}
0 & \stackrel{!}{=} & \de\mf{v}-\frac{1}{4}\left\{ \left\{ \mf{v},\de\mc{J}\right\} ,\mc{J}\right\} +\frac{1}{4}\left\{ \left\{ \de\mf{v},\mc{J}\right\} ,\mc{J}\right\} -\frac{1}{8}\left\{ \mc{J},\left\{ \mc{J},\de\mf{v}\right\} \right\} -\de\mf{v}=\\
 & = & -\frac{1}{4}\left\{ \left\{ \mf{v},\de\mc{J}\right\} ,\mc{J}\right\} +\frac{1}{8}\left\{ \left\{ \de\mf{v},\mc{J}\right\} ,\mc{J}\right\} =\\
 & = & -\frac{1}{8}\left\{ \left\{ \mf{v},\de\mc{J}\right\} ,\mc{J}\right\} +\frac{1}{8}\left\{ \de\left\{ \mf{v},\mc{J}\right\} ,\mc{J}\right\} =\\
 & = & -\frac{1}{8}\left\{ \left\{ \mf{v},\de\mc{J}\right\} ,\mc{J}\right\} +\frac{1}{8}\de\left\{ \left\{ \mf{v},\mc{J}\right\} ,\mc{J}\right\} +\frac{1}{8}\left\{ \left\{ \mf{v},\mc{J}\right\} ,\de\mc{J}\right\} =\\
 & = & -\frac{1}{8}\left\{ \mf{v},\left\{ \de\mc{J},\mc{J}\right\} \right\} -\frac{1}{2}\de\mf{v}=\\
 & = & \frac{1}{8}\Big(\big\{\left[\mc{J},_{\de}\mc{J}\right]_{(1)}^{\Delta},\mf{v}\big\}-4\de\mf{v}\Big)=\\
 & = & \frac{1}{8}\big\{\left[\mc{J},_{\de}\mc{J}\right]_{(1)}^{\Delta}-4\oo,\mf{v}\big\}\end{eqnarray}
where we used \begin{eqnarray}
\de\mf{v} & = & \left\{ \oo,\mf{v}\right\} \end{eqnarray}
The integrability condition is thus (explaining the normalization
of $\mc{N}$ of above) as promised in (\ref{eq:relation-between-Nij-und-der-big})\begin{equation}
\boxed{\mc{N}\equiv\left[\mc{J},_{\de}\mc{J}\right]_{(1)}^{\Delta}-4\oo\stackrel{!}{=}0}\label{eq:integrability-big-derived}\end{equation}
The derived bracket $\left[\mc{J},_{\de}\mc{J}\right]_{(1)}^{\Delta}$
indeed contains the term $4\oo=4\basis^{M}p_{M}$ \rem{(compare ...) Querverweis!!Achtung! Derived bracket ist nicht die volle, sondern nur die von big bracket abgeleitete!}
which therefore is exactly cancelled. 

Precisely the same calculation can be performed by calculating with
the complete algebraic bracket $\left[\,,\,\right]^{\Delta}$ instead
of the Poisson-bracket, its first order part. Similarly to above,
we have \begin{eqnarray}
\mc{J}\mf{v} & \equiv & \frac{1}{2}[\mc{J},\mf{v}]^{\Delta}\\
\dann[\mc{J},[\mc{J},\mf{v}]^{\Delta}]^{\Delta} & = & 4\mc{J}^{2}\mf{v}=-4\mf{v}\end{eqnarray}
 In combination with (\ref{eq:Dorf-ist-der}) this is enough to redo
the same calculation and get as integrability condition (using $\left[\mc{J}\bs{,}\mc{J}\right]\equiv-[\de\mc{J},\mc{J}]^{\Delta}$)\begin{equation}
\boxed{\mc{N}\equiv\left[\mc{J}\bs{,}\mc{J}\right]-4\oo\stackrel{!}{=}0}\label{eq:integrability-derived}\end{equation}
which also proves that the derived bracket bracket of the big bracket
(which is not necessarily geometrically well defined) coincides in
this case with the complete derived bracket\begin{eqnarray}
\left[\mc{J},_{\de}\mc{J}\right]_{(1)}^{\Delta} & = & \left[\mc{J}\bs{,}\mc{J}\right]\end{eqnarray}
As discussed in (\ref{eq:relation-Buttin-derived}) and (\ref{eq:relation-Buttin-derived-on-Tensor-level}),
throwing away the $\de$-closed part corresponds to taking Buttin's
bracket instead of the derived one. Remember that $\oo=\de x^{k}p_{k}=-\de(\de x^{k}\wedge\pe_{k})$,
s.th. $\de\oo=0$. We can thus equally write \begin{eqnarray}
\mc{N} & = & \left[\mc{J}\bs{,}\mc{J}\right]_{B}\label{eq:Nijenhuis-Buttin}\end{eqnarray}

\rem{Spinor-Chapter hier versteckt} 

}
\Teil{Bibliography}{
\addcontentsline{toc}{section}{\refname}
\providecommand{\href}[2]{#2}\begingroup\raggedright\endgroup

}
\end{document}